\documentclass[twocolumn]{aastex701}
\usepackage{amsmath}	% Advanced maths commands
\usepackage{subcaption}  % in your preamble
\usepackage{comment}

\newcommand{\g}{G}
\received{2025 December 10}
\revised{2026 October 6}
\accepted{2026 October 9}
\published{2026 November 17}
\begin{document}

\title{Mean-motion Resonances in the Accretion Disks of Active Galactic Nuclei}

\author[orcid=0000-0001-9310-7808]{Marguerite Epstein-Martin}
\affiliation{Department of Astronomy, Columbia University, 550 West 120th Street, New York, NY 10027, USA}
\email[show]{mae2153@columbia.edu}  

\author[orcid=0000-0002-4337-9458]{Nicholas Stone}  
\affiliation{Department of Astronomy, University of Wisconsin--Madison, Madison, WI 53706, USA}
\affiliation{Racah Institute of Physics, The Hebrew University, Jerusalem, 91904, Israel}
\email[hide]{}  

\author[orcid=0000-0002-7733-4522]{Juliette Becker} 
\affiliation{Department of Astronomy, University of Wisconsin--Madison, Madison, WI 53706, USA}
\email[hide]{}

\begin{abstract}
Mean-motion resonances (MMRs) are a generic outcome of convergent migration for bodies embedded in accretion disks around a central mass. Long studied in planetary systems, the same phenomenon should occur for stellar-mass black holes (BHs) in the disks of active galactic nuclei (AGN). In this work, we derive simple analytic criteria describing when BH pairs are driven out of resonance, and use them to chart MMR stability across AGN parameter space, accounting for disruption from general-relativistic apsidal precession, hydrodynamic turbulence, and stellar stirring. Across plausible AGN disk models, we find three MBH mass regimes: (i) for $M/  M_\odot\gtrsim 10^{7.5}$, first-order resonances are generically unstable; (ii) for $M/ M_\odot\lesssim 10^{6.5}$, stable MMRs are always present; and (iii) for $10^{6.5}\lesssim M / M_\odot \lesssim 10^{7.5}$, stability depends on disk mass flux, the summed mass of the orbiters, and the nuclear-cusp slope. When present, stable MMRs commonly occur between an inner anti-trap and an outer trap set by thermal torques, a region where embedded objects migrate outward in the disk. These results imply that high-mass AGN allow convergent migration to proceed to LIGO-Virgo-KAGRA (LVK)-band mergers largely without resonant chains, whereas low- and intermediate-mass AGN can host MMRs, with the potential to reshape merger pathways. 
\end{abstract}



\section{Introduction}\label{sec:introduction}

The groundbreaking discovery of gravitational waves (GWs) by Advanced LIGO in 2015 inaugurated a new era in high-energy astrophysics \citep{2016abbott}.  Since the first observed compact binary merger, the LIGO-Virgo-KAGRA (LVK) collaboration has published over 200 additional GW signals \citep{2023abbott, 2025LVKa}, the overwhelming majority of which are from binary mergers of stellar-mass black holes (BHs).  In contrast to the small minority of binary neutron star mergers, whose event rates and origins had been precalibrated by short gamma-ray bursts \citep{2012coward}, the detection of so many binary BHs came as a surprise.  

A decade later, there is still no consensus on the astrophysical origins of the LVK BH mergers, and half a dozen scenarios still contend to explain this population.  These scenarios can be broken into roughly three categories.  First, isolated binary evolution channels, such as a common-envelope phase \citep{1993tutukov, 2016belczynski}, chemically homogeneous evolution following tidal synchronization \citep{2016mandel}, or stable mass transfer \citep{2022vanson}.  Second, vacuum gravitational dynamics channels, such as chaotic three-body scatterings in dense star clusters \citep{2000portegieszwart, 2016rodriguez} or Kozai-Lidov cycles in hierarchical triples \citep{2012antonini, 2017antonini}.  In this paper, we focus on the third type of binary-BH origin: hydrodynamic assembly and merger of binaries in the gas-rich environments of active galactic nuclei \citep[AGN;][]{2014mckernan, 2016bellovary, 2017stone, 2017bartos}.

In the ``AGN channel'' for binary-BH mergers, there are multiple ways for the BH binaries to assemble.  They may be formed {\it in situ} via Toomre instability \citep{2017stone}, they may be captured by gas drag \citep{2017bartos} from a preexisting binary population in the nuclear star cluster (NSC), or they may be assembled via single-single capture within the AGN disk itself.  These single-single captures are mediated by various forms of dissipation \citep{2020tagawa, 2023rowan, 2024wang} during close passages between isolated BHs.  BHs embedded in the AGN typically move on quasi-circular orbits\footnote{The minority of BHs embedded on retrograde orbits are an exception to this \citep{2021secunda}.}, so these close passages require {\it convergent migration}.  

Protoplanetary disk-bearing systems offer a useful analog to the AGN channel \citep{2012mckernan}. Exchange of angular momentum between embedded planets and the gaseous disk leads to planetary migration \citep{1980goldreich}; the exact same forms of gravito-hydrodynamical angular-momentum transport lead to BH migration in AGN.  Migration rates are mass dependent \citep{2013paardekooper}, so BHs of different masses may overtake one another, setting the stage for single-single capture.  This can occur in the bulk of the disk, or alternatively at special locations known as migration traps, where the sign of the migration torque flips and a region of inward migration sits exterior to one of outward migration \citep{2016bellovary}.  Regardless, AGN disk conditions that favor single-single capture will likely lead to large enhancements in the LVK merger rate (relative to AGN channel mergers limited to primordial binaries), along with the tantalizing possibility of hierarchical mergers and BH growth into the pair-instability mass gap \citep{2019yang, 2021tagawa}.

Further consideration of the analogy between protoplanetary and AGN disks also leads to an equivalent AGN channel for extreme-mass-ratio inspirals \citep[EMRIs;][]{2018amaroseoane}.  EMRIs are a future class of GW sources produced when a stellar-mass compact object (typically but not necessarily a BH) inspirals into a massive BH in a galactic nucleus.  EMRI waveforms do not extend to the high GW frequencies observable by the LVK collaboration, but would be visible to space-based GW interferometers like the future {\it LISA} \citep{2023amaroseoane} or {\it Taiji/TianQin} \citep{2016luo, 2020ruan} missions.  While most theoretical calculations of EMRI formation focus on scattering of BHs onto high-eccentricity ``loss cone'' orbits \citep{2005hopman}, orbital migration may lead to ``wet EMRI'' production in AGN disks \citep{2007levin, 2021panA}.  These wet EMRIs would have relatively unique signatures \citep{2022pan} and may even dominate overall EMRI rates \citep{2021panB, 2023derdzinski}, but their properties depend on the outcomes of AGN migration.

In planetary systems, migration among planetary pairs is broadly expected to be slow and convergent, theoretically guaranteeing capture into mean-motion resonances \citep[MMRs;][]{1982henrard, 2015batygin}, and simulations of disk-driven migration routinely produce compact chains of planets locked in first-order resonances \citep[e.g.,][]{2002lee, 2007terquem, 2006cresswell, 2008cresswell}. Observations only partially bear this out. Within our own solar system, Jupiter and Saturn may have once occupied a resonant state \citep{2007morbidelli}, before undergoing instabilities that drove orbits to their current locations \citep{2010batygin}. Likewise, \citet{1995malhotra} proposed that this process may have been the origin of the 3:2 MMR between Neptune and Pluto. In the larger exoplanet population, however, resonances are the exception. The period-ratio distribution of adjacent pairs in the {\it Kepler} multitransiting systems is broadly smooth \citep{2011lissauer, 2014fabrycky}, with a small but significant excess of planetary pairs just wide of the first-order resonances 2:1 and 3:2 (by 1\%-2\%), and a shortage just narrow of them (see Figure 4 of \citealt{2014fabrycky}). 

The shortage of resonant pairs in the {\it Kepler} data relative to theoretical expectations is now understood to stem principally from postdisk dispersal dynamics. While the disk is present, its eccentricity damping stabilizes resonant chains; once it is gone, the compact chains produced by migration are prone to dynamical instability \citep{2017izidoro, 2021izidoro, 2022goldberg}. Consistent with this evolutionary picture, planets in young ($\lesssim 100 \, \rm Myr$) systems are markedly more likely to lie near resonance than those in mature systems \citep{2024dai}.

Resonances can also be prevented or broken while the disk is still present. Nonaxisymmetric disk potentials and rapid migration reduce the probability of capture into resonances \citep{2006Quillen, 2015batygin}, and where a pair does capture, the competition between migration-driven eccentricity excitation and disk damping can render the resonant libration overstable, so that the pair escapes and settles just wide of commensurability \citep{2014goldreich, 2015delisle, 2015deck}. Most relevant to the present work, stochastic gravitational forcing by turbulent density fluctuations perturbs the orbits of embedded bodies \citep{2004laughlin, 2004nelson} and, when sufficiently strong, unbinds resonances outright \citep{2004nelson, 2017batygin}.
Turbulent forcing combined with smooth migration has been shown to help reproduce the period-ratio distribution of the {\it Kepler} sample \citep{2012rein}. While more recent work indicates that turbulence is unlikely to account for the overall scarcity of resonances \citep{2017batygin}, it does appear to shape the architecture of the pairs that remain near them: \citet{2023goldberg} find that near-resonant {\it Kepler} pairs sit systematically far from the resonant forced equilibrium, with resonant angles that circulate beyond $\approx 0.6\%$ wide of exact commensurability -- a state inconsistent with placement by disk damping or tides alone, but reproduced when stochastic stirring by turbulent eddies is added to the dissipative evolution.

The few-body numerical integrations of \citet{2019secunda, 2020secunda} have demonstrated that stellar-mass BHs can likewise capture into MMR while migrating toward traps in AGN disks\footnote{More recent works have also found evidence for the capture of stellar-mass BHs into MMR with {\it intermediate-mass} BHs migrating inwards due to either GW emission \citep{2025reved} or gas torques \citep{2025peng}, although we do not consider intermediate-mass BH secondaries in this paper.}. As in the planetary case, however, the strength and stability of these resonances is an open question, and the AGN environment differs from the protoplanetary one in ways that bear directly on the stochastic channel.  Protoplanetary disks are cold and weakly ionized,  with turbulent viscosities observationally constrained to $\alpha_{\rm disk} \sim 10^{-4}$--$10^{-3}$ \citep{2023rosotti}. In contrast, AGN disks are hot, well ionized, and turbulent throughout, with standard models adopting $\alpha_{\rm disk} \sim 0.01$--$0.3$ \citep{sirko2003, 2005thompson}. Migration theory developed for protoplanetary disks must therefore be taken in its high-viscosity, turbulent limit, and stochastic forcing, whose amplitude scales with $\alpha_{\rm disk}$, should be correspondingly more important for embedded BHs than for planets. Additional sources of instability, absent from the protoplanetary context, also emerge for BHs in MMR in AGN: weak gravitational perturbations from stars in the disk or the surrounding NSC \citep{2017stone}, and the increased importance of general-relativistic (GR) apsidal precession near the massive black hole \citep[MBH;][]{2010seto}.

Whether stable resonant locking occurs in AGN disks is critical for determining the source of LVK BH mergers: A pair of BHs captured into MMR is held at a fixed period ratio and cannot be driven to the close approaches on which single-single capture depends. If an $\mathcal{O}(1)$ fraction of migrating BHs were to capture into long-lived resonant chains, the maximum binary-BH merger rate from the AGN channel would be substantially reduced, hierarchical growth suppressed, and the mass function of wet EMRIs altered \citep{2022pan}. Conversely, if resonances break readily, convergent migration proceeds to close encounters largely unimpeded.

In this paper, we systematically explore the landscape of MMR stability in AGN disks. Our approach treats resonance survival as a competition between the restoring dynamics of the resonance and the processes that act to disrupt it, evaluated analytically so that we can survey the broad parameter space of AGN, which spans many orders of magnitude in MBH mass and gas accretion rate. Two classes of disruption enter. The first is deterministic: Near the MBH, differential relativistic apsidal precession competes with resonant libration and can detune the resonance outright. The second is stochastic: Turbulent density fluctuations in the disk and gravitational scattering by stars in the nuclear cluster both diffuse a resonant pair's period ratio away from commensurability, while convergent migration acts to restore it. Whether a resonance survives the stochastic channel therefore depends on the ratio of the diffusion rate to the migration rate, and the latter is itself set by the full complement of torques acting on an embedded BH: Type I/II gas torques, GW emission, and thermal torques, which can reverse the sign of migration and create migration traps.

The paper is organized to incorporate these interrelated processes in turn. In \S\ref{sec:analytical_model}, we construct the resonant Hamiltonian and derive the condition under which GR precession breaks the resonance. In \S\ref{sec:migration}, we assemble the migration torques and their timescales. In \S\ref{sec:stochastic_forces}, we quantify the two sources of stochastic perturbation -- stellar scattering and disk turbulence -- and combine them into a single effective diffusion coefficient. In \S\ref{sec:criterion_resonance_disruption}, we derive an analytic criterion for resonance disruption under stochastic forcing. In \S\ref{sec:discussion}, we apply our criteria across AGN parameter space and discuss the implications for GW production. Finally, in \S \ref{sec:conclusions}, we conclude with a summary of our main findings.

\section{Mean-motion Resonances: Setup of the Integrable Problem}\label{sec:analytical_model}

We begin our analysis by introducing the integrable approximation for first-order resonant motion (i.e., the orbital period ratio of $P_2/P_1\simeq \left(k+1\right)/k$). This approach has been widely used in the planetary dynamics literature (e.g., \citealt{1984sessin, 2013deck, Batygin2013, 2015deck}) but is less common in the context of AGN, although their geometries are analogous.  The system of interest includes two coplanar, embedded objects of mass $m_1$ and $m_2$ orbiting a central mass $M$, with periods near the first-order ratio, $k+1:k$. In this regime, the Hamiltonian can be approximated as
\begin{equation}\label{eqn:H}
\begin{split}
\mathcal{H} = &-\frac{G M m_1}{2 a_1} - \frac{G M m_2}{2 a_2} - \frac{G m_1 m_2}{a_2} \\&
 \times [ f_{k+1, 27}(\alpha) e_1 \cos{\left(\theta - \varpi_1\right)} \\&
+ \left( f_{k+1, 31} (\alpha) - 2\delta_{k, 1}\alpha\right) e_2 \cos{(\theta - \varpi_2)} ] \,,
\end{split}
\end{equation}
where $\theta = \left(k + 1\right) \lambda_2 - k \lambda_1$, and $a_i$, $e_i$, $\lambda_i$, and $\varpi_i$ are the semimajor axes, eccentricities, mean longitudes, and longitudes of periapsis of the orbiters, respectively. The quantities $f_{k+1, 27}$ and $f_{k+1, 31}$ are dimensionless functions of Laplace coefficients \citep{MurrayDermott1999} evaluated at $\alpha \equiv a_1/a_2$. At the resonance center, $a_1 = \left[k/\left(k+1 \right)\right]^{2/3} a_2$. Expanding around this equilibrium point allows us to approximate our coefficients as functions of $k$:
\begin{equation}
\begin{split}
&f_{k+1, 27} \approx -0.8 k \\ 
&f_{k+1, 31}\approx 0.8 k\,,
\end{split}
\end{equation}
as shown in \citet{2011quillen}. Finally, the term including the Kronecker delta function, $\delta_{k,1}$, comes from the indirect part of the disturbing function and contributes only to the 2:1
resonance ($k = 1$).

Keplerian orbital elements are not canonical, so we transform Equation~\eqref{eqn:H} into Poincaré action-angle variables: 
\begin{equation}
\begin{aligned}
\Lambda_i &= m_i \sqrt{G M a_i} 
    &&\quad \lambda_i\\
\Gamma_i &= \Lambda_i \left(1 - \sqrt{1 - e_i^2} \right) \approx \Lambda_i \frac{e_i^2}{2} 
    &&\quad \gamma_i = -\varpi_i\,
\end{aligned}
\end{equation}
where we have approximated $\Gamma_i \approx \Lambda_i e_i^2/2$, an assumption valid for nearly circular orbits. We adopt this near-circular limit throughout the paper. The Keplerian and resonant components of the Hamiltonian are rewritten as
\begin{equation}
\mathcal{H}_{\rm kep} = - \frac{G^2 M^2m_1^3}{2 \Lambda_1^2} - \frac{G^2 M^2m_2^3}{2 \Lambda_2^2}\,
\end{equation}
\begin{equation}
\begin{aligned}
\mathcal{H}_{\rm res} =\ 
& - \frac{G^2 M m_1 m_2^3}{\Lambda_2^2} \Bigg( 
    f_{k+1, 27} \sqrt{\frac{2\Gamma_1}{\Lambda_1}} \cos{(\theta + \gamma_1)} \\
& \quad\quad\quad\quad + 
    f_{k+1, 31}' \sqrt{\frac{2\Gamma_2}{\Lambda_2}} \cos{(\theta + \gamma_2)} 
\Bigg)\,,
\end{aligned}
\label{eq:res_hammy_transformed}
\end{equation}
where we have defined $f_{k+1, 31}' \equiv f_{k+1, 31} - 2 \delta_{k,1}\alpha$ for simplicity. 

Following the procedure laid out in \citet{Batygin2013}, we further simplify the Hamiltonian by employing an additional canonical transformation using the action-angle variables
\begin{equation}
\begin{aligned}
\mathcal{K} &= \Lambda_1 + \frac{k}{k+1}\Lambda_2
    &&\quad \kappa = \lambda_1\,, \\
\Theta &= \Lambda_2/(k+1) 
    &&\quad \theta = (k+1) \lambda_2 - k\lambda_1\,.
\end{aligned}
\end{equation}
We then expand the Hamiltonian around the nominal resonant location, carrying out the expansion to second order in $\delta\Lambda_i = \Lambda_i - \left[\Lambda_i\right]$, where $\left[\Lambda_i\right] = m_i\left[n_i\right]  \left[a_i^2\right]$ is the nominal resonance value and $\left[n_i\right] = \sqrt{G M/\left[a_i \right]^3}$ is the nominal mean motion. Using the resonant relationship $k \left[n_1\right] = (k+1)\left[n_2\right]$, the Keplerian Hamiltonian becomes
\begin{equation}
    \begin{split}
        \mathcal{H}_{\rm kep} =& 4 \left[n_1\right]\mathcal{K} + 3 k \left[h_1\right] \mathcal{K}\Theta -\frac{3}{2}\Theta^2  \\
        & \times \left(\left[h_1\right] k^2 +  \left[h_2\right]\left(k + 1\right)^2 \right) -\frac{3}{2} \mathcal{K}^2\left[h_1\right]\,,
    \end{split}
    \label{MMRhammy}
\end{equation}
where $\left[h_i\right] = \left[n_i\right]/\left[\Lambda_i\right] = 1/(m_i \left[a_i\right]^2)$. The resonant contribution to $\mathcal{H}$ takes the form
\begin{equation}
    \mathcal{H}_{\rm res} = -\zeta \sqrt{2\Gamma_1} \cos{(\gamma_1 +\theta)} -\beta \sqrt{2\Gamma_2} \cos{(\gamma_2 +\theta)}\,,
    \label{eq:basichammy}
\end{equation}
where the prefactors $\zeta$ and $\beta$ are 
\begin{equation}
    \begin{split}
        &\zeta = \frac{G^2 M m_1 m_2^3}{\left[\Lambda_2\right]^2} \frac{f_{k+1, 27}}{\sqrt{\left[\Lambda_1\right]}} = \frac{f_{k +1, 27}}{f'_{k +1, 31}}\sqrt{\frac{\left[ \Lambda_2\right]}{\left[ \Lambda_1\right]}} \beta\,.
    \end{split}
\end{equation}
Critically, $\kappa$ is no longer in $\mathcal{H}$, thus $\mathcal{K}$ is a constant of motion and we can drop the first and last terms in $\mathcal{H_{\rm kep}}$. We are left with three sets of conjugate coordinate pairs, too many for integrability. 

The Hamiltonian can be reduced to a single degree of freedom by exploiting a symmetry in the resonant problem: To leading order, the disturbing function depends only on a linear combination of the eccentricity vectors of the original orbits \citep{2019hadden, 2025tomayo}. In this sense, the geometry and dynamics are equivalent to the circular restricted case \citep{2006Quillen}, with the test particle’s eccentricity replaced by a generalized composite eccentricity. This much simpler coordinate system is defined by a canonical rotation (e.g., \citealt{1984sessin, 1986wisdom, Batygin2013}), with associated action-angle coordinates: 
\begin{equation}
\begin{aligned}
\Phi &= \frac{\left(\zeta^2\Gamma_1+\beta^2\Gamma_2+2\zeta\beta\sqrt{\Gamma_1\Gamma_2}\cos (\gamma_1 - \gamma_2)\right)}{\zeta^2 + \beta^2} , \\[4pt]
\phi &= \arctan\!\left(\frac{\zeta\sqrt{2\Gamma_1}\sin\gamma_1+\beta\sqrt{2\Gamma_2}\sin\gamma_2}{\zeta\sqrt{2\Gamma_1}\cos\gamma_1+\beta\sqrt{2\Gamma_2}\cos\gamma_2}\right)\,.
\end{aligned}
\end{equation}
Rewritten in terms of these coordinates, the Hamiltonian is reduced to a form that contains only a single angle:
\begin{equation}
\begin{split}
  \mathcal{H} =& 3k\left[h_1\right]\mathcal{K} \Theta - \frac{3}{2}\left(\left[h_1\right] k^2 + \left[h_2\right]\left(k + 1\right)^2\right) \Theta^2 \\ &- \sqrt{\zeta^2 + \beta^2}\sqrt{2 \Phi} \cos{(\phi + \theta)}\,.
\end{split}
\end{equation}
Employing a final contact transformation yields the action-angle variables 
\begin{equation}
\begin{aligned}
\Psi &= \Phi\,,
    &&\quad \psi = \phi + \theta\,, \\
\Omega &= \Theta - \Psi\,,
    &&\quad w = \theta\,.
\end{aligned}
\end{equation}
In these coordinates the Hamiltonian is independent of $w$, making $\Omega$ the final conserved quantity. Our system is thus reduced to a single degree of freedom. Dropping constant terms, the Hamiltonian is recast in integrable form as
\begin{equation}
\begin{split}
  \mathcal{H} =& -3 \left( \left[h_2\right] (k+1)^2 \Omega - k\left[h_1\right]  (\mathcal{K} - k \Omega)\right) \Psi \\
  &- \frac{3}{2} \left(\left[h_1\right] k^2 +  \left[h_2\right](k+1)^2\right)\Psi^2\\
  & - \sqrt{\zeta^2 + \beta^2} \sqrt{2 \Psi} \cos{(\psi)}\,.
\end{split}
\end{equation}

We can simplify this expression further by
rescaling the actions by a constant factor $\eta$, chosen so that the coefficient of $\Psi^2$ is half that of the harmonic term $\sqrt{2\Psi}$. This gives
\begin{equation}
    \eta = \left( \frac{\zeta^2 + \beta^2}{9 ([h_1]k^2 + [h_2](k + 1)^2)^2}\right)^{1/3}\,.
\end{equation}
In the ``compact limit'' approximation \citep{1986henon, 2013deck} -- i.e., $\langle a \rangle \equiv a_1\simeq a_2$ (equivalently, $k +1 \simeq k$) -- the same rescaling can be written in terms of the masses as
\begin{equation}
    \eta = \left( \frac{\zeta^2\langle a\rangle^4 m_1^2 m_2(m_1 + m_2)}{9 \left( m_2k^2 +  m_1(k + 1)^2 \right )^2}\right)^{1/3}\,.
\end{equation}
This approximation agrees well with $N$-body integrations, particularly for $k \geq 2$ \citep{2013deck}. The rescaled actions are defined as
\begin{equation}
\begin{aligned}
    \widetilde{\mathcal{H}} = \mathcal{H}/\eta  && \widetilde{\Psi} = \Psi/\eta && \widetilde{\mathcal{K}} = \mathcal{K}/\eta && \widetilde{\Omega} = \Omega/\eta\,.
\end{aligned}
\end{equation}
By choosing to measure time in units of $2 \,m_1m_2 \langle a \rangle^2 / (3 \eta \, (m_2 k^2 + m_1 (k + 1)^2)$, we divide $\mathcal{H}$ by the same factor and obtain 
\begin{equation}\label{eqn:Hfinal}
 \mathcal{H} = 3 (\delta + 1) \widetilde{\Psi} - \widetilde{\Psi}^2 - 2\sqrt{2 \widetilde{\Psi}} \cos{\left(\psi \right)}\,.
\end{equation}

Although cumbersome to derive, Equation~\eqref{eqn:Hfinal} has the form of the second-fundamental model of resonance as described in \citet{1983henrard}. The phase-space morphology is governed entirely by the resonant proximity parameter $\delta$. For $\delta < 0 $ the system behaves like a forced harmonic oscillator, with circulating $\psi$. When $\delta \geq 0$, the system is pendulum-like, and librates within the emergent, resonant island. 

Following \citet{2017batygin}, it is useful to define the Hamiltonian variables in terms of physical quantities. $\delta$ is
\begin{equation}
    \delta = \frac{1}{3} \left( \frac{15}{4} \frac{k M}{m_1 + m_2}\right)^{2/3} \left(\sigma_c^2 - \frac{\Delta \alpha}{k}\right) 
\end{equation}
where $\Delta \alpha = \Delta (a_1/a_2)$ denotes the deviation of the semimajor-axis ratio from its nominal resonant value. The action-angle coordinates are  
\begin{equation}
    \begin{aligned}
        \widetilde{\Psi}  = \frac{\sigma_c^2}{2} \left( \frac{15}{4} \frac{k M}{m_1 + m_2}\right)^{2/3},\\
        \psi = (k + 1) \lambda_2 - k \lambda_1 - \varpi_c\,,
    \end{aligned}
\label{eqn:tildePsi_and_tildepsi}
\end{equation}
where the composite eccentricity $\sigma_c$ and generalized longitude of periapsis $\varpi_c$ are defined by 
\begin{equation}
\sigma_c= \sqrt{
    e_{1}^{2}+e_{2}^{2}- 2e_{1}e_{2}\cos(\varpi_1 - \varpi_2)
}\,,
\end{equation}
and
\begin{equation}
\varpi_c = \arctan{\left[\frac{e_1 \sin{\varpi_1} - e_2 \sin{\varpi_2}}{e_1 \cos{\varpi_1} - e_2 \cos{\varpi_2}}\right]}\,.
\end{equation}
The evolution of the resonant system is now fully specified given quantities $q\equiv (m_1 + m_2)/M$, $k$, $e_{1,2}$ and $\varpi_{1,2}$. Three features of this reduced-resonance system are worth emphasizing as they are unique to the compact, nearly circular limit. First, the mass dependence of the orbiters enters only through the ratio of their sum ($m_1 + m_2$) to the MBH mass $M$. This mass symmetry is a consequence of the compact-limit reduction, and usefully collapses the parameter space, as systems with the same $q$ are dynamically equivalent \citep{1986henon, 2013deck, 2025tomayo}. Second, the eccentricities enter only through their vector difference,
$\sigma_c$, because to leading order in $e$ an eccentric orbit is a circle offset
from the central mass by $a e$, so that equal shifts of both eccentricity vectors
leave the relative geometry of the two orbits unchanged \citep[\S7.1
of][]{2025tomayo}. Third, the reduced problem has a single intrinsic eccentricity scale, set by the width of the resonance ($\Delta\widetilde{\Psi}$). At the onset of libration ($\delta = 0$), the resonance half-width is $\Delta\widetilde{\Psi}/2 = 2$ \citep{2015batygin}. Through Equation~\eqref{eqn:tildePsi_and_tildepsi}, the composite eccentricity at this
half-width is
\begin{equation}\label{eq:Delta_sigma}
   \Delta\sigma_c \simeq \left(\frac{4}{15}\frac{m_1 + m_2}{k M}\right)^{1/3}
   \simeq 0.64\,\left(q/k\right)^{1/3}\,.
\end{equation}  
The quantity $\Delta \sigma_c$ is the largest kick to $\sigma_c$ a resonant
pair can absorb and remain in resonance -- and it is one of the quantities against which we
measure stochastic forcing (see \S\ref{subsec:e_diffusion}). The same scale bounds the validity of the near-circular treatment adopted here. The resonant interaction can change $\sigma_c$ by at most $\sim\Delta\sigma_c$ over a libration cycle, so Equation~\eqref{eqn:Hfinal}, in which the eccentricity and the resonant angle evolve together, is the correct description only when the pair's eccentricity is itself no larger than this, $\sigma_c \lesssim \Delta\sigma_c\sim q^{1/3}$ \citep[\S5.6 of][see also \citealt{2007ferrazmello, 2013deck}]{2025tomayo}.\footnote{For $\sigma_c \gg \Delta\sigma_c$ the resonance changes the eccentricity by only a small fraction of its value, so $\sigma_c$ is effectively constant over a libration cycle and the resonance reduces to the classical pendulum model \citep{MurrayDermott1999}, whose width grows as $\sqrt{\sigma_c}$ rather than remaining fixed at Equation~(\ref{eq:Delta_sigma}). This threshold lies well inside the orbit-crossing eccentricity $\sim 1/k$ that bounds the compact limit itself
\citep{2025tomayo}.}

%Notably, the mass dependence of the orbiters enters only through their sum ($m_1 + m_2$). This mass symmetry is a consequence of the compact-limit reduction \citep{2017batygin} and usefully collapses the parameter space, as systems with the same $m_1 + m_2$ are dynamically equivalent. 

\begin{figure}
    \centering
    \includegraphics[width=1\linewidth]{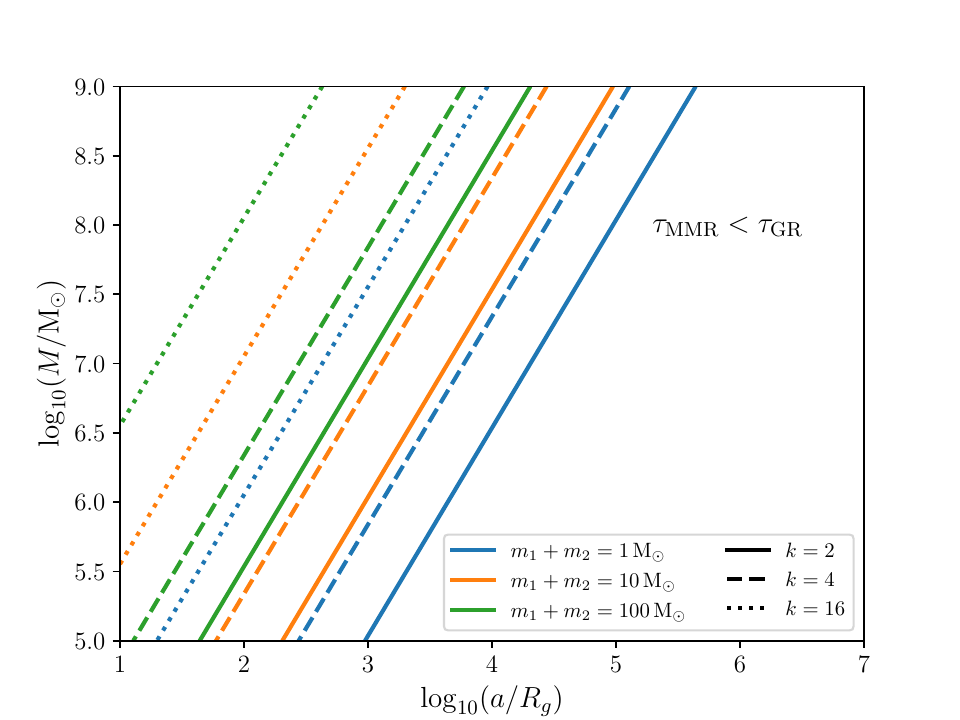}
    \caption{The boundary $\omega_{\text{lib}}/ \Delta\dot{\varpi}_{\text{GR}} \approx 1$, which separates the interior parameter space where GR dominates the dynamics (and the resonance is broken) from the exterior region where GR precession is sufficiently weak that the MMR can exist. Colors denote different combined masses of the objects participating in the resonance, and line styles denote the $k$ in the $k+1:k$ first-order resonance. Larger $k$ (more closely spaced resonances) shifts the GR disruption boundary to smaller semimajor axes $a$ by reducing the {\it differential} relativistic apsidal precession $\Delta \dot{\varpi}_{\rm GR}$ between a resonant pair.}
    \label{fig:timescales}
\end{figure}

\subsection{Mean-motion Resonance under General-relativistic Precession}\label{subsec:poincare}
The classic first-order post-Newtonian (1PN) GR Hamiltonian describes the precession in $\varpi$ due to general relativity \citep{Richardson1988, Adams2006, Migaszewski2009,Volpi2024}. This Hamiltonian can be written as
\begin{equation}
\mathcal{H}_{1PN} = -\frac{3 \g^2 M^2 m_{i}}{a_{i}^2 c^2 \sqrt{1-e_{i}^2}},
\label{eq:GR}
\end{equation}
where $c$ is the speed of light. Following Equation~\eqref{eq:GR}, we can write the precession rate of a low-eccentricity planetary orbit due to GR: 
\begin{equation}
\dot{\varpi}_{\text{GR}} = \frac{3 \left( G M\right)^{3/2}}{a_i^{5/2} c^2}.
\end{equation}

For two bodies near a first-order MMR (with period ratio $k+1:k$), with a Hamiltonian defined by Equation~\eqref{MMRhammy}, the relevant GR frequency will be the differential precession, defined as
\begin{equation}
\begin{aligned}
\Delta\dot{\varpi}_{\text{GR}}&=  |\dot{\varpi}_{\text{GR,1}} - \dot{\varpi}_{\text{GR,2}}| 
%\\&= 3 \left( G \centralmass \right)^{3/2} c^{-2} \bigg( a_1^{-5/2} - a_2^{-5/2}  \bigg)\\
\\&\simeq
\frac{3\,(GM)^{3/2}}{c^{2}\,a_1^{5/2}}\,\bigg( 1 - \left(\frac{k+1}{k}\right)^{-5/3}\bigg) \label{eq:diffApsidal}
\end{aligned}
\end{equation}

Linearizing the reduced second-fundamental-model resonant Hamiltonian (Equation~\ref{eqn:Hfinal}) above \citep[see also][]{Batygin2013} about its stable fixed point gives the small-amplitude libration frequency \citep[][see their Equation 8.47 for the analogous pendulum result]{MurrayDermott1999}: 
\begin{equation}
\begin{aligned}
    \omega_{\text{lib}} =& \frac{3 [n_2]}{2} \left(\frac{M}{m_1+m_2}\right)^{-2/3} \\ & \quad \quad \times \left[ \frac{ \big( 3 |f_{k+1, 27}| \big)^{-1/3} } { \big( (k+1)^5 (k) \big)^{1/9} } \right]^{-1}.
\end{aligned}
\end{equation}

%When $\dot{\omega}_{\text{MMR}}/ \Delta\dot{\omega}_{\text{GR}} \approx 1$, precession due to GR 
%will be commensurate with MMR libration, and the resonance will break. 
Because first-order resonant dynamics depend on the relative eccentricity vector between participating bodies, differential apsidal precession directly forces the generalized resonant angle. As a result, when $\omega_{\text{lib}}/ \Delta\dot{\varpi}_{\text{GR}} \approx 1$, the resonance is likely to be disrupted \citep[following an analogous criterion in][]{Adams2006}.

In Figure~\ref{fig:timescales}, we plot the boundaries in the $\log_{10}(a/R_g)$–$\log_{10}(M/\rm M_\odot)$ parameter space, where $R_g \equiv G M/c^2$ is the gravitational radius of the MBH. The contours in Figure~\ref{fig:timescales} separate the regions where the dynamics are dominated by GR precession versus those dominated by MMR libration.
For larger central masses $M$ and less massive resonant objects $m_i$, GR tends to dominate the dynamics to more distant radii,
pushing outwards the region where stable MMRs can persist.  

Notably, these boundaries shrink to smaller radii when considering more closely spaced first-order MMRs (larger $k$): As we can see from Equation \ref{eq:diffApsidal}, as $k\to \infty$, the {\it differential} apsidal precession rate $\Delta \dot{\varpi}_{\rm GR} \to 0$.  This effect favors the survival of closely spaced MMRs (large $k$) at small radii, although beyond some ($e$-dependent) value of $k$, resonance overlap will act on its own to destabilize these closely spaced resonances.

\section{Migration}\label{sec:migration}
\begin{figure}
    \centering
\includegraphics[width=0.95\linewidth]{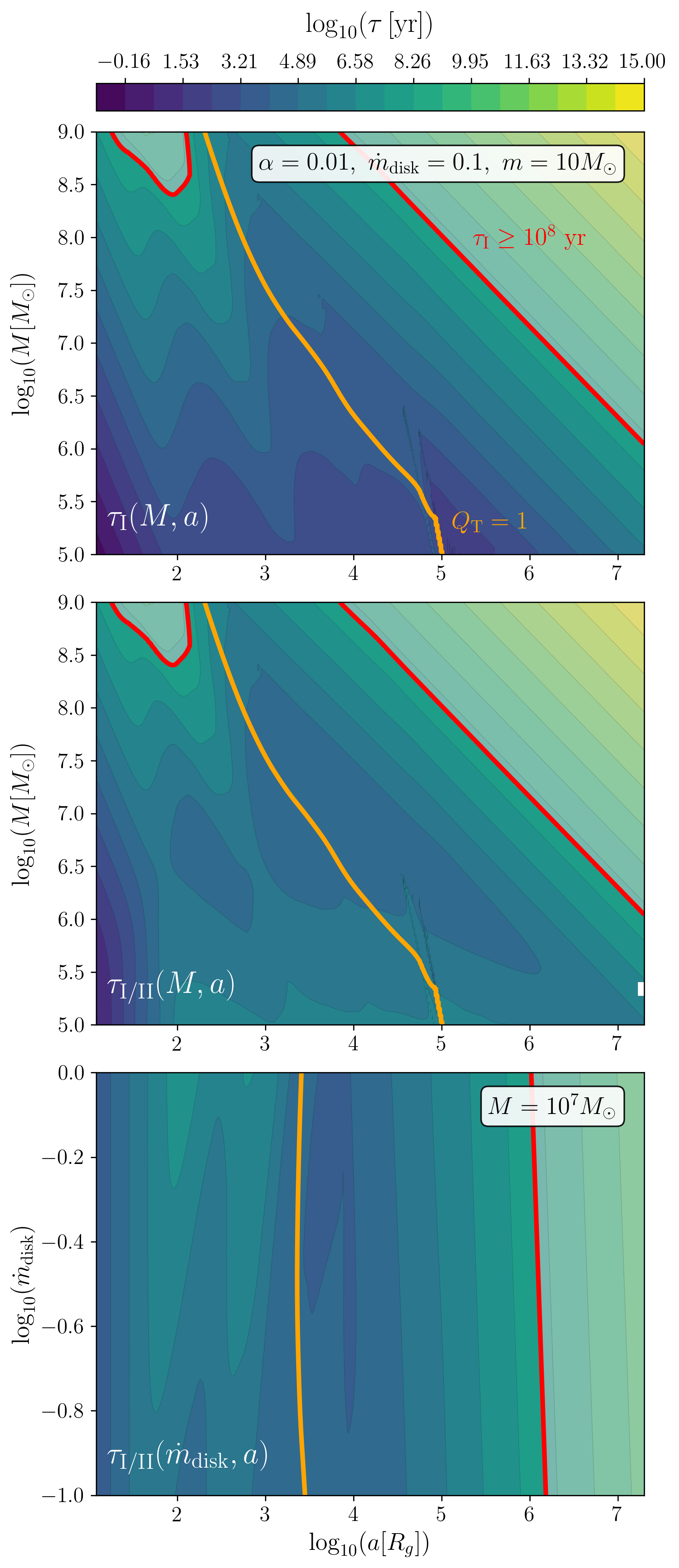}
    \caption{Colored contours show the migration timescale \(\tau\) as a function of distance from the MBH \((a/R_g)\) and disk parameters. The red curve marks \(\tau=10^8\) yr, the maximum AGN lifetime. \textit{Top panel:} \(\tau_{\rm I}\) assuming Type I migration only, with \(\alpha_{\rm disk}=0.01\), \(\dot m_{\rm disk}=0.1\) (the disk mass flux in units of the Eddington rate), and a migrating mass \(m=10\,M_\odot\). \textit{Middle panel:} \(\tau_{\rm I/II}\) computed from Equation~(\ref{eq:typeI-II}), i.e., the Type I rate rescaled by the local surface density; both top and middle panels vary the central mass \(M\). \textit{Bottom panel:} \(\tau_{\rm I/II}\) as a function of \(\dot m_{\rm disk}\) and \(a/R_g\) for fixed \(M=10^7\,M_\odot\). The orange line indicates the minimum radius where marginal stability is imposed; to the right of this line the \citet{sirko2003} models assume stellar feedback maintains \(Q_{\rm T}=1\).}
    \label{fig:migration_typeI/II}
\end{figure}

We anticipate that in AGN disks, as in their planetary system analogs, the dynamics described in the previous section will be modulated by the exchange of angular momentum and orbital energy between the embedded secondaries and the surrounding disk. One such exchange occurs via local excitation of spiral density waves \citep{1979goldreich}. Torques from the overdense waves interior and exterior to the embedded object do not precisely cancel \citep{1997ward}, resulting in a residual net torque \citep{2010paardekooper} of
\begin{equation}\label{eqn:Gamma_I_paardekooper}
\Gamma_{\rm{I}} = C_{\rm{I}}\frac{H}{a}\Gamma_{0}\,,
\end{equation}
where $\Gamma_0$ is the approximate magnitude of the torque from a single spiral arm:
\begin{equation}\label{eqn:Gamma_0}
\Gamma_0 = \left(\frac{m}{M}\right)^2\Sigma a^4 n^2\left(\frac{H}{a}\right)^{-3}\,. 
\end{equation}
The variables $H$ and $\Sigma$ represent the scale height and surface density of the disk, respectively, and $n$ is the Keplerian frequency at a given distance $a$ from the MBH. The dimensionless prefactor $C_{\rm I}$ is usually of order unity, and depends on disk density and temperature gradients.  Recent calibrations from 3D hydrodynamical simulations \citep{2017jimenez} give
\begin{equation}\label{eqn:GammaI_tot_jimenez}
\begin{split}
\Gamma_{\rm I} =& 
 \Biggl[
   C_{\rm L} 
   + \Bigl(
      0.46 
      \\&+ 0.96 \frac{\text{d} \ln \Sigma}{\text{d} \ln a} 
      - 1.8  \frac{\text{d} \ln T}{\text{d} \ln a} 
     \Bigr)\gamma_{\rm ad}^{-1}
 \Biggr]\frac{H}{a}\,\Gamma_0 \,.
\end{split}
\end{equation}
In Equation~\eqref{eqn:GammaI_tot_jimenez}, $\gamma_{\rm ad}$ is the adiabatic index, which we take to be $5/3$, and $C_{\rm L}$, the Lindblad torque prefactor (other dimensionless terms in Equation \ref{eqn:GammaI_tot_jimenez} encode the corotation torque), is given by \citep{2017jimenez}:
\begin{equation}\label{eqn:C_L}
\begin{split}
C_{\rm L} = & \Bigl(-2.34 - 0.1 \frac{\text{d} \ln \Sigma}{\text{d} \ln a} + 1.5  \frac{\text{d} \ln T}{\text{d} \ln a} \Bigr) f_{\rm ad} (x) \,.
\end{split}
\end{equation}
The function $f_{\rm ad}$ interpolates between the locally isothermal and adiabatic limits of disk thermodynamics:
\begin{equation}
\begin{aligned}
f_{\rm ad}(x) &= \frac{(x/2)^{1/2} + 1/\gamma_{\rm ad}}{(x/2)^{1/2} + 1}
\simeq 
\begin{cases}
\dfrac{1}{\gamma_{\rm ad}}, & x \ll 1, \\[6pt]
1, & x \gg 1 \,.
\end{cases}
\end{aligned}
\end{equation}
Here the dimensionless variable $x \equiv \chi / (H^2 n)$ is the ratio of thermal diffusivity ($\chi$) to $H^2 n$, and the $x \ll 1$ ($x \gg 1$) limits represent adiabatic (isothermal) disk conditions. Under the assumption of heat transport by optically thick radiative diffusion, the thermal diffusivity is
\begin{equation}
\chi = \frac{16 \gamma_{\rm ad} (\gamma_{\rm ad} -1) \sigma_{\rm sb} T^4}{3 \kappa \rho^2H^2 n^2}\,,
\end{equation}
where $\sigma_{\rm sb}$ is the Stefan-Boltzmann constant, and $T$, $\rho$, and $\kappa$ are the local temperature, density, and opacity of the AGN disk, respectively. 

In the AGN context, we must also account for GW emission from the compact secondary's orbital motion around the MBH. Unlike in planetary systems, where GR effects manifest  -- at most -- as a conservative form of precession, embedded objects in AGN disks can reach radii where GW back-reaction is dynamically competitive with, and eventually dominates over, gas torques. The GW torque contribution can be expressed in fractional form as
\begin{equation}\label{eqn:Gamma_GW}
\frac{\Gamma_{\rm GW}}{\Gamma_0} = \frac{32}{5} \left( \frac{R_{\rm g}}{a} \right)^{5/2} \left(\frac{H}{a} \right)^3 \frac{M}{\Sigma a^2}
\end{equation}
where %$c_{\rm s} = H n$ is the disk sound speed, and 
we have assumed a quasi-circular inspiral \citep{1964peters}. The GW torque is strictly negative (i.e., causing inward migration) and grows rapidly at small radii, $\Gamma_{\rm GW} \propto a^{-7/2}$. Because $\Gamma_{\rm GW}$ itself is set by the binary dynamics (i.e., it is independent of disk hydrodynamics), we can combine it linearly with $\Gamma_{\rm I}$ to get a total torque $\Gamma_{\rm I} + \Gamma_{\rm GW}$. For the rest of this paper, we include the GW torque contribution implicitly whenever we calculate migration torques and corresponding timescales.

Here we convert torque into a radial migration timescale $\tau_{\rm I}$ by balancing orbital angular momentum ($\Lambda$) against applied torque ($\Gamma_{\rm I}$):
\begin{equation}\label{eqn:tau_mig}
\tau_{\rm I} = \frac{\Lambda}{2|\Gamma_{\rm I}|} = \frac{1}{2 n}\frac{M^2}{m \, \Sigma a^2} \left(\frac{H}{a}\right)^3 \left|\frac{\Gamma_{\rm I}}{\Gamma_0} \right|^{-1} \,.
\end{equation}

The Type I prescription is appropriate for low-mass orbiters that do not appreciably perturb the local disk. However, a massive object can partially deplete the gas near its orbit, lowering the effective surface density and reducing its coupling to the local disk. This regime is known as Type II migration \citep{1986lin}. In the classic Type II picture, a sufficiently massive object opens a gap in the disk and local angular-momentum exchange stops; the embedded object instead locks to the viscous evolution of the disk itself. However, hydrodynamic simulations \citep{2014duffel, 2018kanagawa} have shown that even when the torque from an embedded object exceeds the viscous torque of the gas, some gas leaks through the gap. The resulting migration timescale, even for a massive migrator, is thus given by the Type I migration timescale scaled by a reduced surface density $\Sigma_{\rm min} = \Sigma / (1 + 0.04 K)$, where $K = \left(\frac{m}{M}\right)^2\left(\frac{H}{a}\right)^{-5}\alpha_{\rm disk}^{-1}$ \citep{2018kanagawa} and $\alpha_{\rm disk}<1$ is the Shakura-Sunyaev disk viscosity parameter \citep{1973shakura}. The migration timescale for low- and high-mass orbiters is parameterized by a smooth transition from the unperturbed, Type I limit ($K \ll 1$) to the deep-gap limit ($K \gg 1$) according to the equation 
\begin{equation}
\tau_{\rm I/II} = \frac{\Sigma}{\Sigma_{\rm min}} \tau_{\rm I}\,.
\label{eq:typeI-II}
\end{equation}

In order to evaluate the torques and migration timescales, we must adopt a disk model to supply the local thermodynamic and structural gradients. The classic, steady-state $\alpha$-disk is known to break down in the outer regions of AGN, becoming unstable to self-gravity \citep{schlosman1989}. To prevent runaway fragmentation and disk depletion, many AGN disk models invoke self-limiting star formation, whose feedback regulates the disk. The stability of the disk is represented by the Toomre parameter $Q_{\rm T}$ \citep{toomre1964}, where $Q_{\rm T} < 1$ indicates instability. Two widely used frameworks are those of \citet{sirko2003} and \citet{2005thompson}, which construct modified viscous disks by requiring that the disk maintain marginal stability and setting $Q_{\rm T} = 1$. These models differ most significantly in how they treat mass transport: \citet{2005thompson} allow $\dot{M}_{\rm disk}$ to vary with radius in response to {\it in situ} star formation, whereas \citet{sirko2003} enforce a steady inflow rate. Although the \citet{2005thompson} model may be more physically motivated, here we adopt the \citet{sirko2003} model so that we can investigate dependence on $\dot{M}_{\rm disk}$ explicitly. With this choice, we are able to generate the heat maps in Figure~\ref{fig:migration_typeI/II} using the \texttt{pAGN} package \citep{2024gangardt}. 

Figure~\ref{fig:migration_typeI/II} shows the migration timescales as contour maps: $\tau_{I}$ (top panel) and $\tau_{\rm I/II}$ (middle and bottom panels). The top panel shows timescales in the ($M$, $a$) plane, adopting $m = 10 \, \rm{M_\odot}$, $\alpha_{\rm disk} = 0.01$, and a disk mass-accretion rate $\dot{m}_{\rm disk} = 0.1$, where $\dot{m}_{\rm disk} = \dot{M}_{\rm disk}/\rm \dot{M}_{Edd}$ is the dimensionless ratio of the disk mass flux over the Eddington rate, $\mathrm{\dot{M}_{Edd}}=\mathrm{L}_{\mathrm{Edd}}/ c^2 \eta_{\rm rad}$, where $\mathrm{L_{Edd}} = 4 \pi G c M/\kappa_e$ is the Eddington luminosity, $\kappa_e\simeq 0.34 \, \rm cm^2 g^{-1}$ is the electron scattering opacity, and $\eta_{\rm rad} = 0.1$ is the radiative efficiency. The middle panel shows $\tau_{\rm I/II}$ in the same ($M$, $a$) plane and with the same system parameters. In the bottom panel, $\tau_{\rm I/II}$ is plotted in the ($\dot{m}_{\rm disk}$, $a$) plane, at fixed MBH mass $M = 10^7 \, \rm M_\odot$, again with $\alpha_{\rm disk} = 0.01$. The primary difference between the top and middle panels arises near $\log_{10}(M/M_\odot) \simeq 5$, where $\tau_{\rm I/II}$ is markedly longer than the corresponding $\tau_{\rm I}$. This occurs because the gap parameter $K \propto (m/M)^2$ becomes nonnegligible at fixed $m$ as $M$ decreases. We also mark in red where the migration time reaches the upper limit of a typical disk lifetime $t_{\rm AGN} \sim 10^8 \, \rm yr$ \citep{2001Martini}, and gray out regions with $\tau > t_{\rm AGN}$; systems in these zones are not expected to undergo significant migration-driven evolution in a single AGN episode.  

The bottom panel of Figure~\ref{fig:migration_typeI/II} shows that beyond $a \sim 10^{4}\,R_g$, $\tau_{\rm I/II}$ is nearly independent of $\dot m_{\rm disk}$. This follows from the structure of the self-gravitating outer disk \citep[see the analytic scalings of][]{Grishin2024}: Marginal stability fixes the density, $\rho = n^2/2\pi G$, independently of $\dot m_{\rm disk}$ and $\alpha$, and the mass flux enters only through the temperature required to carry it, $c_s \propto (\dot m_{\rm disk}/\alpha)^{1/3}$, so that $H = c_s/n$ and $\Sigma = 2\rho H$ both scale as $(\dot m_{\rm disk}/\alpha)^{1/3}$. The migration time $\tau_{\rm I} \propto (H/a)^2/\Sigma$ therefore varies only as $(\dot m_{\rm disk}/\alpha)^{1/3}$ in the outer disk, in contrast to the viscously heated inner zones where $\Sigma$ and $H$ depend more strongly on $\dot m_{\rm disk}$.

In all panels in Figure~\ref{fig:migration_typeI/II}, an orange line marks the innermost radius where marginal stability ($Q_{\rm T} = 1$) is invoked. To the right of this line, the \citet{sirko2003} model that we use explicitly assumes that disk stability is maintained by stellar feedback. We flag this region because its structure depends on poorly constrained parameters and physical assumptions. In particular, the mass, number density, and type of embedded objects necessary to stabilize the outer AGN disk remains an open question \citep{gilbaum2022, 2025epstein-martin} -- with consequences for the structure of the disk and anticipated migration torques. We discuss these uncertainties in more detail in \S\ref{subsec:caveats}. 

Convergent migration is a necessary precondition for capture into MMRs. In the three panels of Figure~\ref{fig:migration_typeI/II}, convergent drift for equal-mass pairs is confined to radii of $\sim 10^2$ or $10^4 \, R_{\rm g}$. This does not preclude convergence for unequal-mass pairs. Because $\tau_{\rm I} \propto m^{-1}$, a sufficiently high-mass body migrates faster and can overtake a lower-mass body on an initially interior orbit. This dynamic holds true in the $\tau_{\rm I/II}$ case unless $K \gg 1$ (i.e., the deep-gap regime), at which point migration rates decrease with increased object mass. However, once we incorporate thermal torques, the net torque can reverse sign and produce stable zero-torque zones: migration traps. These traps enforce convergence: Objects outside the trap drift inward, while those inside drift outward, so bodies of any mass are funneled toward the same radius.\footnote{Here we assume that convergent migration suffices for resonance capture. In fact, both convergence and adiabatic migration are required: The migration timescale must exceed the resonance libration period. See \citet{2015batygin}, their Equation~(43), for the adiabatic criterion. Although we defer a quantitative assessment to future work, we return to this constraint in \S\ref{subsec:caveats} where we examine it qualitatively. Additionally, we direct the reader to \citet{2026moncrieff}, who examine the adiabaticity of resonant capture for embedded BHs with N-body integrations.} We discuss the thermal-torque prescription in the next section.  

\subsection{Thermal Torques and Migration Traps}\label{subsec:thermaltorques}
\begin{figure}
    \centering
    \includegraphics[width=0.95\linewidth]{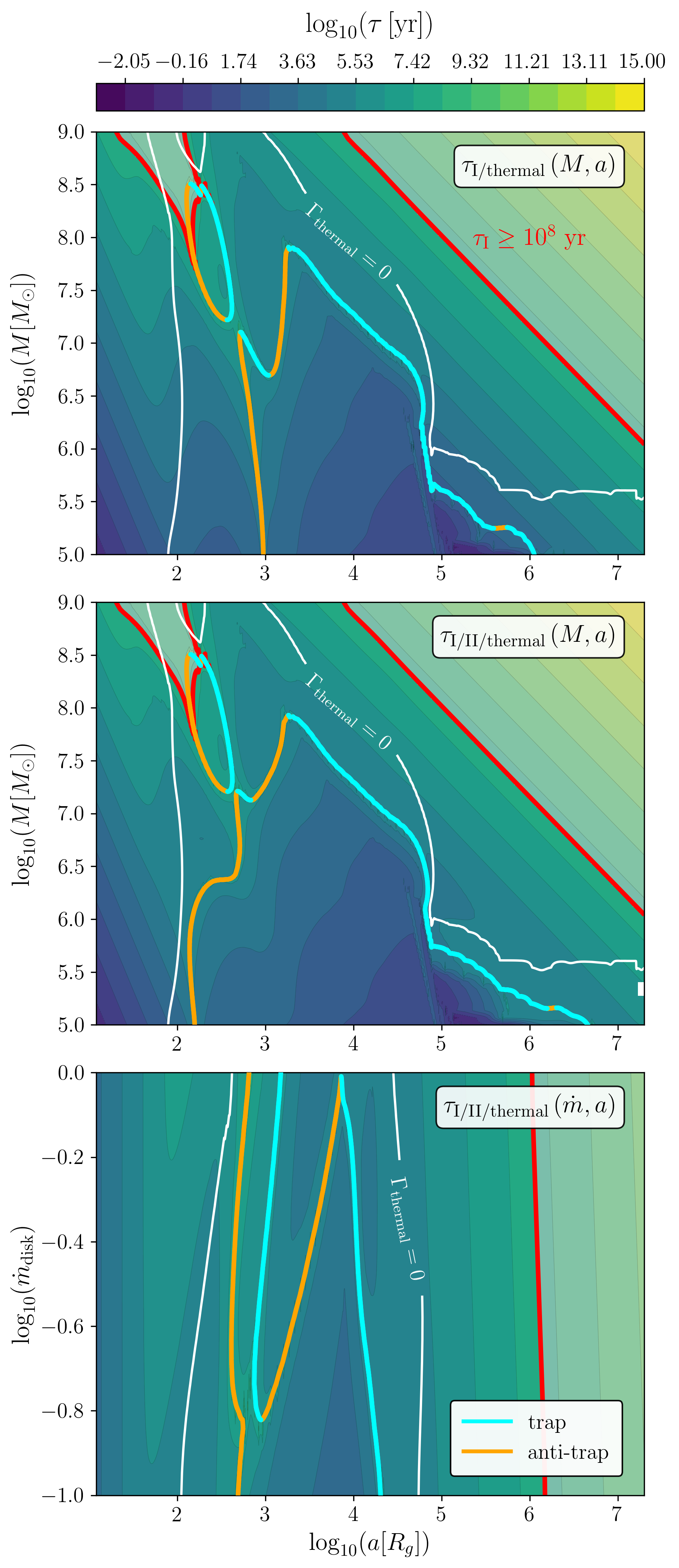}
    \caption{Thermal-torque migration maps, analogous to Figure~\ref{fig:migration_typeI/II}. Colored contours show the migration timescale \(\tau\) versus distance from the MBH \((a[R_g])\); the red curve marks \(\tau=10^8\) yr. The axes and assumed parameters in each panel are identical to those in Figure~\ref{fig:migration_typeI/II}. These maps differ in the torque used to compute \(\tau\): \textit{Top panel:} we exclude the possibility of gap opening and show $\tau_{\rm I/thermal}\propto (\Gamma_{\rm I}+\Gamma_{\rm GW}+\Gamma_{\rm thermal})^{-1}$, with \(\Gamma_{\rm thermal}\) given by Equation~\eqref{eq:gamma_therm_final}. \textit{Middle and bottom panels:} gap opening is now accounted for and we show \(\tau_{\rm I/II/thermal} \propto \Gamma_{\rm{tot}}^{-1}\), where \(\Gamma_{\rm tot}=\Gamma_{\rm I/II}+\Gamma_{\rm GW}+\Gamma_{\rm thermal}\), and \(\Gamma_{\rm I/II}\) follows Equation~(\ref{eq:typeI-II}). Cyan (orange) lines mark \emph{migration traps} (\emph{anti-traps}), defined by zero total torque, and white contours indicate \(\Gamma_{\rm thermal}=0\).}
    \label{fig:tau_thermal}
\end{figure}

In addition to the gravitational perturbations associated with Type I/II torques, embedded objects perturb the temperature of local gas, which alters its density profile and creates gravitational back-reaction. \citet{Masset2017} formalized the resulting ``thermal torque'' using linear theory, with
analytic predictions that agree with simulations \citep{2014lega, 2015benitez-llambay, 2017chrenko, 2020hankla}. In the \citet{Masset2017} framework, the thermal response is governed by two quantities: (i) the diffusion length $\lambda$, which sets how far temperature perturbations spread under finite thermal diffusivity:
\begin{equation}\label{eq:lambda}
\lambda = \sqrt{\frac{2 \chi}{3 \gamma_{\rm ad} n }} \,;
\end{equation}
and (ii) the corotation offset $x_c$, which quantifies the headwind shift, or the small radial displacement between the body’s orbital radius and the radius where the unperturbed gas has the same angular frequency as the body:
\begin{equation}\label{eq:x_C}
x_c = - \frac{H^2}{3 \gamma_{\rm ad} a} \frac{{\rm d} \ln p_{\rm tot}}{{\rm d} \ln a}\,,
\end{equation}
where $p_{\rm tot} \approx \rho H^2 n^2$ is the total midplane pressure and ${\rm d} p_{\rm tot}/{\rm d}r \sim \rho H^2 n^2/r$ \citep{2024gangardt}. Physically, Keplerian shear, together with a nonzero corotation offset, displaces the hot/underdense and cold/overdense lobes asymmetrically. The resulting net thermal torque scales 
with the ratio $\lambda/x_c$. 

For embedded objects with accretion luminosity $L$, the net thermal torque combines a heating torque (outward, positive) and a cold thermal torque (inward, negative), which is present even for nonluminous bodies. Their sum is written as \citep{Masset2017}
\begin{equation}\label{eq:Gamma_therm}
\Gamma_{\rm therm} = 1.61 \frac{\gamma_{\rm ad} -1 }{\gamma_{\rm ad}} \frac{x_c}{\lambda} \left(\frac{L}{L_c} - 1\right)\Gamma_0 \,,
\end{equation}
where $L$ is the body's luminosity and 
\begin{equation}\label{eq:L_c}
L_c = \frac{4 \pi G m \rho}{\gamma_{\rm ad}}\chi
\end{equation}
is the threshold luminosity where the cold and hot torques cancel. 

Equation~\eqref{eq:Gamma_therm} is appropriate so long as the acoustic timescale is longer than the diffusion timescale within the Bondi radius $R_B^2/(4 \chi) \ll R_B/c_s$. From this inequality, we can define a critical thermal mass \citep{2021guilera},
\begin{equation}
m_{\rm c} \simeq \frac{c_s \chi}{G}\,, 
\end{equation}
which corresponds to a dimensionless ratio $\mu_{\rm th} \equiv m_{\rm c}/m$. From high-resolution simulations, \citet{2020velasco} found that, for $\mu_{\rm th} \lesssim 1 $, the embedded object is subjected to a thermal force with a
magnitude in good agreement with the linear theory developed by
\citet{Masset2017}. When $\mu_{\rm th} > 1$, the ratio of the heating torque to its linear estimate decays slowly. \citet{2020velasco} give an approximate expression that fits
the numerical results:
\begin{equation}
\label{eq:gamma_therm_final}
\begin{aligned}
\Gamma_{\rm therm} = &\Gamma_{\rm therm, hot}\frac{4 \mu_{\rm th}}{1 + 4 \mu_{\rm th}} \\ &+ \Gamma_{\rm therm, cold}\frac{2 \mu_{\rm th}}{1 + 2 \mu_{\rm th}} \,,
\end{aligned}
\end{equation}
where $\Gamma_{\rm therm,hot}$ and $\Gamma_{\rm therm,cold}$ refer to the positive and negative terms in Equation~\eqref{eq:Gamma_therm}, respectively. 

An additional caveat to the original work of \citet{Masset2017} was emphasized by \citet{Grishin2024}. Equations~\eqref{eq:Gamma_therm} and \eqref{eq:gamma_therm_final} assume thermal perturbations are transported by radiative diffusion, an assumption which is only valid when the optical depth across the diffusion length exceeds unity: $\tau_\lambda \equiv \lambda /l \gtrsim 1$, where $l = (\kappa_e \rho)^{-1}$ is the photon mean free path and $\kappa_e \simeq 0.34 \, \rm cm^2g^{-1}$ is the electron scattering opacity. Following their implementation, we account for the breakdown toward optically thin conditions with a smooth attenuation factor $1 - \exp{(-\sqrt{\alpha_{\rm disk}} \tau_{\lambda})}$, so that the thermal torque is suppressed when $\tau_{\lambda}$ is small. 

With this analytical framework in hand, we update the migration timescales in Figure~\ref{fig:migration_typeI/II} to include the thermal torque. We plot the results in Figure~\ref{fig:tau_thermal}. As before, the top and middle panels map the $(M, a)$ plane, adopting $m = 10 \, \rm M_\odot$, $\alpha_{\rm disk} = 0.01, $ and $\dot{m}_{\rm disk} = 0.1$; the bottom panel presents $(\dot{m}_{\rm disk}, a)$ at fixed $M = 10^7 \, \rm M_\odot$, $\alpha_{\rm disk} = 0.01$, and $m = 10\, \rm M_\odot$. Note that we assume the orbiter is accreting at the Eddington rate\footnote{This assumption is motivated by the generally super-Eddington Bondi-Hoyle accretion rates onto embedded BHs \citep{2017stone}, which only fall below the Eddington rate for the outer regions of high-mass disks, where migration times are usually too long for MMR capture.  We thus follow \citet{gilbaum2022} in assuming that feedback results in a time-averaged Eddington accretion rate, though we note that this is an uncertain area of accretion physics.} such that its luminosity is $L = 4 \pi \,G\,c \,m/ \kappa_{\rm e}$. Contours show the migration timescale $\tau$, now computed from the total torque. In the top panel, we use $\tau_{\rm I/thermal} \propto |\Gamma_{\rm I} + \Gamma_{\rm thermal} + \Gamma_{\rm GW}|^{-1}$, while in the middle and bottom panels, we use $\tau_{\rm I/II/thermal} \propto |\Gamma_{\rm I/II} + \Gamma_{\rm thermal} + \Gamma_{\rm GW}|^{-1}$, where $\Gamma_{\rm I/II}$ is the gap-reduced gas torque from Equation~\eqref{eq:typeI-II}. Our plots also include a white contour that indicates where $\Gamma_{\rm thermal} = 0$, i.e., where the direction of the torque changes sign. 

Our results broadly align with the power-law disk models of \citet{Grishin2024}. In their framework, positive thermal torques often overwhelm Type I torques and carve a band of outward migration bounded by a trap and anti-trap. Their trap generally lies between $10^3$ and $10^5 \, R_{\rm g}$, shifting inward with increasing $M$. They also find that traps disappear at high MBH masses: For $\dot{m}_{\rm disk} \sim 0.1$ and 1, traps are confined to $M \lesssim 10^8 \, $ and $M \lesssim 10^7 \, \rm{M_\odot}$, respectively. These results look very similar to the outer migration trap presented in the top panel of Figure~\ref{fig:tau_thermal}. However, because \citet{Grishin2024} assume a simplified power-law approximation to the \citet{sirko2003} disk, their model misses a secondary trap near $a\simeq 10^3 \, R_g$, which widens the parameter space for traps up to $10^{8.5} \, \rm M_\odot$. This secondary feature is captured in the more detailed disk calculations given by \citet{2024gangardt} and \citet{Gilbaum2025}. Having used the \texttt{pAGN} package of \citet{2024gangardt}, our maps track their results. 

In the middle panel, migration traps extend to larger radii.  This occurs because reducing the effective surface density weakens the Type I torque and allows the thermal term to dominate over an increased range of $a$. In the bottom panel, we show changes in the outer and inner traps with $\dot{m}_{\rm disk}$, and find that as $\dot{m}_{\rm disk}$ increases, the outer trap--anti-trap pair narrows and vanishes for $M = 10^7\, \rm M_\odot$ by $\dot{m}_{\rm disk}\simeq1$, in agreement with \citet{Grishin2024}. For the same $M$, the inner trap appears at $\log_{10}(\dot{m}_{\rm disk}) \simeq -0.8$ and persists to $\log_{10}(\dot{m}_{\rm disk}) \simeq 0$, remaining near $a \simeq 10^3 \, R_g$. 

\subsection{Eccentricity Damping} \label{subsec:e_damping}

In the classic Type I regime, disk torques also damp the eccentricity of an embedded object. In the linear, small-$e$ limit, \citet{2004tanaka} found
\begin{equation}\label{eq:TW04}
\begin{aligned}
\left. \frac{\dot{e}}{e} \right|_{\rm TW04} &\equiv \tau_{e, \rm TW04}^{-1} \simeq - \frac{3}{4} n \left(\frac{m}{M}\right) \left(\frac{\Sigma a^2}{M}\right) \left(\frac{H}{a}\right)^{-4} \,.
\end{aligned}
\end{equation}
Relative to Type I migration, this implies $\tau_{e, \rm TW04}\sim \tau_{\rm I}\left(\frac{H}{a}\right)^2$. In the thin-disk regime relevant here, this makes $\tau_{e, \rm TW04}$ several orders of magnitude shorter than the migration timescales shown in Figure~\ref{fig:migration_typeI/II}. This is confirmed in 3D simulations by \citet{2008cresswell}, which also showed that once $e \gtrsim H/a$ the effective damping time increases. As we work in the small-$e$ limit, the \citet{2004tanaka} prescription suffices here. 

Note that, as in the case of $\tau_{\rm mig}$, we expect the eccentricity damping timescale to lengthen as the local gas density is reduced. By the same reasoning as Equation~\eqref{eq:typeI-II}, $\tau_{e,\rm I/II} = \Sigma/\Sigma_{\rm min} \tau_{e, \rm TW04}$, maintaining the proportionality constant $(H/a)^2$ between $\tau_{\rm I/II}$ and $\tau_{e,\rm I/II}$. 

In \S\ref{subsec:thermaltorques} we noted that heating and cooling also modify migration torques. As shown in \citet{2017chrenko, 2017eklund, 2019fromenteau}, heating and cooling will likewise alter eccentricity evolution. \citet{2017eklund} showed in simulations of luminous, accreting embryos that $e$ can rise to values comparable to the disk aspect ratio on short timescales. \citet{2019fromenteau} extended this with linear theory and provided a small-$e$, orbit-averaged damping rate:
\begin{equation}\label{eq:thermal_eDampingRate}
    \frac{\dot{e}}{e} \simeq \frac{1.46} {t_{\rm thermal}}\left(\frac{L}{L_c} - 1\right)\,,
\end{equation}
where 
\begin{equation}
    t_{\rm thermal} = \frac{c_s^2 n \sqrt{\chi / n}}{2 (\gamma_{\rm ad} - 1) G^2 M \rho} \,,
\end{equation}
valid for $ea \ll \lambda$ and $\lambda \ll H$. For nonluminous (``cold'') bodies with $L<L_c$, the linear thermal contribution increases the damping rate relative to the classical isothermal/adiabatic result by roughly a factor of $\sim H/\lambda$. For luminous bodies with $L> L_c$, the sign in Equation~\eqref{eq:thermal_eDampingRate} reverses, and the thermal contribution drives eccentricity growth. Equation~\eqref{eq:thermal_eDampingRate} is approximately $H/\lambda$ times larger than the \citet{2004tanaka} damping rate, indicating that thermal torques should nearly always dominate their nonthermal counterparts except where $L \simeq L_c$, i.e., the white line shown in Figure~\ref{fig:tau_thermal}. 

In the luminous case, once $ea \sim \lambda$, growth transitions toward a headwind/dynamical friction-like regime and then saturates as the motion becomes transonic. Practically, \citet{2019fromenteau} find that the saturation eccentricity lies between $\lambda/a$ and $H/a$.

\section{Stochastic Forces}\label{sec:stochastic_forces}

Dissipative migration of stellar-mass objects through the AGN disk will frequently create conditions that appear favorable for resonance capture.  However, the dynamically complex environments of AGN create a number of challenges for MMR survivability. In particular, stochastic forcing from (i) gravitational fluctuations driven by %magnetorotational instability (MRI) 
turbulence in the disk and (ii) impulsive kicks from stars in the surrounding nuclear cluster can disrupt resonances. The former has been extensively studied in the planetary context \citep{Johnson2006, 2008adams, 2009rein, 2017batygin}; in this section, we build on the logic outlined in those works. 

In general, a perturbative, stochastic acceleration $\boldsymbol{F}$ (i.e., a force per unit mass) follows a stationary stochastic Gaussian process with an autocorrelation function given by 
\begin{equation}
\langle F(t) F(t') \rangle = \langle F^2\rangle \exp{\left(- \frac{|t - t'|}{\tau_c}\right)}
\end{equation}
where $\tau_c$ is the autocorrelation time. A quantity $A$ driven by such an acceleration undergoes a random walk, and the mean-square displacement $\langle \Delta A^2\rangle $ grows linearly with $t$ for $t \gg \tau_c$. Thus the evolution of $A$ can be characterized by a constant diffusion coefficient $\mathcal{D}_A = \langle (\Delta A)^2 \rangle /(2t)$. \citet{2009rein} and \citet{2013okuzumi} derived a general set of celestial mechanics equations describing the response of planets to a stochastic acceleration $\boldsymbol{F}$, computing diffusion coefficients in $a$ and $e$ according to
 \begin{equation}\label{eq:gen_diffusion_coefficients}
 \frac{\mathcal{D}_a}{a^2} \propto \mathcal{D}_e \propto\frac{\langle F^2_\phi \rangle\tau_c }{a^2 n^2}\,,
 \end{equation}
where $\boldsymbol{F} = (F_r, F_\phi)$ has been decomposed into its radial and azimuthal components, whose magnitudes are taken to be comparable.

This general analytic framework can be applied to the specific forcing mechanisms expected in AGN disks, and in the following subsections, we translate $\langle F_\phi^2\rangle$ and $\tau_c$ into quantities appropriate for magnetorotational instability (MRI) turbulence and stellar flybys in the nuclear cluster. We calculate diffusion coefficients $\mathcal{D}_{i, \,\rm turb}$ and $\mathcal{D}_{i, \,\rm scat}$, where $i$ is a stand-in for the orbital elements $a$ and $e$, and we compare their relative strengths as functions of MBH mass $M$ and orbital radius $a$, as shown in Figure~\ref{fig:Dturb_vs_Dscat}. 

\subsection{Turbulence}\label{subsec:turbulence} 

MRI turbulence is driven in differentially rotating disks via coupling between the gas and magnetic fields. This interaction excites gas density fluctuations, which give rise to stochastic gravitational forces. From Poisson's equation, a local density perturbation $\delta \rho$ acting on a characteristic scale $l$ produces a fluctuating acceleration of order $F\sim 4 \pi G \delta \rho l $. In a geometrically thin, MRI-driven disk the characteristic length is of order $H$ and we can replace fluctuations in $\delta \rho$ with column density $\delta \Sigma = 2 H\, \delta \rho$, so that $F\sim G\delta\Sigma$. 

The amplitude of these fluctuations can be written in terms of known disk quantities by relating the standard $\alpha$-disk kinematic viscosity, $\nu_{\rm kin}=\alpha c_s H$, to angular-momentum transport by turbulent eddies. If $\delta v$ is the characteristic turbulent velocity fluctuation and $\tau_c$ is the eddy correlation time, the corresponding turbulent diffusivity is $\nu_{\rm kin}\sim\delta v^2\tau_c$. Taking $\tau_c\sim n^{-1}$ and using $H=c_s/n$ gives $\delta v/c_s\sim\sqrt{\alpha}$, consistent with the turbulent velocity dispersion adopted by \citet{2025trani}. For weakly compressible MRI turbulence, the fractional density fluctuations
scale similarly,
$\delta\Sigma/\Sigma\sim\delta v/c_s\sim\sqrt{\alpha}$ \citep{2011okuzumi,2013okuzumi}. Adopting this scaling \citep[as in][]{2017batygin}, we can calculate the variance in $\boldsymbol{F}$ as
\begin{equation}\label{eqn:Fturb_squared}
\langle F^2\rangle \sim \alpha \left(\frac{\Sigma a^2}{M}\right)^2 \left(n^2 a\right)^2\,.
\end{equation}
Under the assumption that the azimuthal and radial components of $\boldsymbol{F}$ are comparable, we can write $\langle F_\phi^2\rangle  \simeq \langle F_r^2\rangle \simeq \langle F^2\rangle/2$.

Density fluctuations contributing to stochastic forcing lie within $H$ of the orbiter \citep{2007oishi}. Consequently, when the mass of the orbiter is large enough to reduce the local surface density, the stochastic acceleration $\boldsymbol{F}$ is also reduced. The lower surface density is parameterized according to Equation~\eqref{eq:typeI-II}, where $\Sigma_{\rm min}$ denotes the locally depleted, azimuthally averaged surface density within $\sim H$ of the body due to partial gap opening by its tidal torques. Accordingly, we replace $\Sigma$ with $\Sigma_{\rm{min}}$ in Equation~\eqref{eqn:Fturb_squared} when evaluating the diffusion coefficients. 

For MRI-driven turbulence, $\tau_c \sim n^{-1}$ is inferred from simulations (e.g., \citealt{2004sano}). From Equation~\eqref{eq:gen_diffusion_coefficients}, the diffusion coefficients are
\begin{equation}\label{eq:D_turb}
\begin{split}
\frac{\mathcal{D}_{a, \,\rm turb}}{a^2} &\sim 2 \,\mathcal{D}_{e, \,\rm turb} \sim \frac{\alpha}{2}\left(\frac{\Sigma_{\rm min} a^2}{M}\right)^2 n \\
& \simeq \frac{1.8 \times 10^{-8}}{\rm year}\left(\frac{\alpha}{0.01}\right)\left(\frac{\Sigma_{\rm min}}{50 \, \rm g/cm^2}\right)^{2} \\
& \quad  \times \left(\frac{a}{\rm 16 \, pc}\right)^{5/2} \left(\frac{M}{10^8 \, \rm M_\odot}\right)^{-3/2}
\,,
\end{split}
\end{equation}
where we have incorporated approximate prefactors from Equation (8) of \citet{2017batygin}.

\subsection{Stellar Flybys}\label{subsec:stellar_flybys}

AGN episodes occur when large quantities of interstellar gas flow into a galactic nucleus.  Prior to this inflow, most galactic nuclei are populated by an MBH and its surrounding NSC.  The quasi-spherical, quasi-isotropic distribution of stars in the NSC creates a population of scatterers that may stochastically influence the orbits of embedded objects migrating in resonance.  While strong (large-angle) scatterings are quite rare, the cumulative stochastic forcing from many weak (distant, small-angle) two-body scatterings may in some circumstances force objects out of MMR.

Consider a spherically symmetric NSC with a 3D number density profile 
\begin{equation}\label{eqn:n_star}
    n_\star(a) = n_{\rm infl} \left( \frac{a}{a_{\rm infl}} \right)^{-\gamma},
\end{equation}
where we define the influence radius as the radial coordinate enclosing a mass in stars equal to the mass of the MBH.  Dynamical modeling of {\it Hubble Space Telescope} observations of nearby galactic nuclei finds that $a_{\rm infl} \approx  16~{\rm pc}~ (M / 10^8 M_\odot)^{0.69}$, with some scatter \citep{2016stone}. The stellar density profile is thus normalized by
\begin{equation}\label{eqn:n_infl}
\begin{split}
n_{\rm infl} &= \frac{M}{\langle m_\star \rangle} \frac{3-\gamma}{4\pi a_{\rm infl}^3}  = \\
& \simeq \frac{2.43 \times 10^3}{\rm{pc}^{3}} \left(\frac{3 - \gamma}{1.25}\right)  \\
& \quad \quad \quad \times \left(\frac{\langle m_\star \rangle}{\rm M_{\odot}}\right)^{-1}\left(\frac{M}{10^8 \, \rm M_\odot}\right)^{-1.07}
\end{split}
\end{equation}
 where $\langle m_\star \rangle$ is the mean stellar mass. Assuming a power-law present-day stellar mass function (PDMF) $dN/dm_\star \propto m_\star^{-\delta_{\rm IMF}}$ with $\delta_{\rm IMF} \in [1.7, 2.35]$ \citep{2010bartko, 2013lu} and bounds $m_{\star, \rm min} = 0.1 \,  \rm M_\odot$, $m_{\star, \rm max} = 120 \, \rm M_\odot$, the mean stellar mass, $\langle m_\star\rangle$, varies between $0.36-1.78\, \rm M_\odot$. For simplicity, we adopt $\langle m_\star\rangle = 1 \, \mathrm M_\odot$. Additionally, we take $\gamma = 7/4$, as is appropriate for a collisionally relaxed single-mass system \citep{1976bahcall}, but we also consider $\gamma = 1$ as an alternative case for an unrelaxed ``core'' profile, as is prevalent for high-mass MBHs \citep{2005lauer}.

The energy relaxation time for a test star embedded in this NSC is, approximately,
\begin{equation}
t_{\rm rel} =c_{\rm rel}\frac{\sigma^3(a)}{G^2 n_\star(a) \langle m_\star^2 \rangle \ln\Lambda}\\
\end{equation}
where $\sigma(a) \approx \sqrt{GM / a}/\sqrt{1+\gamma}$ is the (1D) stellar velocity dispersion, $\langle m_\star^2 \rangle \simeq 23 \rm \, M_\odot^2$ is the second moment of the stellar PDMF 
\footnote{For a power-law $dN/dm_\star \propto m_\star^{-\delta_{\rm IMF}}$ with $m_\star \in [0.1, 120] \, \rm M_\odot$, choosing $\delta_{\rm IMF} \approx 1.88$ yields $\langle m_\star\rangle \approx 1 \, \rm M_\odot$, $\langle m_\star^2\rangle \approx 23 \, \rm M_\odot^2$.}
, $\ln \Lambda \approx \ln(0.4 M/ \langle m_\star\rangle)$ is the Coulomb logarithm, and $c_{\rm rel}\approx 0.34$ is a dimensionless number of order unity.  This relaxation time can be interpreted in terms of a second-order energy-space diffusion coefficient, $t_{\rm rel} = E^2 / \mathcal{D}_{E, \, \rm scat}$. For quasi-circular orbits, we take $|E| \simeq GM/(2a)$, and a change of variables gives
\begin{equation}
\mathcal{D}_{a, \,\rm scat} = \left(\frac{da}{dE}\right)^2 \mathcal{D}_{E, \,\rm scat} = \frac{a^2}{t_{\rm rel}}\,,
\end{equation}
and from Equation~\eqref{eq:D_turb}, diffusion in eccentricity is related to semimajor axis by 
\begin{equation}
\begin{split}
2\mathcal{D}_{e, \,\rm scat} &\sim \mathcal{D}_{a, \, \rm scat}/a^2 \sim t_{\rm rel}^{-1}\\
& \propto M^{0.69 \gamma - 2.57} \, a^{3/2 - \gamma} \kappa_\star \langle m_\star\rangle
\,,
\end{split}
\end{equation}
where $\kappa_\star = \langle m_\star^2 \rangle / \langle m_\star \rangle^2$. For our adopted PDMF, $\kappa_\star \approx 23$. For fiducial parameters $M = 10^8 \, \rm M_\odot$, $a = 16 \, \rm pc$ (i.e., $a \simeq a_{\rm infl}$), $\gamma = 7/4$, and $\ln \, \Lambda \simeq 17.5$, we obtain $t_{\rm rel}^{-1} \sim (1.7 \times 10^{10} \, \rm yr)^{-1} \simeq 6 \times 10^{-11} \, yr^{-1}$. Note that for both nuclear cusps, $\gamma = 7/4$ and $1$, $\mathcal{D}_{e, \rm scat}$ decreases with MBH mass and decreases (weakly, $\propto a^{-1/4}$) with radius for $\gamma = 7/4$ while it increases with radius for $\gamma = 1$.

\subsection{Relative Importance of Turbulence and Stellar Flybys}

\begin{figure}
    \centering
\includegraphics[width=0.98\linewidth]{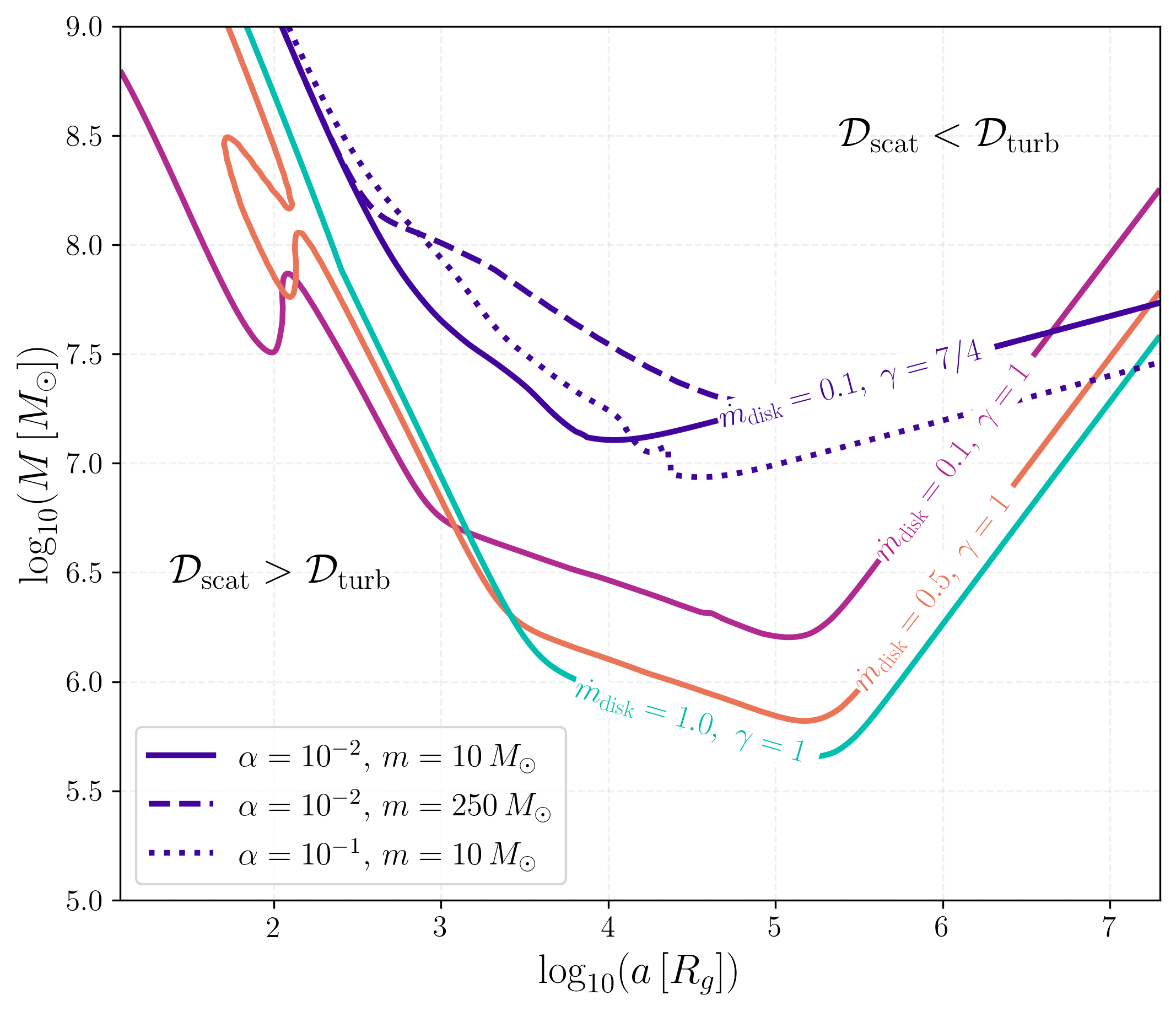}
    \caption{Isodiffusion contours where $\mathcal{D}_{\rm scat} = \mathcal{D}_{\rm turb}$ in the ($M/M_\odot$, $a/R_g$) plane. The dark purple contour shows the fiducial model ($\alpha = 0.01$, $m = 10 ~\rm{M}_\odot$, $\dot{m}_{\rm disk} = 0.1$, $\gamma = 7/4$). Dashed and dotted curves vary one disk parameter relative to fiducial: dashed uses $m = 250 ~\rm{M}_\odot$; dotted uses $\alpha = 0.1$. Solid magenta, orange, and aquamarine curves adopt a core profile for the NSC stars ($\gamma = 1$) with $\dot{m}_{\rm disk} = 0.1, ~0.5$ and 1.0, respectively.   $\mathcal{D}_{\rm scat}$ obeys a simple power law with semimajor axis $a$, but $\mathcal{D}_{\rm turb}$ has a more complex evolution driven by the AGN disk model.  Scattering achieves greater importance at the smallest and largest radii, where disk turbulence is weakened by declines in $\Sigma$ arising from radiation-pressure dominance and $Q_{\rm T}=1$ self-regulation, respectively.}
    \label{fig:Dturb_vs_Dscat}
\end{figure}

Having defined two independent sources of stochasticity -- MRI turbulence and stellar flybys from the nuclear cluster -- it is useful to compare their relative strength across a broader range of AGN-relevant parameters. Diffusion in $a$ and $e$ scales identically in the flyby and turbulence contexts:
\begin{equation}
\begin{split}
\frac{\mathcal{D}_{a, \, \rm scat}}{\mathcal{D}_{a, \, \rm turb}} = & \frac{\mathcal{D}_{e, \, \rm scat}}{\mathcal{D}_{e, \, \rm turb}} \simeq \frac{2 n^{-1}}{\alpha t_{\rm rel}} \left(\frac{M}{\Sigma_{\rm min} a^2}\right)^2\,.
\end{split}
\end{equation}
In Figure~\ref{fig:Dturb_vs_Dscat}, we evaluate the relative importance of these distinct processes by plotting the locus $\mathcal{D}_{\rm scat}/\mathcal{D}_{\rm turb} = 1$, across the $(M, a)$ plane for model parameters indicated in the caption.  

Global trends in $M$ and $a$ space can be identified across all of the curves shown in Figure~\ref{fig:Dturb_vs_Dscat}. In the lower-left quadrant of the figure (small $a$ and $M$), stellar scattering is always dominant. This follows from Equation~\eqref{eq:D_turb}, $\mathcal{D}_{e, \rm turb} \propto \Sigma_{\rm min}^2a^{5/2}M^{-3/2}$, which scales strongly with $a$. Note too that for fixed $a/R_g$, decreasing $M$ strengthens scattering, because $t_{\rm rel}$ decreases (smaller $a_{\rm infl}$ and larger $n_\star$), and in general stellar scattering dominates at low $M$. 

In addition to the  broad trends in $M$ and $a$, the balance between turbulence and scattering is also impacted by model-specific parameters including the nuclear cusp ($\gamma$), the disk viscosity ($\alpha$), the mass flux through the disk ($\dot{m}_{\rm disk}$), and the mass of the embedded object ($m$). A steeper cusp (larger $\gamma$) raises the stellar density at small radii, and the relative scattering diffusion rate increases as a result. In Figure~\ref{fig:Dturb_vs_Dscat}, the solid purple ($\gamma = 7/4$) and magenta ($\gamma = 1$) curves are differentiated only by their respective $\gamma$ and illustrate this dependence clearly -- with the scattering-dominant region limited to lower MBH masses in the $\gamma = 1$ case. 

Turbulent diffusion carries an explicit linear dependence on $\alpha$, but $\alpha$ also impacts turbulence implicitly through $\Sigma$. For a steady disk, decreasing $\alpha$ raises $\Sigma$, partially offsetting the explicit $\mathcal{D}_{\rm turb} \propto \alpha$ scaling. To clarify this point, it is useful to turn to the scaling relations given in \citet{Grishin2024}: In the outer, marginally stable disk $\Sigma \propto \alpha^{-1/3}$; deeper in, the inverse scaling steepens to $\alpha^{-1}$. This behavior is visible in the dotted dark purple curve ($\alpha = 10^{-1}$): In the outer disk $\mathcal{D}_{\rm turb}$ expands relative to the fiducial solid curve ($\alpha = 10^{-2}$), but inside $10^4 \, R_g$ the dotted contour lies at or above the solid line. 

The disk mass supply rate also influences $\mathcal{D}_{\rm turb}$ implicitly through $\Sigma$. A higher $\dot{m}_{\rm disk}$ raises $\Sigma$ and amplifies turbulent diffusion. This is especially true where Toomre regulation ties $\Sigma$ closely to the mass supply rate. Note that the magenta, orange, and aquamarine contours in Figure~\ref{fig:Dturb_vs_Dscat} differ only in their $\dot{m}_{\rm disk}$ values, which are $0.1$, $0.5$, and $1.0$ respectively. For $a > 10^3 \, R_g$, increasing $\dot{m}_{\rm disk}$ expands the region of parameter space where turbulence dominates scattering. 

In Figure~\ref{fig:Dturb_vs_Dscat}, the dashed purple line assumes $m = 250 \, \rm{M}_\odot$. This variable enters our diffusion model through $\Sigma_{\rm min}$, the locally depleted surface density near the orbiter. A more massive body opens a deeper partial gap, reducing the disk surface density and hence stochastic gravitational forcing from turbulence. The impact is most evident in the intermediate disk ($10^3 \leq a / R_g \leq 10^5$), where turbulent diffusion would otherwise dominate. 

\begin{figure}
    \centering
\includegraphics[width=0.98\linewidth]{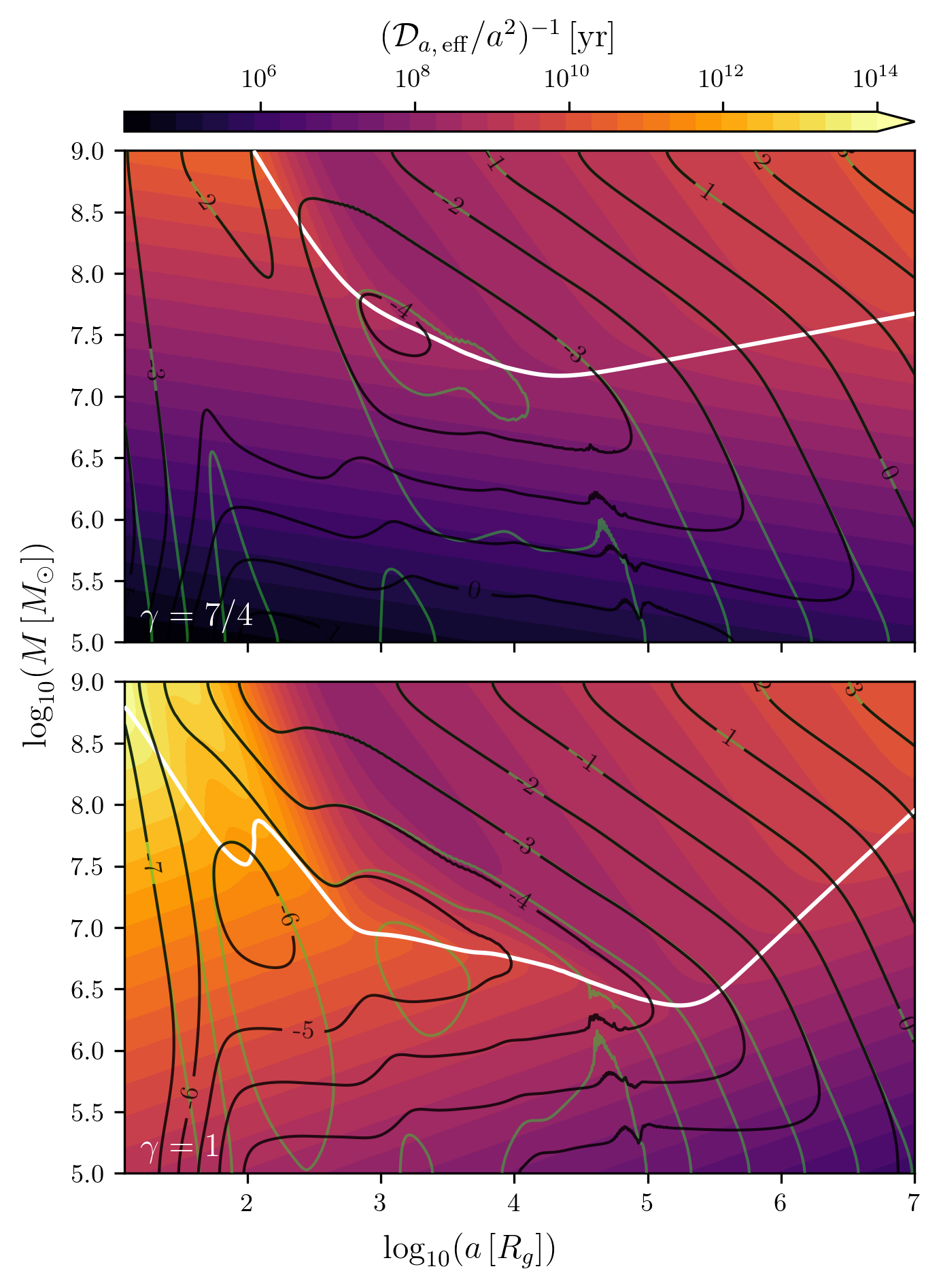}
    \caption{Inverse effective diffusion, $(\mathcal{D}_{a,\mathrm{eff}}/a^2)^{-1}$, across the $(\log_{10} (M\,[M_\odot]),\ \log_{10} (a\, [R_g])$ plane. Filled contours show $(\mathcal{D}_{a,\mathrm{eff}}/a^2)^{-1}$ (in years), computed from turbulent and scattering contributions added in quadrature. White contours indicate where $\mathcal{D}_{\rm scat} = \mathcal{D}_{\rm turb}$. \textit{Top:} scattering term evaluated for a stellar cusp with slope $\gamma=7/4$. Black contours show where the product of migration timescale and diffusion, $\log_{10}(\tau_{\rm I/II}\,\mathcal{D}_{a,\mathrm{eff}}/a^2)\in[-7,3]$. Green contours assume only Type I migration. \textit{Bottom:} $\gamma=1$.}
    \label{fig:InvDe_eff}
\end{figure}

Given that neither stellar scattering nor turbulence dominates across all parameter space, it is convenient to combine their effects into a single effective diffusion coefficient. Turbulence and stellar scattering are statistically independent, sourced by different components (gas versus stars). Here we propose a conservative effective diffusion $\mathcal{D}_{\rm eff}$, adding the turbulent and scattering coefficients in quadrature. In the remainder of this paper, we therefore use
\begin{equation}
\frac{\mathcal{D}_{a,\rm eff}}{a^2} \sim 2\mathcal{D}_{ e,\rm eff} \sim 2 \sqrt{\mathcal{D}_{e,\rm turb}^2 + \mathcal{D}_{e, \rm scat}^2}\,,
\end{equation}
to evaluate the stability of MMRs to stochastic forcing. We illustrate the effective semimajor-axis diffusion timescale, $(\mathcal{D}_{a,\rm eff}/a^2)^{-1}$, in Figure~\ref{fig:InvDe_eff} for fiducial disk parameters ($\alpha=0.01$, $\dot{m}_{\rm disk}=0.1$) and an embedded mass $m=10\,\mathrm{M_\odot}$. The top panel adopts a nuclear-cusp slope $\gamma=7/4$, and the bottom panel $\gamma=1$. White curves mark where scattering and turbulence contribute equally, $\mathcal{D}_{\rm scat}=\mathcal{D}_{\rm turb}$, corresponding to the solid purple and magenta contours in Figure~\ref{fig:Dturb_vs_Dscat}. In the $\gamma=7/4$ case, diffusion is strongest at low $a$ and $M$, generally weakening toward larger radii and higher MBH masses, with an island of enhanced diffusion at intermediate radii ($\sim10^{3}$–$10^{4}\,R_g$) and high $M$. A similar island appears for $\gamma=1$, though the overall peak shifts toward smaller $M$ and larger $a$. Black contours show the dimensionless product of the migration time and the diffusion coefficient, $\log_{10}(\tau_{\rm I/II}\,\mathcal{D}_{a,\rm eff}/a^2)$. Green contours display the same quantity computed under pure Type I migration (no gap opening). The interplay between migration and diffusion implied by these contours is examined in the next section.

\section{Criterion for Resonance Disruption}\label{sec:criterion_resonance_disruption}

With analytic representations of an embedded object's interactions with other bodies in the disk (\S\ref{sec:analytical_model}), the MBH (\S\ref{subsec:poincare}), the NSC (\S\ref{subsec:stellar_flybys}), and the disk itself (\S\ref{sec:migration} and \S\ref{subsec:turbulence}), we are now positioned to evaluate the stability of MMRs in AGN. 

Following \citet{2017batygin},
we track the evolution of the resonant offset $\xi \equiv n_2/n_1 - k/(k+1)$ and the composite eccentricity $\sigma_c$, which together quantify the system's distance from exact resonance. We model their evolution as a competition between turbulent forcing and disk-driven damping, yielding an analytically tractable Ornstein-Uhlenbeck (OU) process for the evolving variable $X_t \in \{\xi,\sigma_c\}$,
\begin{equation}
\mathrm{d}X_t = -\kappa X_t \mathrm{d}t + \sigma\mathrm{d}W_t ,
\end{equation}
where $\kappa$ is the linear damping rate toward $X=0$, $\sigma$ sets the strength of the stochastic forcing, and $W_t$ is a Wiener process. As $t\to\infty$, $X_t$ approaches a stationary Gaussian with variance $\sigma^2/(2\kappa)$.\footnote{A similar approach was taken by \citet{2025trani} in their investigation of the eccentricities and inclinations of BHs embedded in AGN disks. Modeling the turbulent velocity field acting on embedded BHs as a stochastic process and treating the resulting evolution of the eccentricity and inclination as 2D OU processes, they show that these elements relax to steady-state Rayleigh distributions whose variances are set by the local disk properties and the ratio of the damping rate to the orbital frequency.} 
Comparing the steady-state rms, $\langle X^2\rangle^{1/2}=\sigma/\sqrt{2\kappa}$, to the intrinsic resonance half-width $\Delta X_{\rm res}$ provides a quantitative disruption criterion: resonance breaks when $\langle X^2\rangle^{1/2}\gtrsim \Delta X_{\rm res}$.

In the following subsections, we apply this criterion to diffusion in $\xi$ (\S\ref{subsec:a_diffusion}) and in $\sigma_c$ (\S\ref{subsec:e_diffusion}), treating the two channels independently. Although $\sigma_c$ is one of the two quantities that set the resonant dynamics (\S\ref{sec:analytical_model}), we find that disruption along this axis is rare. The diffusion coefficients in the two channels are comparable, but eccentricity is damped on the timescale $\tau_e \sim (H/a)^2\,\tau_{\rm mig}$, far shorter than the migration timescale that damps $\xi$, and the resonance is wider in $\sigma_c$ than in $\xi$. The semimajor-axis channel therefore sets the stability boundaries in practice. When we map resonance stability across the AGN parameter space in \S\ref{sec:discussion} we consider disruption in $\xi$ alone.

\subsection{Diffusion of Semimajor Axes}\label{subsec:a_diffusion}

Diffusion in $\xi$ inherits from diffusion in $(a_1,a_2)$ via the multivariate It\^o chain rule. Assuming that the semimajor axes execute independent Gaussian diffusion with a common coefficient \(\mathcal D_{a,{\rm eff}}\), diffusion in $\xi$ is given by
\begin{equation}
\label{eq:Dxi}
\begin{aligned}
\mathcal D_\xi
& = \mathcal D_{a,{\rm eff}} \left[\left(\frac{\partial \xi}{\partial a_1}\right)^2 
+ \left(\frac{\partial \xi}{\partial a_2}\right)^2 \right]\\
& \, \simeq \left(\frac{3}{2}\right)^{\!2}\,
\frac{2\,\mathcal D_{a,{\rm eff}}}{\langle a\rangle^2} = 9 \mathcal{D}_{e, \rm eff}
 \,,
 \end{aligned}
\end{equation}
where, in the second line of Equation~\eqref{eq:Dxi}, we invoke the compact limit, $a_1\approx a_2\approx \langle a\rangle$.

Disk-driven damping of \(\xi\) follows from
\begin{equation}
\label{eq:dxi_dt}
\frac{\mathrm{d}\xi}{\mathrm{d} t} \;\simeq\; \frac{3\,\xi}{2}\!
\left(\frac{\dot a_1}{a_1}-\frac{\dot a_2}{a_2}\right)
\;=\; -\,\frac{3\,\xi}{2}\,\tau_{\rm rel}^{-1},
\end{equation}
where we define migration timescales $\tau_i \equiv a_i/\dot a_i$ and $\tau_{\rm rel}^{-1}\equiv \tau_2^{-1}-\tau_1^{-1}$. For fixed disk conditions and comparable $m_1,m_2$, $\tau_{\rm rel}\propto |m_1-m_2|^{-1}$, implying $\tau_{\rm rel}\gtrsim \tau_{\rm mig}(m_1{+}m_2)$. In what follows, we adopt the conservative replacement $\tau_{\rm rel}\to \tau_{\rm mig}(m_1{+}m_2)$, thereby maximizing resonance survival.

Combining Equations \ref{eq:Dxi} and \ref{eq:dxi_dt} yields an OU stochastic differential equation:
\begin{equation}
\mathrm{d}\xi = \sqrt{2\,\mathcal{D}_\xi}\,\mathrm{d}W_t - \frac{3}{2}\,\frac{\xi}{\tau_{\rm mig}}\,\mathrm{d}t \,,
\end{equation}
%
%
%\begin{equation}
%\mathrm{d}\xi = \frac{3}{2}\sqrt{\frac{2 \, \mathcal{D}_{a,\, \rm{eff}}}{\langle a \rangle^2}} \mathrm{d}W_t - \frac{3}{2}\,\frac{\xi}{\tau_{\rm mig}}\mathrm{d}t \,,
%\end{equation}
%
so that the steady-state rms is
\begin{equation}
\label{eq:delta_xi}
    \delta\xi = \sqrt{\frac{2}{3}\,\mathcal{D}_\xi\,\tau_{\rm mig}}\,,
\end{equation}
%
%
%\begin{equation}
%\label{eq:delta_xi}
%\delta\xi = \sqrt{\frac{3}{2}\frac{ \mathcal D_{a, \, \rm eff}\,\tau_{\rm mig}}{\langle a\rangle^2}}\,.
%\end{equation}
%

The resonance width $\Delta \xi$, is maximized at the onset of libration (i.e., when the resonance first appears at $\delta = 0$) and is given by \citep{2015batygin}
\begin{equation}
\label{eqn:Delta_xi}
\Delta \xi \simeq 5 \left[ \frac{\sqrt{k} \; (m_1 + m_2)}{M}\right]^{2/3} = 5\, k^{1/3}q^{2/3} \,.
\end{equation}
Here $\Delta \xi$ is the period-ratio offset at which a pair on near-circular orbits exits the resonance, whereas $\delta \xi$ is a measure of the width of the distribution in $\xi$ under stochastic evolution. As argued in \citet{2017batygin}, disruption of the resonance occurs when the ratio of these quantities exceeds unity:
\begin{equation}
\label{delta_xi_Delta_xi}
\begin{aligned}
\frac{\delta\xi}{\Delta\xi}
\simeq {}&
1.3\times10^{3}\,
\sqrt{\mathcal{D}_\xi\,\tau_{\rm mig}}
\\
&\times
\left(\frac{k}{2}\right)^{-1/3}
\left(\frac{q}{10^{-6}}\right)^{-2/3}
\gtrsim 1\,.
\end{aligned}
\end{equation}
We can estimate where this condition is met by turning back to Figure~\ref{fig:InvDe_eff}, whose contours give $\log_{10}(\tau_{\rm mig}\,\mathcal{D}_{a,\rm eff}/a^2)$. Using $\mathcal{D}_\xi \simeq (9/2)\,\mathcal{D}_{a,\rm eff}/\langle a\rangle^2$ (Equation~\ref{eq:Dxi}), Equation~\eqref{delta_xi_Delta_xi} places the disruption threshold at $\tau_{\rm mig}\,\mathcal{D}_{a,\rm eff}/\langle a\rangle^2 \gtrsim 1.3\times10^{-7}$ for $m_1 + m_2 \simeq 10\,{\rm M_\odot}$ and $M \simeq 10^7\,{\rm M_\odot}$, i.e., the $-7$ contour of Figure~\ref{fig:InvDe_eff}. Diffusion approaches this threshold only for $\gamma = 1$ and $a \lesssim 10^2\,R_g$, assuming a Type I/II torque prescription.
%
%\begin{equation} \label{delta_xi_Delta_xi}
%\begin{split}
%\frac{\delta \xi}{\Delta \xi} \simeq & \, 1.9 \times 10^3 \, \sqrt{\left(\frac{\mathcal{D}_{a, \rm eff} \tau_{\rm mig}}{\langle a\rangle ^2}\right)} \left(\frac{k}{2}\right)^{-1/3}  \\ 
%& \times \left(\frac{M}{10^7\, \rm M_\odot}\right)^{2/3} \left(\frac{m_1 + m_2}{10 \, M_\odot}\right)^{-2/3}  \gtrsim 1 \,.
%\end{split}
%\end{equation}
%
%In practice, it is convenient to track the related diagnostic $x \equiv \log_{10}(\tau_{\rm mig}\,\mathcal{D}_{\xi})$, which is the quantity displayed in Figure~\ref{fig:InvDe_eff}. Equation~\eqref{delta_xi_Delta_xi} then places the disruption threshold at $x \gtrsim -6$ for $m_1 + m_2 \simeq 10\,{\rm M_\odot}$ and $M \simeq 10^7\,{\rm M_\odot}$. Turning to Figure~\ref{fig:InvDe_eff}, we find $\tau_{\rm mig}\,\mathcal{D}_{\xi}$ large enough to approach this threshold only for $\gamma = 1$ and $a \lesssim 10^3\,R_g$, assuming a Type I/II torque prescription. % $\tau_{\rm mig}\mathcal{D}_{e,{\rm eff}} \sim 10^x$: resonance breaking is expected near $x \sim 10^{-6} \, (q/10^6)^{-2/3}$, where $q \equiv(m_1 +m_2)/M$.  Turning to Figure~\ref{fig:InvDe_eff}, for $m_1{+}m_2\simeq10 \,{\rm M_\odot}$ and $M\simeq10^{7}\,{\rm M_\odot}$ we find $\mathcal{D}_{e,\rm eff}^{-1}$ large enough (and $\tau_{\rm mig}$ short enough) to approach $x \sim 10^{-6}$ only for $\gamma =1$ and $a\lesssim10^3\, R_g$, assuming a Type I/II torque prescription.

\begin{figure}
    \centering
\includegraphics[width=0.98\linewidth]{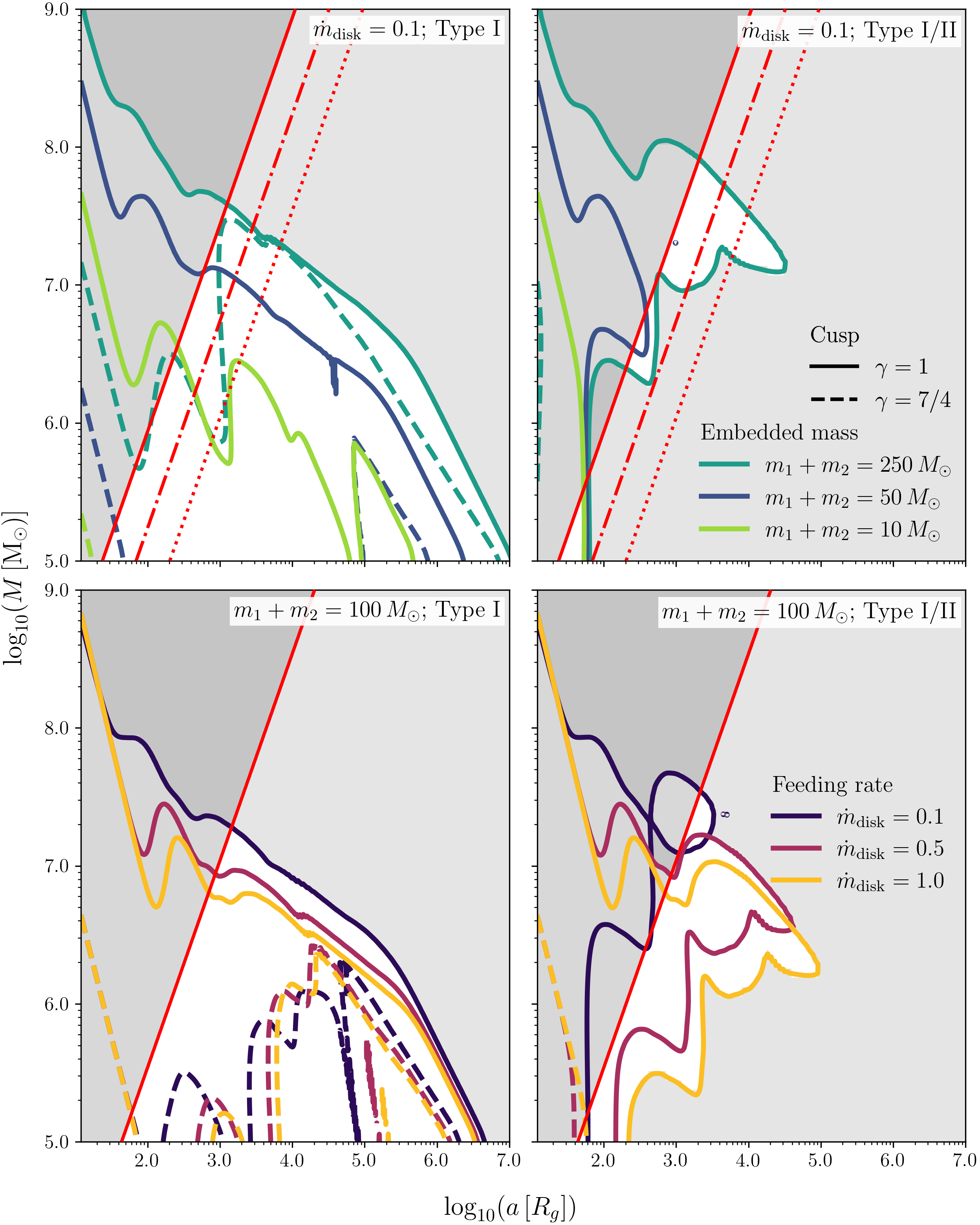}
    \caption{Contours indicate where the resonance destabilization condition $\delta \xi = \Delta \xi$ is met in the ($\log_{10}(M/\rm M_\odot)$, $\log_{10}(a/R_g)$) plane. %This contour marks the threshold for stochastic resonance disruption; 
    Gray shading above and to the right of the contours indicates regions where MMRs are expected to break due to stochastic forcing. Overplotted in red are loci where deterministic GR forcing from apsidal precession is expected to break resonances. All curves assume $k = 2$. \\ 
    \textit{Top left:} Eddington ratio $\dot{m}_{\rm disk} = 0.1 \, [\dot{\mathrm{M}}_{\rm Edd}]$. Colors denote the summed embedded-object masses: aquamarine, $250 \, \rm M_\odot$; dark blue, $50 \, \rm M_\odot$; green, $10 \, \rm M_\odot$. Solid curves assume a nuclear stellar cusp with slope $\gamma = 1$; dashed curves show $\gamma = 7/4$. The solid red line corresponds to GR-induced resonance breaking for the largest embedded mass ($250 \, \rm M_\odot$), the dashed-dotted line to $50\, \rm M_\odot$, and the dotted line to $10 \, \rm M_\odot$. \\
    \textit{Top right:} same color/line scheme as top left, but migration rates assume the Type I/II prescription. \\
    \textit{Bottom left and right}: fixed embedded-object mass $m = 100 \, \rm M_\odot$. Colors now indicate the disk mass flux in Eddington units: dark purple, $\dot{m}_{\rm disk} = 0.1$; magenta, $\dot{m}_{\rm disk} = 0.5$; orange, $\dot{m}_{\rm disk} = 1.0$. Solid versus dashed lines retain the $\gamma =1$ versus $\gamma = 7/4$ convention. The solid red line still indicates where GR forces dominate resonant coupling, but is now set according to the fixed mass $m_1 + m_2 = 100 \, \rm M_\odot$.}
    \label{fig:delta_chi_Delta_chi}
\end{figure}

We explicitly illustrate the resonance-breaking criterion in Figure~\ref{fig:delta_chi_Delta_chi}, plotting the locus $\delta\xi=\Delta\xi$ in the $(\log_{10}M,\ \log_{10}a)$ plane for $k=2$. Only the semimajor-axis criterion is shown; as demonstrated in \S\ref{subsec:e_diffusion} below, resonance disruption due to diffusion in $\sigma_c$ is subdominant by several orders of magnitude and does not alter the plotted boundaries. Regions shaded gray (above and to the right of the curves) are expected to undergo stochastic resonance breaking, while those below are stable. The top row compares Type I (left) to Type I/II (right) migration prescriptions; line colors encode either the total embedded mass (top row) or the Eddington ratio (bottom row), while solid versus dashed styles denote $\gamma=1$ versus $\gamma=7/4$. Red curves overplot the independent condition for secular GR-induced disruption; where these lie below the stochastic boundary, GR precession alone is sufficient to unbind the resonance. 

Broadly, in regions where turbulent forcing dominates -- typically at larger $a\gtrsim10^{3}\,R_{\rm g}$ and toward higher $M$ -- the effective diffusion rate $\mathcal{D}_{a,{\rm eff}}/a^2$ is large relative to $\tau_{\rm mig}$ and resonances are more fragile. Steeper cusps ($\gamma = 7/4$) further shrink the stable region, particularly between $a \sim 10^{2-4} \, R_{\rm g}$ where diffusion timescales are minimized and $\tau_{\rm mig}$ is relatively large. At the smallest radii, GW-driven migration can restabilize resonances, but in this inner zone secular GR forcing itself typically destabilizes them.

When we include gap opening, the region of parameter space that supports stable resonances contracts markedly. This is most evident at low MBH mass $M$, where the gap–opening parameter $K\propto(m/M)^2$ is large and the surface density is depleted around the embedded mass. Increasing the AGN Eddington ratio $\dot m_{\rm disk}$ partly offsets this by raising $\Sigma$ and the scale height $H$, thereby reducing $K$. When we include Type II torques and neglect thermal torques, we therefore expect MMRs to largely disappear, surviving only in small islands at intermediate radii and for sufficiently large embedded masses.

In Figure~\ref{fig:delta_chi_Delta_chi_w_thermal} we recompute $\tau_{\rm mig}$ including thermal torques. The layout mirrors Figure~\ref{fig:delta_chi_Delta_chi}: The top row varies the total embedded mass, the bottom row varies the disk mass flux, and we compare Type I (left) to Type I/II (right) prescriptions for angular-momentum exchange. Solid and dashed lines differentiate the nuclear-cusp slope, and dotted curves mark the location of traps/anti-traps where $\tau_{\rm mig}\to\infty$. Recall that our disruption criterion depends on the magnitude of the migration timescale and is agnostic to the torque sign. Notably, in the Type II regime at low $M$, the inclusion of thermal torques stabilizes resonances interior to the migration trap, where the hot (outward) thermal torque dominates. We conclude that, so long as a migrator is sufficiently luminous, resonance locking can remain robust in this band. 

At migration traps (or anti-traps) the net torque is $\Gamma_{\rm tot}(a)=0$, so a single body has $\dot a=0$ and hence $\tau_{\rm mig}\to \infty$, implying resonance breaking. However, our disruption criterion used $\tau_{\rm mig}$ as a convenient proxy for the \textit{relative} migration timescale, which is overly conservative near traps. Unless both bodies sit exactly at the same zero-torque radius, their torques differ slightly and $\tau_{\rm rel}$ will be finite. $\tau_{\rm rel}$ is set by the local slope $\partial \Gamma / \partial a$ and the pair's separation across the trap. Consequently, where trap--anti-trap curves intersect regions of wide-spread resonant stability -- where diffusion is modest and migration is fast --  long-lived resonant locking will persist, even though the single-body $\tau_{\rm mig}$ formally diverges at the trap center. 

One additional parameter defines the disk models but is not varied in Figures~\ref{fig:delta_chi_Delta_chi} and~\ref{fig:delta_chi_Delta_chi_w_thermal}: the viscosity parameter $\alpha$, which we fix at $0.01$ but which may be an order of magnitude larger in AGN disks. We estimate, however, that $\alpha$ has only a limited effect on the boundaries of MMR stability. Although the turbulent forcing ($\mathcal{D}_{a,\rm turb} \propto \alpha\Sigma_{\rm min}^2$) and the migration time ($\tau_{\rm I} \propto (H/a)^2/\Sigma$) each depend on $\alpha$ through $\Sigma$ and $H$, the disruption criterion depends only on their product, $\delta\xi^2 \propto \mathcal{D}_{a,\rm turb}\tau_{\rm mig} \propto \alpha\,\Sigma_{\rm min}(H/a)^2$. In a steady $\alpha$-disk, $\nu\Sigma = \dot M/3\pi$ implies $\alpha\Sigma(H/a)^2 = \dot M/(3\pi n a^2)$ regardless of the opacity or pressure regime, so that, absent gap opening, $\delta\xi^2 \propto \dot m_{\rm disk}\,(M/m)\,n^{-1}$ and the dependence on $\alpha$ cancels. Where stellar scattering dominates instead, $\delta\xi^2 \propto \mathcal{D}_{\rm scat}\tau_{\rm I}$ retains only the weak $\alpha$-dependence of $\tau_{\rm I}$ ($\propto \alpha^{-1/3}$ in the outer disk). The viscosity otherwise enters through gap opening ($K \propto \alpha^{-1}$, so that a larger $\alpha$ weakens gaps and moves the Type I/II stability bounds toward those of the Type I regime) and through the balance between turbulence and scattering, shown in Figure~\ref{fig:Dturb_vs_Dscat}. There, increasing $\alpha$ by an order of magnitude (solid versus dotted purple) modestly enlarges the turbulence-dominated region in the outer disk, where $\Sigma \propto \alpha^{-1/3}$,  but shrinks it in the inner disk, where $\Sigma \propto \alpha^{-1}$ to $\alpha^{-4/5}$ (see the analytical scaling relations derived in Equations (2)--(4) of \citet{Grishin2024}).  The resonance width, libration frequency, GR boundary, and scattering diffusion are independent of the disk parameters altogether.

Note that our treatment emphasizes an absolute disruption criterion. However, in regions of parameter space where resonance breaking is predicted, the OU process reduces -- to first order -- to pure diffusion in $\xi$. A simple break-time estimate then follows by setting the rms equal to the resonance width, $\sqrt{2\mathcal D_\xi \, t_{\rm break}}=\Delta\xi$, yielding
\begin{equation}
\begin{split}
t_{\rm break}  \simeq  \frac{(\Delta \xi)^2}{2 \mathcal{D}_{\xi}}&\simeq 2\times10^{2}  \, \mathrm{yr} \,\left(\frac{k}{2}\right)^{2/3} \\
& \quad \times \left(\frac{q}{10^{-6} \, \rm }\right)^{4/3}  \left(\frac{\mathcal{D_\xi}}{10^{-9}\, \rm yr^{-1}}\right)^{-1}\,.
\end{split}
\end{equation}
Once a migrator stochastically exits one resonance, it may drift and subsequently be captured into a more closely spaced first-order resonance (larger $k$) or a resonance of different order. Thus while individual resonances may break, the time-averaged probability of being in \textit{some resonance} over the disk lifetime can remain substantial. Quantifying this occupancy -- and its implications for merger rates -- will be the subject of future work. 

\subsection{Diffusion in Eccentricity}\label{subsec:e_diffusion}

If the two components of the
composite eccentricity vector experience diffusion with coefficient $\mathcal{D}_{e,\rm eff}$, then the combined diffusion amplitude is $\mathcal{D}_{\sigma_c} = 2\,\mathcal{D}_{e,\rm eff}$, so the stationary dispersion in $\sigma_c$ is
\begin{equation}
   \delta\sigma_c \simeq \sqrt{\mathcal{D}_{\sigma_c}\,\tau_e}
\end{equation}
where $\tau_e$ is the eccentricity damping timescale defined in \S\ref{subsec:e_damping}. The corresponding resonance width in
eccentricity is the $\Delta\sigma_c$ of Equation~(\ref{eq:Delta_sigma}), the
change in $\sigma_c$ that carries a pair from the resonant equilibrium to the
edge of the resonant island. Combining these, we obtain a second criterion for resonance
disruption, this time in eccentricity:
\begin{equation}\label{eq:sigma_criterion}
\begin{split}
   \frac{\delta\sigma_c}{\Delta\sigma_c}
    &\simeq \left(\frac{15}{4}\frac{kM}{m_1+m_2}\right)^{1/3} \sqrt{\mathcal{D}_{\sigma_c}\,\tau_e}\\ &
   \simeq 2 \times 10^{2} \left(\frac{k}{2}\right)^{1/3}
   \left(\frac{q}{10^{-6}}\right)^{-1/3} \sqrt{\mathcal{D}_{\sigma_c}\,\tau_e} \;\gtrsim\; 1\,.
\end{split}
\end{equation}

Up to order-unity factors, one finds $\mathcal{D}_{\sigma_c} \sim \mathcal{D}_\xi$. If $\tau_e$ follows the classic \citet{2004tanaka} scaling (Equation~\ref{eq:TW04}), then $\tau_e \sim (H/a)^2\,\tau_{\rm I/II}$ and $\delta \sigma_c \ll \delta \xi$. Combining this relationship with the two widths $\Delta \xi \simeq 5 (\sqrt{k}q)^{2/3}$ (Equation~\ref{eqn:Delta_xi}) and $\Delta \sigma_c \simeq (q/k)^{1/3}$, the ratio of the two criteria is 
\begin{equation}
\begin{split}
\frac{\delta\sigma_c/\Delta\sigma_c}{\delta\xi/\Delta\xi}
   &
   \sim 5\,\frac{H}{a}\,k^{2/3}\,q^{1/3}
   \\ & 
   \sim 10^{-4}\left(\frac{H/a}{10^{-3}}\right)\left(\frac{k}{2}\right)^{2/3}
   \left(\frac{q}{10^{-6}}\right)^{1/3}\,.
\end{split}
\end{equation}
Together these imply that resonance breaking proceeds far more rapidly in the $\xi$ channel than through eccentricity.

Including thermal effects does not qualitatively change this conclusion. Where cold thermal torques dominate, eccentricity damping is strengthened and resonances are more robust to diffusion. Where hot thermal torques dominate, eccentricity can grow toward the disk aspect ratio, $e \sim H/a$ \citep{2017eklund, 2019fromenteau}, and by Equation~(\ref{eq:Delta_sigma}) this reaches the width of the island wherever $H/a \gtrsim 0.64\,(q/k)^{1/3} \approx 5 \times 10^{-3}$ for $q\simeq 10^{-6}$ and $k =2$ -- i.e., in the outer regions of the \citet{sirko2003} disk. In such regions, the pair also leaves the near-circular regime in which our resonant model applies (\S\ref{sec:analytical_model}). Since higher eccentricities make resonant capture less likely \citep{2015batygin}, our estimate of MMR occurrence in these regions can be considered a conservative upper bound.

\section{Discussion}\label{sec:discussion}

\begin{figure}
    \centering
\includegraphics[width=0.98\linewidth]{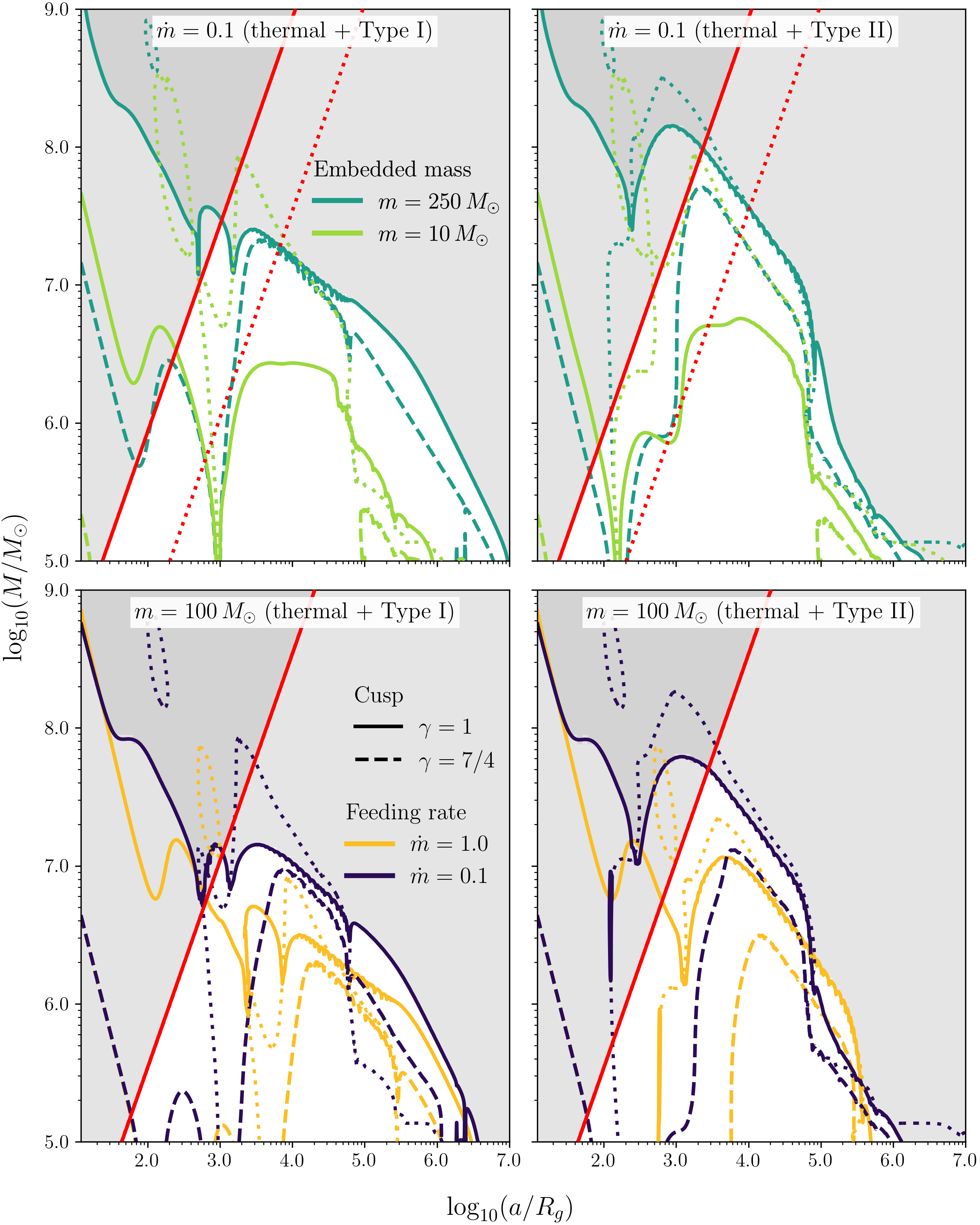}
    \caption{Contours where $\delta \xi = \Delta \xi$ in the ($\log_{10}(M\, [\rm M_\odot])$, $\log_{10}(a\, [R_g])$) plane, now including thermal torques in the migration prescription. As before, gray shading above the contours denotes regions where MMRs are expected to break. All panels assume $k = 2$. Resonance-breaking contours are shown as solid (dashed) lines for cusp slope $\gamma = 1$ ($\gamma = 7/4$). Dotted curves indicate the locations of thermal migration traps/anti-traps. \\
    \textit{Left column:} thermal + Type I migration. \\
    \textit{Right column:} thermal + Type I/II migration. \\
    \textit{Top row (varying embedded mass)}: colors denote total embedded mass: aquamarine, 250 $\rm M_\odot$; green, 10 $\rm M_\odot$. Overplotted in red are GR resonance-breaking loci: solid red for 250 $\rm M_\odot$ and dotted for 10 $\rm M_\odot$. \\
    \textit{Bottom row (varying the Eddington ratio):} colors denote $\dot{m}_{\rm disk}$: orange, $\dot{m}_{\rm disk} = 1.0$; purple, $\dot{m}_{\rm disk} = 0.1$, in Eddington units.}
    \label{fig:delta_chi_Delta_chi_w_thermal}
\end{figure}

Do MMRs exist for embedded stellar-mass objects in AGN disks?  Yes, sometimes.  Are MMRs common?  The answer to this question is more nuanced, and ultimately model dependent, as we have seen in Figures~\ref{fig:delta_chi_Delta_chi} and \ref{fig:delta_chi_Delta_chi_w_thermal}.  If we neglect the role of thermal torques (Figure~\ref{fig:delta_chi_Delta_chi}), then MMRs are quite rare, and exist only over quite narrow radial ranges for AGN in the mass window $10^5 \lesssim M/M_\odot \lesssim 10^7$.  But if we consider thermal torques (Figure~\ref{fig:delta_chi_Delta_chi_w_thermal}), then the picture for MMR survival becomes more favorable, and MMRs may persist over 1-3 orders of magnitude in radius $a/R_{\rm g}$ within the same mass window ($10^5 \lesssim M/M_\odot \lesssim 10^7$).  Because the thermal torque depends on uncertain details of accretion feedback into AGN disks, we have considered the durability of resonances with and without it (and will discuss this further in \S \ref{subsec:caveats}).

Why does it matter whether or not MMRs are stable in AGN?  Stars and stellar-mass BHs in AGN generally experience differential migration, and will frequently overtake one another when moving through the AGN disk.  {\it Convergent} differential migration will thus often create the conditions for resonance capture, and if an $\mathcal{O}(1)$ fraction of migrating BHs do indeed capture into MMRs, the implications for GW production in AGN will be significant.  As an extreme case, we can imagine a scenario in which all migrating BHs capture into resonant chains or convoys, and in which migration traps are wholly absent from AGN.  In this situation, it would become {\it impossible} for AGN to synthesize new binary-BH systems out of individual embedded BHs.  As binary assembly from single-single capture (mediated by either gas dynamical friction or minidisk interactions during close flybys) has the greatest potential for producing a large population of binary BHs in AGN, this scenario would greatly reduce the maximum LVK-band merger rate from the AGN channel.\footnote{This ``worst case'' scenario would not completely deactivate the AGN channel for LVK-band mergers, as binary BHs could still be captured from a preexisting NSC population or formed {\it in situ} by Toomre instability.} Moreover, stable MMR capture would completely foreclose the possibility of repeated mergers and hierarchical growth of BHs up to, for example, characteristic masses forbidden by single-star evolution, which today is viewed as one of the most promising ``smoking guns'' of the AGN channel.  This would also alter the mass function of wet EMRIs visible in the {\it LISA} band relative to the situation in which MMRs are rare \citep{2022pan}.

In the previous sections, however, we have seen that MMRs in AGN disks are generally fragile.  In particular, leading-order post-Newtonian precession will overwhelm apsidal libration and destroy MMRs at sufficiently small radii $a$ (Figure~\ref{fig:timescales}). MMRs can persist down to smaller dimensionless radii $a/R_{\rm g}$ for less extreme mass ratios $q=(m_1 + m_2)/M$, favoring resonance survival for (i) IMBH primaries and (ii) relatively large stellar-mass BHs\footnote{This factor is noteworthy given the recent LVK observations of $m\sim 100M_\odot$ BHs in the pair-instability mass gap forbidden to single-star evolution.  Such BHs are likely the product of repeated mergers and hierarchical growth, which may only be possible in AGN disks due to the challenge that GW recoil kicks pose for merger product retention.}.  Ultimately, any MBH-BH combination will only be capable of producing stable resonances outside of a critical radius
\begin{align}
    R_{\rm GR} =& \frac{2}{(3 f_{k+1,27})^{1/3}} \frac{k^{8/9}}{(k+1)^{14/9}} \left(1-\left( \frac{k+1}{k}\right)^{-5/3} \right) \notag \\
    & \times R_{\rm g} \left( \frac{M}{m_1+m_2} \right)^{2/3}
\end{align}
If we imagine a stable pair of binary BHs migrating inwards in resonance with each other, the MMR will break once they reach $R \approx R_{\rm GR}$.  Interestingly, however, $R_{\rm GR}$ is a decreasing function of the MMR resonance index $k$ (as what ultimately matters is not the single-body apsidal precession rate $\dot{\varpi}_{\rm GR}$ but rather the differential apsidal precession rate $\Delta \dot{\varpi}_{\rm GR}$), so if the migration is still convergent at $R_{\rm GR}$, the two objects may cascade through a sequence of increasingly closely spaced first-order MMRs (increasing $k$). 

While deterministic precession always eliminates MMRs at $a< R_{\rm GR}$, stochastic forcing will sometimes, though not always, have the same effect at larger radii. We have seen that stochastic perturbations are most destabilizing to MMRs around larger MBHs.  For the largest MBHs, disk turbulence is the dominant form of stochastic perturbation, while for the smallest MBHs, it is instead stellar scattering that dominates.  The transition between these two types of perturbations is radius dependent, and further depends on AGN parameters (see Figure~\ref{fig:Dturb_vs_Dscat}), but in the radii of greatest interest the transition from scattering-dominated to turbulence-dominated stochasticity occurs for $M / \rm M_\odot \sim 10^{6-7.5}$.

Accounting for both deterministic (Figure~\ref{fig:timescales}) and stochastic (Figure~\ref{fig:delta_chi_Delta_chi}) destabilization of MMRs, we see that MMRs are generally fragile in most AGN.  For low-mass migrators ($m \sim 1 M_\odot$), MMRs will rarely be stable in any AGN disks.  Higher-mass migrators ($m \sim 100 M_\odot$) are more robust against GR precession and disk turbulence.  As mentioned previously, MMRs only exist in a narrow radial range if thermal torques are neglected.  In the case when thermal torques do exist and are described by standard formulae \citep{Masset2017}, we can identify three distinct AGN regimes:
\begin{enumerate}
    \item High-mass AGN ($M \gtrsim 10^{7.5} M_\odot$): stable MMRs never exist.
    \item Low-mass AGN ($M \lesssim 10^{6.5} M_\odot$): stable MMRs always exist.  They are generally destabilized by stellar scatterings for $a \gtrsim 10^{5-6} R_{\rm g}$ and by GR precession for $a \lesssim 10^{2-3} R_{\rm g}$; stochastic effects may sometimes play a destabilizing role in the $a \sim 10^{2-3} R_{\rm g}$ range as well.  Regions of MMR stability are generally interior to a migration trap and exterior to an anti-trap.
    \item Intermediate-mass AGN ($10^{6.5} \lesssim M/M_\odot \lesssim 10^{7.5}$): in this regime MMRs may or may not be stable depending on system parameters; generally, MMR stability is favored by lower $M$, lower $\dot{m}_{\rm disk}$, and higher $m$ values.  As with the low-mass case, radial zones of MMR stability are generally in between a trap and an anti-trap (i.e., they are regions of outward migration).
\end{enumerate}

There are several observable implications of this varied MMR landscape.  First, we can say that in high-mass AGN, convergent migration will efficiently form LVK-band BH-BH binaries, as MMRs are too unstable to prevent this from occurring.  In low- and intermediate-mass AGN, MMRs will be common in the zone of outward migration that lies between an inner anti-trap and an exterior trap produced by thermal torques.  Whether this affects LVK production rates will depend on the radial origin of most single BHs embedded in AGN disks.  If the majority of such BHs are formed {\it in situ} via Toomre fragmentation, it will usually be at larger radii, from which they can migrate inwards toward migration traps. Across much of the MBH parameter space, traps lie along or exterior to MMR stability bounds, leading to pileups and enhanced likelihood for mergers. Narrow MBH windows exist (just below and above $10^6 \, \rm M_\odot$) where a stable-MMR swath lies exterior to the trap, so resonant capture could delay or suppress mergers. For sufficiently massive embedded objects, gap opening can lower local opacity and thereby weaken thermal torques; such BHs experience Type II migration only, and pass through the trap. We return to these caveats in \S\ref{subsec:caveats}.

BHs and other stellar-mass objects may originate in the inner disk if they are captured by gas drag from preexisting NSC populations.  While retrograde captures will be quickly excited to high eccentricity and lost from the AGN disk \citep{2023generozov}, prograde captures will circularize.  Realistic $\Sigma(r)$ profiles imply that capture rates will fall off as functions of radius, so that a large fraction of gas-captured BHs may originate in the region of outward migration (where MMRs are most likely to be stable).  It is only for this subset of disk-embedded BHs that stable resonant chains can reduce orbit crossings and LVK binary formation rates.  Given the complexity of modeling the joint evolution of AGN disks and their embedded BH populations \citep{2025epstein-martin}, we do not attempt in this paper to determine the overall impact of MMRs on LVK-band merger rates, but note here that we have isolated the particular $M$ and $a/R_{\rm g}$ values where they {\it can} matter.

Because MMR stability appears coextensive with regions of outward migration due to thermal torques, MMRs appear unlikely to have a major impact on wet EMRI rates and properties.  The no-MMR regime of high-mass AGN is not relevant for wet EMRIs, as $M \gtrsim 10^{7.5} M_\odot$ lies well outside the {\it LISA} band.  For low- and intermediate-mass AGN whose EMRIs will be detectable, MMRs could in principle affect EMRI properties (if, for example, they were ubiquitous enough to prevent hierarchical mergers from repeated binary formation) but this appears not to be the case.

The previous three paragraphs have worked under the assumption that thermal torques exist and can be described by standard formulae from the protoplanetary literature.  These formulae rely on a hierarchy of distance scales, $x_{\rm c} < \lambda < H$, as well as the broader assumption that heating from the embedded object only creates a linear perturbation in the disk structure (i.e., $L_{\rm Edd}(m)/H^2 < \sigma T_{\rm eff}^4$, assuming that the accretion luminosity of the embedded object is Eddington limited).  Both the linear perturbation assumption and the $\lambda < H$ assumption will break down at the largest radii \citep{Grishin2024}.  Linearity of the heating perturbation becomes more challenging to maintain for smaller $M$ values ($M \lesssim 10^7$).  If thermal torques are dramatically weakened in these regimes and the situation becomes more Type I-like, then MMRs will be even less robust.

\subsection{Imprints on Gravitational Waveforms}

If high-$k$ MMRs can survive down to $\sim 10 - 100 \, R_{\rm g}$, an EMRI's waveform could be sensitive to the accompanying perturbations. {\it LISA} will track up to $\sim 10^5$ orbital cycles \citep{2023amaroseoane}, making even weak environmental forces measurable. Prior work has shown that gas torques (Type I/II), accretion drag, galactic potentials, light scalar fields, and tidal resonances can all imprint detectable phase shifts \citep[e.g.,][]{2011kocsis, 2011yunes, 2012amaro-seoane, 2017yang, 2019bonga, 2022cardoso, 2023barsanti}. A long-lived high-$k$ resonant companion would impart quasiperiodic torques leading to cumulative phase drift that could, in principle, be identified with {\it LISA}. 

For stellar-mass binaries in the LVK band, maintaining resonance to $\sim 10 - 100 \, R_{\rm g}$ would enable single-single capture and merger deep in the MBH potential, where strong-field effects are unavoidable. The center-of-mass acceleration induces time-varying Doppler shifts and gravitational redshift \citep{2017inayoshi}. Close alignment can yield GW ``echoes'' (a secondary, delayed copy of the signal) and strong-lensing effects \citep{2017meiron, 2022gondan, 2025xu}. Although a resonant cascade through higher $k$ indices is for now speculative, and could be derailed by (for example) higher-order post-Newtonian terms or high-$k$ resonance overlap, the prospect of strong-field signatures for both LVK- and {\it LISA}-band GWs merits further investigation.

\subsection{Caveats and Future Work}\label{subsec:caveats}

 As noted in \S\ref{sec:migration}, there is not yet consensus on the most appropriate AGN disk model -- especially in the outer disk, where the high mass flux needed to power an AGN drives the disk toward self-gravitating collapse. In the class of models most commonly used to describe AGN structure, this tendency is explicitly countered by self-regulating star formation feedback. We use the \citet{sirko2003} framework because it provides a clear parameterization of steady-state disk properties with a radially constant $\dot{m}_{\rm disk}$, isolating the impact of the disk mass flux. This choice is, however, somewhat less physically motivated than the common alternative, proposed by \citet{2005thompson}, which includes star-formation-driven gas consumption, and decreasing $\dot{m}_{\rm disk}$ nearer to the MBH. The decrease in $\dot{m}_{\rm disk}$ across the Thompson-type disk maps to a lower surface density $\Sigma$, thereby reducing turbulent diffusion and potentially enlarging regions where MMRs can survive -- highlighted in Figures~\ref{fig:delta_chi_Delta_chi} and \ref{fig:delta_chi_Delta_chi_w_thermal}. 

Alternative sources of disk-stabilizing pressure support have also been proposed: Both magnetic pressure \citep{2024hopkins, 2025gerling-dunsmore} and accretion onto embedded stellar-mass BHs \citep{2003levin, 2007levin} have been shown to provide support comparable to, or exceeding, feedback from star formation. This additional heating can reshape the disk thermodynamics and surface-density profile, as well as the expected number density of embedded objects \citep{gilbaum2022, 2025epstein-martin}. Beyond setting the disk structure, the mass and number density of embedded bodies contribute to an effective turbulent viscosity. Strong turbulence can modify migration torques -- introducing stochastic kicks -- so that trajectories resemble a random walk rather than smoother Type I/II drift \citep{2010baruteau}. Indeed, \citet{2024wu} find that for $\alpha_{\rm disk}\gtrsim0.1$, turbulent torques dominate. 

Environmental factors -- such as gas metallicity, opacity prescriptions, and nonaxisymmetric gravitational structure from spiral arms or bars -- also alter AGN disk profiles. For example, an enhanced effective viscosity due to external torques tends to lower $\Sigma$, which in turn weakens turbulent diffusion and can promote MMR stability. Quantifying these effects in AGN conditions remains an open problem. We regard a systematic exploration as important future work but beyond the scope of this paper. 

Beyond the disk physics considered above, migration of embedded bodies can independently restructure the disk density profile. In particular, sufficiently massive perturbers open gaps that decouple them from the local disk environment, suppressing thermal torques. We do not impose this criterion in our analysis, but note that the threshold mass is sensitive to both $\alpha_{\rm disk}$ and the MBH mass $M$. \citet{Gilbaum2025} find that for $\alpha_{\rm disk}=0.01$ and $M=10^{6}\,{\rm M_\odot}$, the minimum gap-opening mass can be as low as $\sim5\,{\rm M_\odot}$, rising rapidly with $M$ and peaking at $\sim10^{3}\,{\rm M_\odot}$ for $M\gtrsim10^{8}\,{\rm M_\odot}$. Consequently, in regions of parameter space with low $M$, our thermal-torque predictions may be inapplicable, and a Type I/II torque prescription is more appropriate. For BHs formed in situ in the outer disk, the absence of thermal torques suggests that embedded BHs will migrate inward through the migration trap without encountering other stellar-mass objects -- unless some originate at small radii and are migrating outwards. 

In constructing the Hamiltonian to describe the dynamics of the MMR, we have made several approximations. We assume the system is compact ($a_2/a_1 \sim 1.2 - 2$), that the two objects in orbital resonance have roughly equal mass ($m_1 \approx m_2$) and thus migrate at roughly equal rates, and that the resonance is first order (such that the orbital periods $P_2:P_1$ are described by the ratio $k+1:k$). For orbital architectures that do not match these assumptions, the exact dynamics of the resonance will differ from what we describe here. %For example, first-order resonances technically don’t exist at zero eccentricity \citep{Malhotra2020}, while in contrast, higher-order resonances ($k+2:k$ and beyond) have substantial phase space even at $e\sim0$. 
Similarly, we consider only two-body resonances in this analysis, but capture of migrating objects into three-body Laplace resonances \citep[e.g.,][]{Leleu2021, Weisserman2023, Pucacco2024} could also be possible. Spin-orbit resonances \citep{Li2023} and evection resonances \citep{Bhaskar2022} may add further complexity in the subset of systems where they occur. Secular resonances between the apsidal precession rates of embedded objects \citep{Bhaskar2023} are also beyond our scope, but these couple the slow precession frequencies rather than the orbital frequencies and are correspondingly weaker than MMRs to both the GR precession and stochastic forcing. Thus where MMRs are rare, secular resonances should be rarer.

Our analysis assumes that, absent dissipation and GR precession, convergent, disk-mediated migration generically yields resonant capture. In reality, capture requires adiabatic evolution: The time to cross the resonance must be longer than the libration period, $\Delta \xi \, \tau_{\rm rel} \gtrsim 2 \pi/ \omega_{\rm{lib}}$ \citep[Equation 43 of][]{2015batygin}. Because both $\Delta\xi$ and $\omega_{\rm lib}$ scale as $q^{2/3}$, this condition becomes much more demanding at the small mass ratios relevant to AGN than for planets, and is unlikely to be met by a pair converging at the full Type I rate. Adiabatic capture is instead favored where convergence is slow compared with migration -- for near-equal-mass pairs, or for pairs approaching a migration trap, where the differential migration rate vanishes\footnote{Adiabatic capture may also be favored at ``migration swamps'' \citep{2025mckernan} where the sign of the migratory torque is fixed but its amplitude rapidly grows, as is often the case in the inner radiation-pressure-dominated zones of AGN \citep{gilbaum2022}. } -- consistent with the capture near traps found in the simulations of \citet{2019secunda, 2020secunda} and with the N-body results of \citet{2026moncrieff}. We note, moreover, that the fast-convergence limit $\tau_{\rm rel} \to \tau_{\rm mig}(m_1+m_2)$ adopted in \S\ref{sec:criterion_resonance_disruption} is the limit most favorable to resonance survival (Equation~\ref{eq:delta_xi}) but least favorable to capture. The conditions favoring capture and those favoring long-term survival thus overlap only partially. 

In addition to outright disruption, embedded objects can escape MMRs via resonant metastability \citep{2014goldreich,2015deck}. Recent simulations suggest that realistic turbulence amplifies this pathway by sustaining high-equilibrium eccentricities \citep{2025chen}. Taken together, these caveats imply our criteria likely overestimate capture rates, and the true incidence of long-lived MMRs in AGN disks may be lower. A full assessment of these effects in the AGN context remains an important direction for future work.

\section{Conclusion}\label{sec:conclusions}

In this work, we map where MMRs between embedded compact objects can survive in AGN disks by developing analytic prescriptions for resonance breaking under (i) GR precession and (ii) stochastic forcing due to disk turbulence and stellar flybys. For the former, we find that resonances are intrinsically fragile near the central MBH, wherever the GR-driven differential apsidal precession rate $\Delta \dot{\varpi}_{\rm GR}$ is greater than the libration frequency of the resonance, $\omega_{\rm lib}$. Setting these two quantities equal yields a breaking radius $R_{\rm GR}$ that shifts inward for larger embedded masses ($m$), smaller MBH mass ($M$) and more closely spaced commensurabilities (larger $k$). 

We derive a unified stochastic disruption criterion by coupling diffusion (turbulence plus stellar flybys) with migration-driven damping (Type I/II, GW, and thermal torques), comparing the variance in eccentricity and semimajor axis to the libration width. For diffusion in $a$, this analytic framework lets us survey a broad AGN parameter space and reveals three MBH mass regimes: (i) for $M \gtrsim 10^{7.5} \, \rm M_\odot$, stable first-order resonances do not occur; (ii) for $M\lesssim 10^{6.5} \, \rm M_\odot$, they are always present; and (iii) for $10^{6.5}\, \rm M_\odot \lesssim M \lesssim 10^{7.5} M_\odot$, stability depends on system parameters, including the disk Eddington ratio $\dot{m}_{\rm disk}$, the summed mass of the embedded objects $m_1 + m_2$, and the nuclear-cusp slope $\gamma$. When they exist, radial stability bands typically lie between an inner anti-trap and an outer trap, i.e., where objects are expected to migrate outward. 

Stochastic changes in eccentricity play a secondary role. Under the classic \citet{2004tanaka} prescription, $e$ damps on a timescale short relative to diffusion, and its variance remains below the resonant width. For luminous perturbers, thermal torques can excite $e$ toward the disk aspect ratio $H/a$, which reaches the resonance width only in the outer disk, where higher eccentricities make capture less likely and our estimate of MMR occurrence is a conservative upper bound. Thus the semimajor-axis channel still sets the break criterion. 

Our results suggest that in high-mass AGN ($M \gtrsim 10^{7.5} \, \rm M_\odot$) convergent migration can proceed toward LVK-band mergers largely unimpeded by capture into resonant chains. In low- and intermediate-mass AGN, stable chains are expected primarily within the outward-migration band carved by thermal torques. There they may suppress orbit crossings, with population-level consequences that depend on whether compact objects originate via {\it in situ} formation in the outer disk (exterior to traps) or are captured from the inner disk. 

We note several important sources of uncertainty in our approach, including alternative disk prescriptions, which may impact turbulent diffusion and migration timescales, as both explicitly depend on the disk structure. Moreover, thermal torques require a sufficiently small embedded mass that a gap does not open, as well as a scale hierarchy ($\lambda < H$). The former can break down for small $M$ and large $m$ while the latter breaks down at large radii. If thermal torques are suppressed, the problem reverts toward Type I/II-like behavior and MMRs become much less robust -- persisting only in small islands at intermediate $M$ and $a$.  It is also worth noting the analytic nature of our work, which should be checked by numerical simulations.  The N-body integrations of \citet{2026moncrieff}, which examine resonant capture and binary formation for unequal-mass embedded BHs with self-consistent AGN torques, provide a first such check. They find that the capture probability is set by the ratio of the libration time to the resonance-crossing time, and that resonant blocking of mergers is most effective for near-equal-mass pairs and suppressed at extreme mass ratios, consistent with the adiabatic-capture considerations outlined in \S\ref{subsec:caveats}. %As this paper was being completed, we became aware of a forthcoming investigation (Moncrieff \& Grishin {\it in prep}) which examines similar questions about embedded BH dynamics through numerical orbit integration.

While uncertainties in present AGN disk models preclude precise forecasts of merger suppression by MMRs, our criteria and instability maps offer insight into how embedded populations evolve and identify where resonant stabilization is likely. We offer these results as a baseline for future, more detailed modeling that folds in more realistic disk hydrodynamics, torque prescriptions, and diffusion calibrations as those inputs improve.

\medskip
\section*{Acknowledgments}
We thank Fred Adams and Konstantin Batygin for useful conversations, and the anonymous referee for a highly detailed and constructive series of suggestions. M.E.M. gratefully acknowledges support from \textit{GFSD}. 
The collaboration visit that started this work was supported by the Wisconsin Alumni Research Foundation, award No. AAM3428.  N.C.S. gratefully acknowledges support from  the Israel Science Foundation (Individual Research grant No. 2414/23).

\bibliography{ref}{}

@ARTICLE{2025mckernan,
       author = {{McKernan}, Barry and {Ford}, K.~E. Saavik and {Cook}, Harrison E. and {Delfavero}, Vera and {McPike}, Emily and {Nathaniel}, Kaila and {Postiglione}, Jake and {Ray}, Shawn and {O'Shaughnessy}, Richard},
        title = "{McFACTS I: Testing the LVK AGN Channel with Monte Carlo for AGN Channel Testing and Simulation (McFACTS)}",
      journal = {\apj},
         year = 2025,
        month = sep,
       volume = {990},
       number = {2},
          eid = {217},
        pages = {217},
          doi = {10.3847/1538-4357/adf114},
archivePrefix = {arXiv},
       eprint = {2410.16515},
 primaryClass = {astro-ph.HE},
       adsurl = {https://ui.adsabs.harvard.edu/abs/2025ApJ...990..217M}
}

@ARTICLE{2024wang,
       author = {{Wang}, Yihan and {Zhu}, Zhaohuan and {Lin}, Douglas N.~C.},
        title = "{Stellar/BH population in AGN discs: direct binary formation from capture objects in nuclei clusters}",
      journal = {\mnras},
         year = 2024,
        month = mar,
       volume = {528},
       number = {3},
        pages = {4958-4975},
          doi = {10.1093/mnras/stae321},
archivePrefix = {arXiv},
       eprint = {2308.09129},
 primaryClass = {astro-ph.GA},
       adsurl = {https://ui.adsabs.harvard.edu/abs/2024MNRAS.528.4958W}
}

@ARTICLE{2019yang,
       author = {{Yang}, Y. and {Bartos}, I. and {Gayathri}, V. and {Ford}, K.~E.~S. and {Haiman}, Z. and {Klimenko}, S. and {Kocsis}, B. and {M{\'a}rka}, S. and {M{\'a}rka}, Z. and {McKernan}, B. and {O'Shaughnessy}, R.},
        title = "{Hierarchical Black Hole Mergers in Active Galactic Nuclei}",
      journal = {\prl},
         year = 2019,
        month = nov,
       volume = {123},
       number = {18},
          eid = {181101},
        pages = {181101},
          doi = {10.1103/PhysRevLett.123.181101},
archivePrefix = {arXiv},
       eprint = {1906.09281},
 primaryClass = {astro-ph.HE},
       adsurl = {https://ui.adsabs.harvard.edu/abs/2019PhRvL.123r1101Y}
}

@ARTICLE{2025peng,
       author = {{Peng}, Peng and {Franchini}, Alessia and {Bonetti}, Matteo and {Sesana}, Alberto and {Chen}, Xian},
        title = "{The Fate of EMRI-IMRI Pairs in Active Galactic Nucleus Accretion Disks: Hydrodynamical and Three-body Simulations}",
      journal = {\apj},
         year = 2025,
        month = aug,
       volume = {989},
       number = {1},
          eid = {122},
        pages = {122},
          doi = {10.3847/1538-4357/adef42},
       adsurl = {https://ui.adsabs.harvard.edu/abs/2025ApJ...989..122P}
}

@ARTICLE{2025reved,
       author = {{Reved}, Omri and {Friedland}, Lazar and {Stone}, Nicholas C.},
        title = "{Resonant capture of stars by black hole binaries: extreme eccentricity excitation}",
      journal = {\mnras},
         year = 2025,
        month = feb,
       volume = {537},
       number = {2},
        pages = {661-673},
          doi = {10.1093/mnras/staf051},
archivePrefix = {arXiv},
       eprint = {2410.12911},
 primaryClass = {astro-ph.HE},
       adsurl = {https://ui.adsabs.harvard.edu/abs/2025MNRAS.537..661R}
}

@ARTICLE{2010seto,
       author = {{Seto}, Naoki and {Muto}, Takayuki},
        title = "{Relativistic astrophysics with resonant multiple inspirals}",
      journal = {\prd},
         year = 2010,
        month = may,
       volume = {81},
       number = {10},
          eid = {103004},
        pages = {103004},
          doi = {10.1103/PhysRevD.81.103004},
archivePrefix = {arXiv},
       eprint = {1005.3114},
 primaryClass = {astro-ph.CO},
       adsurl = {https://ui.adsabs.harvard.edu/abs/2010PhRvD..81j3004S}
}

@ARTICLE{2023generozov,
       author = {{Generozov}, A. and {Perets}, H.~B.},
        title = "{Capture of stars into gaseous discs around massive black holes: alignment, circularization, and growth}",
      journal = {\mnras},
         year = 2023,
        month = jun,
       volume = {522},
       number = {2},
        pages = {1763-1778},
          doi = {10.1093/mnras/stad1016},
archivePrefix = {arXiv},
       eprint = {2212.11301},
 primaryClass = {astro-ph.GA},
       adsurl = {https://ui.adsabs.harvard.edu/abs/2023MNRAS.522.1763G}
}

@ARTICLE{2016luo,
       author = {{Luo}, Jun and {Chen}, Li-Sheng and {Duan}, Hui-Zong and {Gong}, Yun-Gui and {Hu}, Shoucun and {Ji}, Jianghui and {Liu}, Qi and {Mei}, Jianwei and {Milyukov}, Vadim and {Sazhin}, Mikhail and {Shao}, Cheng-Gang and {Toth}, Viktor T. and {Tu}, Hai-Bo and {Wang}, Yamin and {Wang}, Yan and {Yeh}, Hsien-Chi and {Zhan}, Ming-Sheng and {Zhang}, Yonghe and {Zharov}, Vladimir and {Zhou}, Ze-Bing},
        title = "{TianQin: a space-borne gravitational wave detector}",
      journal = {Classical and Quantum Gravity},
         year = 2016,
        month = feb,
       volume = {33},
       number = {3},
          eid = {035010},
        pages = {035010},
          doi = {10.1088/0264-9381/33/3/035010},
archivePrefix = {arXiv},
       eprint = {1512.02076},
 primaryClass = {astro-ph.IM},
       adsurl = {https://ui.adsabs.harvard.edu/abs/2016CQGra..33c5010L}
}

@ARTICLE{2020ruan,
       author = {{Ruan}, Wen-Hong and {Guo}, Zong-Kuan and {Cai}, Rong-Gen and {Zhang}, Yuan-Zhong},
        title = "{Taiji program: Gravitational-wave sources}",
      journal = {International Journal of Modern Physics A},
         year = 2020,
        month = jun,
       volume = {35},
       number = {17},
          eid = {2050075},
        pages = {2050075},
          doi = {10.1142/S0217751X2050075X},
archivePrefix = {arXiv},
       eprint = {1807.09495},
 primaryClass = {gr-qc},
       adsurl = {https://ui.adsabs.harvard.edu/abs/2020IJMPA..3550075R}
}

@ARTICLE{2021panA,
       author = {{Pan}, Zhen and {Yang}, Huan},
        title = "{Formation rate of extreme mass ratio inspirals in active galactic nuclei}",
      journal = {\prd},
         year = 2021,
        month = may,
       volume = {103},
       number = {10},
          eid = {103018},
        pages = {103018},
          doi = {10.1103/PhysRevD.103.103018},
archivePrefix = {arXiv},
       eprint = {2101.09146},
 primaryClass = {astro-ph.HE},
       adsurl = {https://ui.adsabs.harvard.edu/abs/2021PhRvD.103j3018P}
}

@ARTICLE{2021panB,
       author = {{Pan}, Zhen and {Lyu}, Zhenwei and {Yang}, Huan},
        title = "{Wet extreme mass ratio inspirals may be more common for spaceborne gravitational wave detection}",
      journal = {\prd},
         year = 2021,
        month = sep,
       volume = {104},
       number = {6},
          eid = {063007},
        pages = {063007},
          doi = {10.1103/PhysRevD.104.063007},
archivePrefix = {arXiv},
       eprint = {2104.01208},
 primaryClass = {astro-ph.HE},
       adsurl = {https://ui.adsabs.harvard.edu/abs/2021PhRvD.104f3007P}
}

@ARTICLE{2023derdzinski,
       author = {{Derdzinski}, Andrea and {Mayer}, Lucio},
        title = "{In situ extreme mass ratio inspirals via subparsec formation and migration of stars in thin, gravitationally unstable AGN discs}",
      journal = {\mnras},
         year = 2023,
        month = may,
       volume = {521},
       number = {3},
        pages = {4522-4543},
          doi = {10.1093/mnras/stad749},
archivePrefix = {arXiv},
       eprint = {2205.10382},
 primaryClass = {astro-ph.GA},
       adsurl = {https://ui.adsabs.harvard.edu/abs/2023MNRAS.521.4522D}
}

@ARTICLE{2005hopman,
       author = {{Hopman}, Clovis and {Alexander}, Tal},
        title = "{The Orbital Statistics of Stellar Inspiral and Relaxation near a Massive Black Hole: Characterizing Gravitational Wave Sources}",
      journal = {\apj},
         year = 2005,
        month = aug,
       volume = {629},
       number = {1},
        pages = {362-372},
          doi = {10.1086/431475},
archivePrefix = {arXiv},
       eprint = {astro-ph/0503672},
 primaryClass = {astro-ph},
       adsurl = {https://ui.adsabs.harvard.edu/abs/2005ApJ...629..362H}
}

@ARTICLE{2022pan,
       author = {{Pan}, Zhen and {Lyu}, Zhenwei and {Yang}, Huan},
        title = "{Mass-gap extreme mass ratio inspirals}",
      journal = {\prd},
         year = 2022,
        month = apr,
       volume = {105},
       number = {8},
          eid = {083005},
        pages = {083005},
          doi = {10.1103/PhysRevD.105.083005},
archivePrefix = {arXiv},
       eprint = {2112.10237},
 primaryClass = {astro-ph.HE},
       adsurl = {https://ui.adsabs.harvard.edu/abs/2022PhRvD.105h3005P}
}

@ARTICLE{2007levin,
       author = {{Levin}, Yuri},
        title = "{Starbursts near supermassive black holes: young stars in the Galactic Centre, and gravitational waves in LISA band}",
      journal = {\mnras},
         year = 2007,
        month = jan,
       volume = {374},
       number = {2},
        pages = {515-524},
          doi = {10.1111/j.1365-2966.2006.11155.x},
archivePrefix = {arXiv},
       eprint = {astro-ph/0603583},
 primaryClass = {astro-ph},
       adsurl = {https://ui.adsabs.harvard.edu/abs/2007MNRAS.374..515L}
}

@ARTICLE{2023amaroseoane,
       author = {{Amaro-Seoane}, Pau and {Andrews}, Jeff and {Arca Sedda}, Manuel and {Askar}, Abbas and {Baghi}, Quentin and {Balasov}, Razvan and {Bartos}, Imre and {Bavera}, Simone S. and {Bellovary}, Jillian and {Berry}, Christopher P.~L. and {Berti}, Emanuele and {Bianchi}, Stefano and {Blecha}, Laura and {Blondin}, St{\'e}phane and {Bogdanovi{\'c}}, Tamara and {Boissier}, Samuel and {Bonetti}, Matteo and {Bonoli}, Silvia and {Bortolas}, Elisa and {Breivik}, Katelyn and {Capelo}, Pedro R. and {Caramete}, Laurentiu and {Cattorini}, Federico and {Charisi}, Maria and {Chaty}, Sylvain and {Chen}, Xian and {Chru{\'s}li{\'n}ska}, Martyna and {Chua}, Alvin J.~K. and {Church}, Ross and {Colpi}, Monica and {D'Orazio}, Daniel and {Danielski}, Camilla and {Davies}, Melvyn B. and {Dayal}, Pratika and {De Rosa}, Alessandra and {Derdzinski}, Andrea and {Destounis}, Kyriakos and {Dotti}, Massimo and {Du{\c{t}}an}, Ioana and {Dvorkin}, Irina and {Fabj}, Gaia and {Foglizzo}, Thierry and {Ford}, Saavik and {Fouvry}, Jean-Baptiste and {Franchini}, Alessia and {Fragos}, Tassos and {Fryer}, Chris and {Gaspari}, Massimo and {Gerosa}, Davide and {Graziani}, Luca and {Groot}, Paul and {Habouzit}, Melanie and {Haggard}, Daryl and {Haiman}, Zoltan and {Han}, Wen-Biao and {Istrate}, Alina and {Johansson}, Peter H. and {Khan}, Fazeel Mahmood and {Kimpson}, Tomas and {Kokkotas}, Kostas and {Kong}, Albert and {Korol}, Valeriya and {Kremer}, Kyle and {Kupfer}, Thomas and {Lamberts}, Astrid and {Larson}, Shane and {Lau}, Mike and {Liu}, Dongliang and {Lloyd-Ronning}, Nicole and {Lodato}, Giuseppe and {Lupi}, Alessandro and {Ma}, Chung-Pei and {Maccarone}, Tomas and {Mandel}, Ilya and {Mangiagli}, Alberto and {Mapelli}, Michela and {Mathis}, St{\'e}phane and {Mayer}, Lucio and {McGee}, Sean and {McKernan}, Berry and {Miller}, M. Coleman and {Mota}, David F. and {Mumpower}, Matthew and {Nasim}, Syeda S. and {Nelemans}, Gijs and {Noble}, Scott and {Pacucci}, Fabio and {Panessa}, Francesca and {Paschalidis}, Vasileios and {Pfister}, Hugo and {Porquet}, Delphine and {Quenby}, John and {Ricarte}, Angelo and {R{\"o}pke}, Friedrich K. and {Regan}, John and {Rosswog}, Stephan and {Ruiter}, Ashley and {Ruiz}, Milton and {Runnoe}, Jessie and {Schneider}, Raffaella and {Schnittman}, Jeremy and {Secunda}, Amy and {Sesana}, Alberto and {Seto}, Naoki and {Shao}, Lijing and {Shapiro}, Stuart and {Sopuerta}, Carlos and {Stone}, Nicholas C. and {Suvorov}, Arthur and {Tamanini}, Nicola and {Tamfal}, Tomas and {Tauris}, Thomas and {Temmink}, Karel and {Tomsick}, John and {Toonen}, Silvia and {Torres-Orjuela}, Alejandro and {Toscani}, Martina and {Tsokaros}, Antonios and {Unal}, Caner and {V{\'a}zquez-Aceves}, Ver{\'o}nica and {Valiante}, Rosa and {van Putten}, Maurice and {van Roestel}, Jan and {Vignali}, Christian and {Volonteri}, Marta and {Wu}, Kinwah and {Younsi}, Ziri and {Yu}, Shenghua and {Zane}, Silvia and {Zwick}, Lorenz and {Antonini}, Fabio and {Baibhav}, Vishal and {Barausse}, Enrico and {Bonilla Rivera}, Alexander and {Branchesi}, Marica and {Branduardi-Raymont}, Graziella and {Burdge}, Kevin and {Chakraborty}, Srija and {Cuadra}, Jorge and {Dage}, Kristen and {Davis}, Benjamin and {de Mink}, Selma E. and {Decarli}, Roberto and {Doneva}, Daniela and {Escoffier}, Stephanie and {Gandhi}, Poshak and {Haardt}, Francesco and {Lousto}, Carlos O. and {Nissanke}, Samaya and {Nordhaus}, Jason and {O'Shaughnessy}, Richard and {Portegies Zwart}, Simon and {Pound}, Adam and {Schussler}, Fabian and {Sergijenko}, Olga and {Spallicci}, Alessandro and {Vernieri}, Daniele and {Vigna-G{\'o}mez}, Alejandro},
        title = "{Astrophysics with the Laser Interferometer Space Antenna}",
      journal = {Living Reviews in Relativity},
         year = 2023,
        month = dec,
       volume = {26},
       number = {1},
          eid = {2},
        pages = {2},
          doi = {10.1007/s41114-022-00041-y},
archivePrefix = {arXiv},
       eprint = {2203.06016},
 primaryClass = {gr-qc},
       adsurl = {https://ui.adsabs.harvard.edu/abs/2023LRR....26....2A}
}

@ARTICLE{2018amaroseoane,
       author = {{Amaro-Seoane}, Pau},
        title = "{Relativistic dynamics and extreme mass ratio inspirals}",
      journal = {Living Reviews in Relativity},
         year = 2018,
        month = dec,
       volume = {21},
       number = {1},
          eid = {4},
        pages = {4},
          doi = {10.1007/s41114-018-0013-8},
archivePrefix = {arXiv},
       eprint = {1205.5240},
 primaryClass = {astro-ph.CO},
       adsurl = {https://ui.adsabs.harvard.edu/abs/2018LRR....21....4A}
}

@ARTICLE{1964peters,
       author = {{Peters}, P.~C.},
        title = "{Gravitational Radiation and the Motion of Two Point Masses}",
      journal = {Physical Review},
         year = 1964,
        month = nov,
       volume = {136},
       number = {4B},
        pages = {1224-1232},
          doi = {10.1103/PhysRev.136.B1224},
       adsurl = {https://ui.adsabs.harvard.edu/abs/1964PhRv..136.1224P}
}

@ARTICLE{1997ward,
       author = {{Ward}, William R.},
        title = "{Protoplanet Migration by Nebula Tides}",
      journal = {\icarus},
         year = 1997,
        month = apr,
       volume = {126},
       number = {2},
        pages = {261-281},
          doi = {10.1006/icar.1996.5647},
       adsurl = {https://ui.adsabs.harvard.edu/abs/1997Icar..126..261W}
}

@ARTICLE{1979goldreich,
       author = {{Goldreich}, P. and {Tremaine}, S.},
        title = "{The excitation of density waves at the Lindblad and corotation resonances by an external potential.}",
      journal = {\apj},
         year = 1979,
        month = nov,
       volume = {233},
        pages = {857-871},
          doi = {10.1086/157448},
       adsurl = {https://ui.adsabs.harvard.edu/abs/1979ApJ...233..857G}
}

@ARTICLE{2014goldreich,
       author = {{Goldreich}, Peter and {Schlichting}, Hilke E.},
        title = "{Overstable Librations can Account for the Paucity of Mean Motion Resonances among Exoplanet Pairs}",
      journal = {\aj},
         year = 2014,
        month = feb,
       volume = {147},
       number = {2},
          eid = {32},
        pages = {32},
          doi = {10.1088/0004-6256/147/2/32},
archivePrefix = {arXiv},
       eprint = {1308.4688},
 primaryClass = {astro-ph.EP},
       adsurl = {https://ui.adsabs.harvard.edu/abs/2014AJ....147...32G}
}

@ARTICLE{2021tagawa,
       author = {{Tagawa}, Hiromichi and {Kocsis}, Bence and {Haiman}, Zolt{\'a}n and {Bartos}, Imre and {Omukai}, Kazuyuki and {Samsing}, Johan},
        title = "{Mass-gap Mergers in Active Galactic Nuclei}",
      journal = {\apj},
         year = 2021,
        month = feb,
       volume = {908},
       number = {2},
          eid = {194},
        pages = {194},
          doi = {10.3847/1538-4357/abd555},
archivePrefix = {arXiv},
       eprint = {2012.00011},
 primaryClass = {astro-ph.HE},
       adsurl = {https://ui.adsabs.harvard.edu/abs/2021ApJ...908..194T}
}

@ARTICLE{2021secunda,
       author = {{Secunda}, Amy and {Hernandez}, Betsy and {Goodman}, Jeremy and {Leigh}, Nathan W.~C. and {McKernan}, Barry and {Ford}, K.~E. Saavik and {Adorno}, Jose I.},
        title = "{Evolution of Retrograde Orbiters in an Active Galactic Nucleus Disk}",
      journal = {\apjl},
         year = 2021,
        month = feb,
       volume = {908},
       number = {2},
          eid = {L27},
        pages = {L27},
          doi = {10.3847/2041-8213/abe11d},
archivePrefix = {arXiv},
       eprint = {2009.03910},
 primaryClass = {astro-ph.HE},
       adsurl = {https://ui.adsabs.harvard.edu/abs/2021ApJ...908L..27S}
}

@ARTICLE{2012antonini,
       author = {{Antonini}, Fabio and {Perets}, Hagai B.},
        title = "{Secular Evolution of Compact Binaries near Massive Black Holes: Gravitational Wave Sources and Other Exotica}",
      journal = {\apj},
         year = 2012,
        month = sep,
       volume = {757},
       number = {1},
          eid = {27},
        pages = {27},
          doi = {10.1088/0004-637X/757/1/27},
archivePrefix = {arXiv},
       eprint = {1203.2938},
 primaryClass = {astro-ph.GA},
       adsurl = {https://ui.adsabs.harvard.edu/abs/2012ApJ...757...27A}
}

@ARTICLE{2022vanson,
       author = {{van Son}, L.~A.~C. and {de Mink}, S.~E. and {Renzo}, M. and {Justham}, S. and {Zapartas}, E. and {Breivik}, K. and {Callister}, T. and {Farr}, W.~M. and {Conroy}, C.},
        title = "{No Peaks without Valleys: The Stable Mass Transfer Channel for Gravitational-wave Sources in Light of the Neutron Star-Black Hole Mass Gap}",
      journal = {\apj},
         year = 2022,
        month = dec,
       volume = {940},
       number = {2},
          eid = {184},
        pages = {184},
          doi = {10.3847/1538-4357/ac9b0a},
archivePrefix = {arXiv},
       eprint = {2209.13609},
 primaryClass = {astro-ph.HE},
       adsurl = {https://ui.adsabs.harvard.edu/abs/2022ApJ...940..184V}
}

@ARTICLE{2025LVKa,
       author = {{The LIGO Scientific Collaboration} and {the Virgo Collaboration} and {the KAGRA Collaboration} and {Abac}, A.~G. and {Abouelfettouh}, I. and {Acernese}, F. and {Ackley}, K. and {Adamcewicz}, C. and {Adhicary}, S. and {Adhikari}, D. and {Adhikari}, N. and {Adhikari}, R.~X. and {Adkins}, V.~K. and {Afroz}, S. and {Agapito}, A. and {Agarwal}, D. and {Agathos}, M. and {Aggarwal}, N. and {Aggarwal}, S. and {Aguiar}, O.~D. and {Ahrend}, I. -L. and {Aiello}, L. and {Ain}, A. and {Ajith}, P. and {Akutsu}, T. and {Albanesi}, S. and {Ali}, W. and {Al-Kershi}, S. and {All{\'e}n{\'e}}, C. and {Allocca}, A. and {Al-Shammari}, S. and {Altin}, P.~A. and {Alvarez-Lopez}, S. and {Amar}, W. and {Amarasinghe}, O. and {Amato}, A. and {Amicucci}, F. and {Amra}, C. and {Ananyeva}, A. and {Anderson}, S.~B. and {Anderson}, W.~G. and {Andia}, M. and {Ando}, M. and {Andr{\'e}s-Carcasona}, M. and {Andri{\'c}}, T. and {Anglin}, J. and {Ansoldi}, S. and {Antelis}, J.~M. and {Antier}, S. and {Aoumi}, M. and {Appavuravther}, E.~Z. and {Appert}, S. and {Apple}, S.~K. and {Arai}, K. and {Araya}, A. and {Araya}, M.~C. and {Arca Sedda}, M. and {Areeda}, J.~S. and {Aritomi}, N. and {Armato}, F. and {Armstrong}, S. and {Arnaud}, N. and {Arogeti}, M. and {Aronson}, S.~M. and {Arun}, K.~G. and {Ashton}, G. and {Aso}, Y. and {Asprea}, L. and {Assiduo}, M. and {Assis de Souza Melo}, S. and {Aston}, S.~M. and {Astone}, P. and {Attadio}, F. and {Aubin}, F. and {AultONeal}, K. and {Avallone}, G. and {Avila}, E.~A. and {Babak}, S. and {Badger}, C. and {Bae}, S. and {Bagnasco}, S. and {Baiotti}, L. and {Bajpai}, R. and {Baka}, T. and {Baker}, A.~M. and {Baker}, K.~A. and {Baker}, T. and {Baldi}, G. and {Baldicchi}, N. and {Ball}, M. and {Ballardin}, G. and {Ballmer}, S.~W. and {Banagiri}, S. and {Banerjee}, B. and {Bankar}, D. and {Baptiste}, T.~M. and {Baral}, P. and {Baratti}, M. and {Barayoga}, J.~C. and {Barish}, B.~C. and {Barker}, D. and {Barman}, N. and {Barneo}, P. and {Barone}, F. and {Barr}, B. and {Barsotti}, L. and {Barsuglia}, M. and {Barta}, D. and {Bartoletti}, A.~M. and {Barton}, M.~A. and {Bartos}, I. and {Basalaev}, A. and {Bassiri}, R. and {Basti}, A. and {Bawaj}, M. and {Baxi}, P. and {Bayley}, J.~C. and {Baylor}, A.~C. and {Baynard}, II, P.~A. and {Bazzan}, M. and {Bedakihale}, V.~M. and {Beirnaert}, F. and {Bejger}, M. and {Belardinelli}, D. and {Bell}, A.~S. and {Bellie}, D.~S. and {Bellizzi}, L. and {Benoit}, W. and {Bentara}, I. and {Bentley}, J.~D. and {Ben Yaala}, M. and {Bera}, S. and {Bergamin}, F. and {Berger}, B.~K. and {Bernuzzi}, S. and {Beroiz}, M. and {Berry}, C.~P.~L. and {Bersanetti}, D. and {Bertheas}, T. and {Bertolini}, A. and {Betzwieser}, J. and {Beveridge}, D. and {Bevilacqua}, G. and {Bevins}, N. and {Bhandare}, R. and {Bhatt}, R. and {Bhattacharjee}, D. and {Bhattacharyya}, S. and {Bhaumik}, S. and {Biancalana}, V. and {Bianchi}, A. and {Bilenko}, I.~A. and {Billingsley}, G. and {Binetti}, A. and {Bini}, S. and {Binu}, C. and {Biot}, S. and {Birnholtz}, O. and {Biscoveanu}, S. and {Bisht}, A. and {Bitossi}, M. and {Bizouard}, M. -A. and {Blaber}, S. and {Blackburn}, J.~K. and {Blagg}, L.~A. and {Blair}, C.~D. and {Blair}, D.~G. and {Bode}, N. and {Boettner}, N. and {Boileau}, G. and {Boldrini}, M. and {Bolingbroke}, G.~N. and {Bolliand}, A. and {Bonavena}, L.~D. and {Bondarescu}, R. and {Bondu}, F. and {Bonilla}, E. and {Bonilla}, M.~S. and {Bonino}, A. and {Bonnand}, R. and {Borchers}, A. and {Borhanian}, S. and {Boschi}, V. and {Bose}, S. and {Bossilkov}, V. and {Bothra}, Y. and {Boudon}, A. and {Bourg}, L. and {Boyle}, M. and {Bozzi}, A. and {Bradaschia}, C. and {Brady}, P.~R. and {Branch}, A. and {Branchesi}, M. and {Braun}, I. and {Briant}, T. and {Brillet}, A. and {Brinkmann}, M. and {Brockill}, P. and {Brockmueller}, E.},
        title = "{GWTC-4.0: Updating the Gravitational-Wave Transient Catalog with Observations from the First Part of the Fourth LIGO-Virgo-KAGRA Observing Run}",
      journal = {arXiv e-prints},
         year = 2025,
        month = aug,
          eid = {arXiv:2508.18082},
        pages = {arXiv:2508.18082},
          doi = {10.48550/arXiv.2508.18082},
archivePrefix = {arXiv},
       eprint = {2508.18082},
 primaryClass = {gr-qc},
       adsurl = {https://ui.adsabs.harvard.edu/abs/2025arXiv250818082T}
}

@ARTICLE{2012mckernan,
       author = {{McKernan}, B. and {Ford}, K.~E.~S. and {Lyra}, W. and {Perets}, H.~B.},
        title = "{Intermediate mass black holes in AGN discs - I. Production and growth}",
      journal = {\mnras},
         year = 2012,
        month = sep,
       volume = {425},
       number = {1},
        pages = {460-469},
          doi = {10.1111/j.1365-2966.2012.21486.x},
archivePrefix = {arXiv},
       eprint = {1206.2309},
 primaryClass = {astro-ph.GA},
       adsurl = {https://ui.adsabs.harvard.edu/abs/2012MNRAS.425..460M}
}

@ARTICLE{2023rowan,
       author = {{Rowan}, Connar and {Boekholt}, Tjarda and {Kocsis}, Bence and {Haiman}, Zolt{\'a}n},
        title = "{Black hole binary formation in AGN discs: from isolation to merger}",
      journal = {\mnras},
         year = 2023,
        month = sep,
       volume = {524},
       number = {2},
        pages = {2770-2796},
          doi = {10.1093/mnras/stad1926},
archivePrefix = {arXiv},
       eprint = {2212.06133},
 primaryClass = {astro-ph.GA},
       adsurl = {https://ui.adsabs.harvard.edu/abs/2023MNRAS.524.2770R}
}

@ARTICLE{2020tagawa,
       author = {{Tagawa}, Hiromichi and {Haiman}, Zolt{\'a}n and {Kocsis}, Bence},
        title = "{Formation and Evolution of Compact-object Binaries in AGN Disks}",
      journal = {\apj},
         year = 2020,
        month = jul,
       volume = {898},
       number = {1},
          eid = {25},
        pages = {25},
          doi = {10.3847/1538-4357/ab9b8c},
archivePrefix = {arXiv},
       eprint = {1912.08218},
 primaryClass = {astro-ph.GA},
       adsurl = {https://ui.adsabs.harvard.edu/abs/2020ApJ...898...25T}
}

@ARTICLE{2017stone,
       author = {{Stone}, Nicholas C. and {Metzger}, Brian D. and {Haiman}, Zolt{\'a}n},
        title = "{Assisted inspirals of stellar mass black holes embedded in AGN discs: solving the `final au problem'}",
      journal = {\mnras},
         year = 2017,
        month = jan,
       volume = {464},
       number = {1},
        pages = {946-954},
          doi = {10.1093/mnras/stw2260},
archivePrefix = {arXiv},
       eprint = {1602.04226},
 primaryClass = {astro-ph.GA},
       adsurl = {https://ui.adsabs.harvard.edu/abs/2017MNRAS.464..946S}
}

@ARTICLE{2012amaro-seoane,
       author = {{Amaro-Seoane}, Pau and {Brem}, Patrick and {Cuadra}, Jorge and {Armitage}, Philip J.},
        title = "{The Butterfly Effect in the Extreme-mass Ratio Inspiral Problem}",
      journal = {\apjl},
         year = 2012,
        month = jan,
       volume = {744},
       number = {2},
          eid = {L20},
        pages = {L20},
          doi = {10.1088/2041-8205/744/2/L20},
archivePrefix = {arXiv},
       eprint = {1108.5174},
 primaryClass = {astro-ph.CO},
       adsurl = {https://ui.adsabs.harvard.edu/abs/2012ApJ...744L..20A}
}

@ARTICLE{2023barsanti,
       author = {{Barsanti}, Susanna and {Maselli}, Andrea and {Sotiriou}, Thomas P. and {Gualtieri}, Leonardo},
        title = "{Detecting Massive Scalar Fields with Extreme Mass-Ratio Inspirals}",
      journal = {\prl},
         year = 2023,
        month = aug,
       volume = {131},
       number = {5},
          eid = {051401},
        pages = {051401},
          doi = {10.1103/PhysRevLett.131.051401},
archivePrefix = {arXiv},
       eprint = {2212.03888},
 primaryClass = {gr-qc},
       adsurl = {https://ui.adsabs.harvard.edu/abs/2023PhRvL.131e1401B}
}

@ARTICLE{2017bartos,
       author = {{Bartos}, Imre and {Kocsis}, Bence and {Haiman}, Zolt{\'a}n and {M{\'a}rka}, Szabolcs},
        title = "{Rapid and Bright Stellar-mass Binary Black Hole Mergers in Active Galactic Nuclei}",
      journal = {\apj},
         year = 2017,
        month = feb,
       volume = {835},
       number = {2},
          eid = {165},
        pages = {165},
          doi = {10.3847/1538-4357/835/2/165},
archivePrefix = {arXiv},
       eprint = {1602.03831},
 primaryClass = {astro-ph.HE},
       adsurl = {https://ui.adsabs.harvard.edu/abs/2017ApJ...835..165B}
}

@ARTICLE{2016bellovary,
       author = {{Bellovary}, Jillian M. and {Mac Low}, Mordecai-Mark and {McKernan}, Barry and {Ford}, K.~E. Saavik},
        title = "{Migration Traps in Disks around Supermassive Black Holes}",
      journal = {\apjl},
         year = 2016,
        month = mar,
       volume = {819},
       number = {2},
          eid = {L17},
        pages = {L17},
          doi = {10.3847/2041-8205/819/2/L17},
archivePrefix = {arXiv},
       eprint = {1511.00005},
 primaryClass = {astro-ph.GA},
       adsurl = {https://ui.adsabs.harvard.edu/abs/2016ApJ...819L..17B}
}

@ARTICLE{2012coward,
       author = {{Coward}, D.~M. and {Howell}, E.~J. and {Piran}, T. and {Stratta}, G. and {Branchesi}, M. and {Bromberg}, O. and {Gendre}, B. and {Burman}, R.~R. and {Guetta}, D.},
        title = "{The Swift short gamma-ray burst rate density: implications for binary neutron star merger rates}",
      journal = {\mnras},
         year = 2012,
        month = oct,
       volume = {425},
       number = {4},
        pages = {2668-2673},
          doi = {10.1111/j.1365-2966.2012.21604.x},
archivePrefix = {arXiv},
       eprint = {1202.2179},
 primaryClass = {astro-ph.CO},
       adsurl = {https://ui.adsabs.harvard.edu/abs/2012MNRAS.425.2668C}
}

@ARTICLE{2016abbott,
       author = {{Abbott}, B.~P. and {Abbott}, R. and {Abbott}, T.~D. and {Abernathy}, M.~R. and {Acernese}, F. and {Ackley}, K. and {Adams}, C. and {Adams}, T. and {Addesso}, P. and {Adhikari}, R.~X. and {Adya}, V.~B. and {Affeldt}, C. and {Agathos}, M. and {Agatsuma}, K. and {Aggarwal}, N. and {Aguiar}, O.~D. and {Aiello}, L. and {Ain}, A. and {Ajith}, P. and {Allen}, B. and {Allocca}, A. and {Altin}, P.~A. and {Anderson}, S.~B. and {Anderson}, W.~G. and {Arai}, K. and {Arain}, M.~A. and {Araya}, M.~C. and {Arceneaux}, C.~C. and {Areeda}, J.~S. and {Arnaud}, N. and {Arun}, K.~G. and {Ascenzi}, S. and {Ashton}, G. and {Ast}, M. and {Aston}, S.~M. and {Astone}, P. and {Aufmuth}, P. and {Aulbert}, C. and {Babak}, S. and {Bacon}, P. and {Bader}, M.~K.~M. and {Baker}, P.~T. and {Baldaccini}, F. and {Ballardin}, G. and {Ballmer}, S.~W. and {Barayoga}, J.~C. and {Barclay}, S.~E. and {Barish}, B.~C. and {Barker}, D. and {Barone}, F. and {Barr}, B. and {Barsotti}, L. and {Barsuglia}, M. and {Barta}, D. and {Bartlett}, J. and {Barton}, M.~A. and {Bartos}, I. and {Bassiri}, R. and {Basti}, A. and {Batch}, J.~C. and {Baune}, C. and {Bavigadda}, V. and {Bazzan}, M. and {Behnke}, B. and {Bejger}, M. and {Belczynski}, C. and {Bell}, A.~S. and {Bell}, C.~J. and {Berger}, B.~K. and {Bergman}, J. and {Bergmann}, G. and {Berry}, C.~P.~L. and {Bersanetti}, D. and {Bertolini}, A. and {Betzwieser}, J. and {Bhagwat}, S. and {Bhandare}, R. and {Bilenko}, I.~A. and {Billingsley}, G. and {Birch}, J. and {Birney}, I.~A. and {Birnholtz}, O. and {Biscans}, S. and {Bisht}, A. and {Bitossi}, M. and {Biwer}, C. and {Bizouard}, M.~A. and {Blackburn}, J.~K. and {Blair}, C.~D. and {Blair}, D.~G. and {Blair}, R.~M. and {Bloemen}, S. and {Bock}, O. and {Bodiya}, T.~P. and {Boer}, M. and {Bogaert}, G. and {Bogan}, C. and {Bohe}, A. and {Bojtos}, P. and {Bond}, C. and {Bondu}, F. and {Bonnand}, R. and {Boom}, B.~A. and {Bork}, R. and {Boschi}, V. and {Bose}, S. and {Bouffanais}, Y. and {Bozzi}, A. and {Bradaschia}, C. and {Brady}, P.~R. and {Braginsky}, V.~B. and {Branchesi}, M. and {Brau}, J.~E. and {Briant}, T. and {Brillet}, A. and {Brinkmann}, M. and {Brisson}, V. and {Brockill}, P. and {Brooks}, A.~F. and {Brown}, D.~A. and {Brown}, D.~D. and {Brown}, N.~M. and {Buchanan}, C.~C. and {Buikema}, A. and {Bulik}, T. and {Bulten}, H.~J. and {Buonanno}, A. and {Buskulic}, D. and {Buy}, C. and {Byer}, R.~L. and {Cabero}, M. and {Cadonati}, L. and {Cagnoli}, G. and {Cahillane}, C. and {Bustillo}, J. Calder{\'o}n and {Callister}, T. and {Calloni}, E. and {Camp}, J.~B. and {Cannon}, K.~C. and {Cao}, J. and {Capano}, C.~D. and {Capocasa}, E. and {Carbognani}, F. and {Caride}, S. and {Diaz}, J. Casanueva and {Casentini}, C. and {Caudill}, S. and {Cavagli{\`a}}, M. and {Cavalier}, F. and {Cavalieri}, R. and {Cella}, G. and {Cepeda}, C.~B. and {Baiardi}, L. Cerboni and {Cerretani}, G. and {Cesarini}, E. and {Chakraborty}, R. and {Chalermsongsak}, T. and {Chamberlin}, S.~J. and {Chan}, M. and {Chao}, S. and {Charlton}, P. and {Chassande-Mottin}, E. and {Chen}, H.~Y. and {Chen}, Y. and {Cheng}, C. and {Chincarini}, A. and {Chiummo}, A. and {Cho}, H.~S. and {Cho}, M. and {Chow}, J.~H. and {Christensen}, N. and {Chu}, Q. and {Chua}, S. and {Chung}, S. and {Ciani}, G. and {Clara}, F. and {Clark}, J.~A. and {Cleva}, F. and {Coccia}, E. and {Cohadon}, P. -F. and {Colla}, A. and {Collette}, C.~G. and {Cominsky}, L. and {Constancio}, M. and {Conte}, A. and {Conti}, L. and {Cook}, D. and {Corbitt}, T.~R. and {Cornish}, N. and {Corsi}, A. and {Cortese}, S. and {Costa}, C.~A. and {Coughlin}, M.~W. and {Coughlin}, S.~B. and {Coulon}, J. -P. and {Countryman}, S.~T. and {Couvares}, P. and {Cowan}, E.~E. and {Coward}, D.~M. and {Cowart}, M.~J.},
        title = "{Observation of Gravitational Waves from a Binary Black Hole Merger}",
      journal = {\prl},
         year = 2016,
        month = feb,
       volume = {116},
       number = {6},
          eid = {061102},
        pages = {061102},
          doi = {10.1103/PhysRevLett.116.061102},
archivePrefix = {arXiv},
       eprint = {1602.03837},
 primaryClass = {gr-qc},
       adsurl = {https://ui.adsabs.harvard.edu/abs/2016PhRvL.116f1102A}
}

@ARTICLE{2023abbott,
       author = {{Abbott}, R. and {Abbott}, T.~D. and {Acernese}, F. and {Ackley}, K. and {Adams}, C. and {Adhikari}, N. and {Adhikari}, R.~X. and {Adya}, V.~B. and {Affeldt}, C. and {Agarwal}, D. and {Agathos}, M. and {Agatsuma}, K. and {Aggarwal}, N. and {Aguiar}, O.~D. and {Aiello}, L. and {Ain}, A. and {Ajith}, P. and {Akcay}, S. and {Akutsu}, T. and {Albanesi}, S. and {Allocca}, A. and {Altin}, P.~A. and {Amato}, A. and {Anand}, C. and {Anand}, S. and {Ananyeva}, A. and {Anderson}, S.~B. and {Anderson}, W.~G. and {Ando}, M. and {Andrade}, T. and {Andres}, N. and {Andri{\'c}}, T. and {Angelova}, S.~V. and {Ansoldi}, S. and {Antelis}, J.~M. and {Antier}, S. and {Appert}, S. and {Arai}, Koji and {Arai}, Koya and {Arai}, Y. and {Araki}, S. and {Araya}, A. and {Araya}, M.~C. and {Areeda}, J.~S. and {Ar{\`e}ne}, M. and {Aritomi}, N. and {Arnaud}, N. and {Arogeti}, M. and {Aronson}, S.~M. and {Arun}, K.~G. and {Asada}, H. and {Asali}, Y. and {Ashton}, G. and {Aso}, Y. and {Assiduo}, M. and {Aston}, S.~M. and {Astone}, P. and {Aubin}, F. and {Austin}, C. and {Babak}, S. and {Badaracco}, F. and {Bader}, M.~K.~M. and {Badger}, C. and {Bae}, S. and {Bae}, Y. and {Baer}, A.~M. and {Bagnasco}, S. and {Bai}, Y. and {Baiotti}, L. and {Baird}, J. and {Bajpai}, R. and {Ball}, M. and {Ballardin}, G. and {Ballmer}, S.~W. and {Balsamo}, A. and {Baltus}, G. and {Banagiri}, S. and {Bankar}, D. and {Barayoga}, J.~C. and {Barbieri}, C. and {Barish}, B.~C. and {Barker}, D. and {Barneo}, P. and {Barone}, F. and {Barr}, B. and {Barsotti}, L. and {Barsuglia}, M. and {Barta}, D. and {Bartlett}, J. and {Barton}, M.~A. and {Bartos}, I. and {Bassiri}, R. and {Basti}, A. and {Bawaj}, M. and {Bayley}, J.~C. and {Baylor}, A.~C. and {Bazzan}, M. and {B{\'e}csy}, B. and {Bedakihale}, V.~M. and {Bejger}, M. and {Belahcene}, I. and {Benedetto}, V. and {Beniwal}, D. and {Bennett}, T.~F. and {Bentley}, J.~D. and {Benyaala}, M. and {Bergamin}, F. and {Berger}, B.~K. and {Bernuzzi}, S. and {Berry}, C.~P.~L. and {Bersanetti}, D. and {Bertolini}, A. and {Betzwieser}, J. and {Beveridge}, D. and {Bhandare}, R. and {Bhardwaj}, U. and {Bhattacharjee}, D. and {Bhaumik}, S. and {Bilenko}, I.~A. and {Billingsley}, G. and {Bini}, S. and {Birney}, R. and {Birnholtz}, O. and {Biscans}, S. and {Bischi}, M. and {Biscoveanu}, S. and {Bisht}, A. and {Biswas}, B. and {Bitossi}, M. and {Bizouard}, M. -A. and {Blackburn}, J.~K. and {Blair}, C.~D. and {Blair}, D.~G. and {Blair}, R.~M. and {Bobba}, F. and {Bode}, N. and {Boer}, M. and {Bogaert}, G. and {Boldrini}, M. and {Bonavena}, L.~D. and {Bondu}, F. and {Bonilla}, E. and {Bonnand}, R. and {Booker}, P. and {Boom}, B.~A. and {Bork}, R. and {Boschi}, V. and {Bose}, N. and {Bose}, S. and {Bossilkov}, V. and {Boudart}, V. and {Bouffanais}, Y. and {Bozzi}, A. and {Bradaschia}, C. and {Brady}, P.~R. and {Bramley}, A. and {Branch}, A. and {Branchesi}, M. and {Brandt}, J. and {Brau}, J.~E. and {Breschi}, M. and {Briant}, T. and {Briggs}, J.~H. and {Brillet}, A. and {Brinkmann}, M. and {Brockill}, P. and {Brooks}, A.~F. and {Brooks}, J. and {Brown}, D.~D. and {Brunett}, S. and {Bruno}, G. and {Bruntz}, R. and {Bryant}, J. and {Bulik}, T. and {Bulten}, H.~J. and {Buonanno}, A. and {Buscicchio}, R. and {Buskulic}, D. and {Buy}, C. and {Byer}, R.~L. and {Davies}, G.~S. Cabourn and {Cadonati}, L. and {Cagnoli}, G. and {Cahillane}, C. and {Bustillo}, J. Calder{\'o}n and {Callaghan}, J.~D. and {Callister}, T.~A. and {Calloni}, E. and {Cameron}, J. and {Camp}, J.~B. and {Canepa}, M. and {Canevarolo}, S. and {Cannavacciuolo}, M. and {Cannon}, K.~C. and {Cao}, H. and {Cao}, Z. and {Capocasa}, E. and {Capote}, E. and {Carapella}, G. and {Carbognani}, F.},
        title = "{GWTC-3: Compact Binary Coalescences Observed by LIGO and Virgo during the Second Part of the Third Observing Run}",
      journal = {Physical Review X},
         year = 2023,
        month = oct,
       volume = {13},
       number = {4},
          eid = {041039},
        pages = {041039},
          doi = {10.1103/PhysRevX.13.041039},
archivePrefix = {arXiv},
       eprint = {2111.03606},
 primaryClass = {gr-qc},
       adsurl = {https://ui.adsabs.harvard.edu/abs/2023PhRvX..13d1039A}
}

@ARTICLE{2016mandel,
       author = {{Mandel}, Ilya and {de Mink}, Selma E.},
        title = "{Merging binary black holes formed through chemically homogeneous evolution in short-period stellar binaries}",
      journal = {\mnras},
         year = 2016,
        month = may,
       volume = {458},
       number = {3},
        pages = {2634-2647},
          doi = {10.1093/mnras/stw379},
archivePrefix = {arXiv},
       eprint = {1601.00007},
 primaryClass = {astro-ph.HE},
       adsurl = {https://ui.adsabs.harvard.edu/abs/2016MNRAS.458.2634M}
}

@ARTICLE{2017antonini,
       author = {{Antonini}, Fabio and {Toonen}, Silvia and {Hamers}, Adrian S.},
        title = "{Binary Black Hole Mergers from Field Triples: Properties, Rates, and the Impact of Stellar Evolution}",
      journal = {\apj},
         year = 2017,
        month = jun,
       volume = {841},
       number = {2},
          eid = {77},
        pages = {77},
          doi = {10.3847/1538-4357/aa6f5e},
archivePrefix = {arXiv},
       eprint = {1703.06614},
 primaryClass = {astro-ph.GA},
       adsurl = {https://ui.adsabs.harvard.edu/abs/2017ApJ...841...77A}
}

@ARTICLE{2016rodriguez,
       author = {{Rodriguez}, Carl L. and {Chatterjee}, Sourav and {Rasio}, Frederic A.},
        title = "{Binary black hole mergers from globular clusters: Masses, merger rates, and the impact of stellar evolution}",
      journal = {\prd},
         year = 2016,
        month = apr,
       volume = {93},
       number = {8},
          eid = {084029},
        pages = {084029},
          doi = {10.1103/PhysRevD.93.084029},
archivePrefix = {arXiv},
       eprint = {1602.02444},
 primaryClass = {astro-ph.HE},
       adsurl = {https://ui.adsabs.harvard.edu/abs/2016PhRvD..93h4029R}
}

@ARTICLE{2000portegieszwart,
       author = {{Portegies Zwart}, Simon F. and {McMillan}, Stephen L.~W.},
        title = "{Black Hole Mergers in the Universe}",
      journal = {\apjl},
         year = 2000,
        month = jan,
       volume = {528},
       number = {1},
        pages = {L17-L20},
          doi = {10.1086/312422},
archivePrefix = {arXiv},
       eprint = {astro-ph/9910061},
 primaryClass = {astro-ph},
       adsurl = {https://ui.adsabs.harvard.edu/abs/2000ApJ...528L..17P}
}

@ARTICLE{2016belczynski,
       author = {{Belczynski}, Krzysztof and {Holz}, Daniel E. and {Bulik}, Tomasz and {O'Shaughnessy}, Richard},
        title = "{The first gravitational-wave source from the isolated evolution of two stars in the 40-100 solar mass range}",
      journal = {\nat},
         year = 2016,
        month = jun,
       volume = {534},
       number = {7608},
        pages = {512-515},
          doi = {10.1038/nature18322},
archivePrefix = {arXiv},
       eprint = {1602.04531},
 primaryClass = {astro-ph.HE},
       adsurl = {https://ui.adsabs.harvard.edu/abs/2016Natur.534..512B}
}

@ARTICLE{1993tutukov,
       author = {{Tutukov}, A.~V. and {Yungelson}, L.~R.},
        title = "{The merger rate of neutron star and black hole binaries.}",
      journal = {\mnras},
         year = 1993,
        month = feb,
       volume = {260},
        pages = {675-678},
          doi = {10.1093/mnras/260.3.675},
       adsurl = {https://ui.adsabs.harvard.edu/abs/1993MNRAS.260..675T}
}

@ARTICLE{2005lauer,
       author = {{Lauer}, Tod R. and {Faber}, S.~M. and {Gebhardt}, Karl and {Richstone}, Douglas and {Tremaine}, Scott and {Ajhar}, Edward A. and {Aller}, M.~C. and {Bender}, Ralf and {Dressler}, Alan and {Filippenko}, Alexei V. and {Green}, Richard and {Grillmair}, Carl J. and {Ho}, Luis C. and {Kormendy}, John and {Magorrian}, John and {Pinkney}, Jason and {Siopis}, Christos},
        title = "{The Centers of Early-Type Galaxies with Hubble Space Telescope. V. New WFPC2 Photometry}",
      journal = {\aj},
         year = 2005,
        month = may,
       volume = {129},
       number = {5},
        pages = {2138-2185},
          doi = {10.1086/429565},
archivePrefix = {arXiv},
       eprint = {astro-ph/0412040},
 primaryClass = {astro-ph},
       adsurl = {https://ui.adsabs.harvard.edu/abs/2005AJ....129.2138L}
}

@ARTICLE{1976bahcall,
       author = {{Bahcall}, J.~N. and {Wolf}, R.~A.},
        title = "{Star distribution around a massive black hole in a globular cluster.}",
      journal = {\apj},
         year = 1976,
        month = oct,
       volume = {209},
        pages = {214-232},
          doi = {10.1086/154711},
       adsurl = {https://ui.adsabs.harvard.edu/abs/1976ApJ...209..214B}
}

@ARTICLE{2010baruteau,
       author = {{Baruteau}, C. and {Lin}, D.~N.~C.},
        title = "{Protoplanetary Migration in Turbulent Isothermal Disks}",
      journal = {\apj},
         year = 2010,
        month = feb,
       volume = {709},
       number = {2},
        pages = {759-773},
          doi = {10.1088/0004-637X/709/2/759},
archivePrefix = {arXiv},
       eprint = {0912.0964},
 primaryClass = {astro-ph.EP},
       adsurl = {https://ui.adsabs.harvard.edu/abs/2010ApJ...709..759B}
}

@ARTICLE{2010bartko,
       author = {{Bartko}, H. and {Martins}, F. and {Trippe}, S. and {Fritz}, T.~K. and {Genzel}, R. and {Ott}, T. and {Eisenhauer}, F. and {Gillessen}, S. and {Paumard}, T. and {Alexander}, T. and {Dodds-Eden}, K. and {Gerhard}, O. and {Levin}, Y. and {Mascetti}, L. and {Nayakshin}, S. and {Perets}, H.~B. and {Perrin}, G. and {Pfuhl}, O. and {Reid}, M.~J. and {Rouan}, D. and {Zilka}, M. and {Sternberg}, A.},
        title = "{An Extremely Top-Heavy Initial Mass Function in the Galactic Center Stellar Disks}",
      journal = {\apj},
         year = 2010,
        month = jan,
       volume = {708},
       number = {1},
        pages = {834-840},
          doi = {10.1088/0004-637X/708/1/834},
archivePrefix = {arXiv},
       eprint = {0908.2177},
 primaryClass = {astro-ph.GA},
       adsurl = {https://ui.adsabs.harvard.edu/abs/2010ApJ...708..834B}
}

@ARTICLE{2016stone,
       author = {{Stone}, Nicholas C. and {Metzger}, Brian D.},
        title = "{Rates of stellar tidal disruption as probes of the supermassive black hole mass function}",
      journal = {\mnras},
         year = 2016,
        month = jan,
       volume = {455},
       number = {1},
        pages = {859-883},
          doi = {10.1093/mnras/stv2281},
archivePrefix = {arXiv},
       eprint = {1410.7772},
 primaryClass = {astro-ph.HE},
       adsurl = {https://ui.adsabs.harvard.edu/abs/2016MNRAS.455..859S}
}

@ARTICLE{2008adams,
       author = {{Adams}, Fred C. and {Laughlin}, Gregory and {Bloch}, Anthony M.},
        title = "{Turbulence Implies that Mean Motion Resonances are Rare}",
      journal = {\apj},
         year = 2008,
        month = aug,
       volume = {683},
       number = {2},
        pages = {1117-1128},
          doi = {10.1086/589986},
       adsurl = {https://ui.adsabs.harvard.edu/abs/2008ApJ...683.1117A}
}

@ARTICLE{2010batygin,
       author = {{Batygin}, Konstantin and {Brown}, Michael E.},
        title = "{Early Dynamical Evolution of the Solar System: Pinning Down the Initial Conditions of the Nice Model}",
      journal = {\apj},
         year = 2010,
        month = jun,
       volume = {716},
       number = {2},
        pages = {1323-1331},
          doi = {10.1088/0004-637X/716/2/1323},
archivePrefix = {arXiv},
       eprint = {1004.5414},
 primaryClass = {astro-ph.EP},
       adsurl = {https://ui.adsabs.harvard.edu/abs/2010ApJ...716.1323B}
}

@ARTICLE{2015batygin,
       author = {{Batygin}, Konstantin},
        title = "{Capture of planets into mean-motion resonances and the origins of extrasolar orbital architectures}",
      journal = {\mnras},
         year = 2015,
        month = aug,
       volume = {451},
       number = {3},
        pages = {2589-2609},
          doi = {10.1093/mnras/stv1063},
archivePrefix = {arXiv},
       eprint = {1505.01778},
 primaryClass = {astro-ph.EP},
       adsurl = {https://ui.adsabs.harvard.edu/abs/2015MNRAS.451.2589B}
}

@ARTICLE{2017batygin,
       author = {{Batygin}, Konstantin and {Adams}, Fred C.},
        title = "{An Analytic Criterion for Turbulent Disruption of Planetary Resonances}",
      journal = {\aj},
         year = 2017,
        month = mar,
       volume = {153},
       number = {3},
          eid = {120},
        pages = {120},
          doi = {10.3847/1538-3881/153/3/120},
archivePrefix = {arXiv},
       eprint = {1701.07849},
 primaryClass = {astro-ph.EP},
       adsurl = {https://ui.adsabs.harvard.edu/abs/2017AJ....153..120B}
}

@ARTICLE{2015benitez-llambay,
       author = {{Ben{\'\i}tez-Llambay}, Pablo and {Masset}, Fr{\'e}d{\'e}ric and {Koenigsberger}, Gloria and {Szul{\'a}gyi}, Judit},
        title = "{Planet heating prevents inward migration of planetary cores}",
      journal = {\nat},
         year = 2015,
        month = apr,
       volume = {520},
       number = {7545},
        pages = {63-65},
          doi = {10.1038/nature14277},
archivePrefix = {arXiv},
       eprint = {1510.01778},
 primaryClass = {astro-ph.EP},
       adsurl = {https://ui.adsabs.harvard.edu/abs/2015Natur.520...63B}
}

@ARTICLE{2019bonga,
       author = {{Bonga}, B{\'e}atrice and {Yang}, Huan and {Hughes}, Scott A.},
        title = "{Tidal Resonance in Extreme Mass-Ratio Inspirals}",
      journal = {\prl},
         year = 2019,
        month = sep,
       volume = {123},
       number = {10},
          eid = {101103},
        pages = {101103},
          doi = {10.1103/PhysRevLett.123.101103},
archivePrefix = {arXiv},
       eprint = {1905.00030},
 primaryClass = {gr-qc},
       adsurl = {https://ui.adsabs.harvard.edu/abs/2019PhRvL.123j1103B}
}

@ARTICLE{2025chen,
       author = {{Chen}, Yi-Xian and {Wu}, Yinhao and {Li}, Ya-Ping and {Lin}, Douglas N.~C. and {Alexander}, Richard and {Nayakshin}, Sergei and {Dai}, Fei},
        title = "{Capture and escape of planetary mean-motion resonances in turbulent discs}",
      journal = {\mnras},
         year = 2025,
        month = jun,
       volume = {540},
       number = {2},
        pages = {1998-2007},
          doi = {10.1093/mnras/staf867},
archivePrefix = {arXiv},
       eprint = {2505.13952},
 primaryClass = {astro-ph.EP},
       adsurl = {https://ui.adsabs.harvard.edu/abs/2025MNRAS.540.1998C}
}

@ARTICLE{2017chrenko,
       author = {{Chrenko}, O. and {Bro{\v{z}}}, M. and {Lambrechts}, M.},
        title = "{Eccentricity excitation and merging of planetary embryos heated by pebble accretion}",
      journal = {\aap},
         year = 2017,
        month = oct,
       volume = {606},
          eid = {A114},
        pages = {A114},
          doi = {10.1051/0004-6361/201731033},
archivePrefix = {arXiv},
       eprint = {1706.06329},
 primaryClass = {astro-ph.EP},
       adsurl = {https://ui.adsabs.harvard.edu/abs/2017A&A...606A.114C}
}

@ARTICLE{2022cardoso,
       author = {{Cardoso}, Vitor and {Destounis}, Kyriakos and {Duque}, Francisco and {Macedo}, Rodrigo Panosso and {Maselli}, Andrea},
        title = "{Gravitational Waves from Extreme-Mass-Ratio Systems in Astrophysical Environments}",
      journal = {\prl},
         year = 2022,
        month = dec,
       volume = {129},
       number = {24},
          eid = {241103},
        pages = {241103},
          doi = {10.1103/PhysRevLett.129.241103},
archivePrefix = {arXiv},
       eprint = {2210.01133},
 primaryClass = {gr-qc},
       adsurl = {https://ui.adsabs.harvard.edu/abs/2022PhRvL.129x1103C}
}

@ARTICLE{2008cresswell,
       author = {{Cresswell}, P. and {Nelson}, R.~P.},
        title = "{Three-dimensional simulations of multiple protoplanets embedded in a protostellar disc}",
      journal = {\aap},
         year = 2008,
        month = may,
       volume = {482},
       number = {2},
        pages = {677-690},
          doi = {10.1051/0004-6361:20079178},
archivePrefix = {arXiv},
       eprint = {0811.4322},
 primaryClass = {astro-ph},
       adsurl = {https://ui.adsabs.harvard.edu/abs/2008A&A...482..677C}
}

@ARTICLE{2013deck,
       author = {{Deck}, Katherine M. and {Payne}, Matthew and {Holman}, Matthew J.},
        title = "{First-order Resonance Overlap and the Stability of Close Two-planet Systems}",
      journal = {\apj},
         year = 2013,
        month = sep,
       volume = {774},
       number = {2},
          eid = {129},
        pages = {129},
          doi = {10.1088/0004-637X/774/2/129},
archivePrefix = {arXiv},
       eprint = {1307.8119},
 primaryClass = {astro-ph.EP},
       adsurl = {https://ui.adsabs.harvard.edu/abs/2013ApJ...774..129D}
}

@ARTICLE{2015deck,
       author = {{Deck}, Katherine M. and {Batygin}, Konstantin},
        title = "{Migration of Two Massive Planets into (and out of) First Order Mean Motion Resonances}",
      journal = {\apj},
         year = 2015,
        month = sep,
       volume = {810},
       number = {2},
          eid = {119},
        pages = {119},
          doi = {10.1088/0004-637X/810/2/119},
archivePrefix = {arXiv},
       eprint = {1506.01382},
 primaryClass = {astro-ph.EP},
       adsurl = {https://ui.adsabs.harvard.edu/abs/2015ApJ...810..119D}
}

@ARTICLE{2014duffel,
       author = {{Duffell}, Paul C. and {Haiman}, Zoltan and {MacFadyen}, Andrew I. and {D'Orazio}, Daniel J. and {Farris}, Brian D.},
        title = "{The Migration of Gap-opening Planets is Not Locked to Viscous Disk Evolution}",
      journal = {\apjl},
         year = 2014,
        month = sep,
       volume = {792},
       number = {1},
          eid = {L10},
        pages = {L10},
          doi = {10.1088/2041-8205/792/1/L10},
archivePrefix = {arXiv},
       eprint = {1405.3711},
 primaryClass = {astro-ph.EP},
       adsurl = {https://ui.adsabs.harvard.edu/abs/2014ApJ...792L..10D}
}

@ARTICLE{2017eklund,
       author = {{Eklund}, Henrik and {Masset}, Fr{\'e}d{\'e}ric S.},
        title = "{Evolution of eccentricity and inclination of hot protoplanets embedded in radiative discs}",
      journal = {\mnras},
         year = 2017,
        month = jul,
       volume = {469},
       number = {1},
        pages = {206-217},
          doi = {10.1093/mnras/stx856},
archivePrefix = {arXiv},
       eprint = {1704.01931},
 primaryClass = {astro-ph.EP},
       adsurl = {https://ui.adsabs.harvard.edu/abs/2017MNRAS.469..206E}
}

@ARTICLE{2025epstein-martin,
       author = {{Epstein-Martin}, Marguerite and {Tagawa}, Hiromichi and {Haiman}, Zolt{\'a}n and {Perna}, Rosalba},
        title = "{Time-dependent models of AGN discs with radiation from embedded stellar-mass black holes}",
      journal = {\mnras},
         year = 2025,
        month = mar,
       volume = {537},
       number = {4},
        pages = {3396-3420},
          doi = {10.1093/mnras/staf237},
archivePrefix = {arXiv},
       eprint = {2405.09380},
 primaryClass = {astro-ph.HE},
       adsurl = {https://ui.adsabs.harvard.edu/abs/2025MNRAS.537.3396E}
}

@ARTICLE{2014fabrycky,
       author = {{Fabrycky}, Daniel C. and {Lissauer}, Jack J. and {Ragozzine}, Darin and {Rowe}, Jason F. and {Steffen}, Jason H. and {Agol}, Eric and {Barclay}, Thomas and {Batalha}, Natalie and {Borucki}, William and {Ciardi}, David R. and {Ford}, Eric B. and {Gautier}, Thomas N. and {Geary}, John C. and {Holman}, Matthew J. and {Jenkins}, Jon M. and {Li}, Jie and {Morehead}, Robert C. and {Morris}, Robert L. and {Shporer}, Avi and {Smith}, Jeffrey C. and {Still}, Martin and {Van Cleve}, Jeffrey},
        title = "{Architecture of Kepler's Multi-transiting Systems. II. New Investigations with Twice as Many Candidates}",
      journal = {\apj},
         year = 2014,
        month = aug,
       volume = {790},
       number = {2},
          eid = {146},
        pages = {146},
          doi = {10.1088/0004-637X/790/2/146},
archivePrefix = {arXiv},
       eprint = {1202.6328},
 primaryClass = {astro-ph.EP},
       adsurl = {https://ui.adsabs.harvard.edu/abs/2014ApJ...790..146F}
}

@ARTICLE{2019fromenteau,
       author = {{Fromenteau}, S{\'e}bastien and {Masset}, Fr{\'e}d{\'e}ric S.},
        title = "{Impact of thermal effects on the evolution of eccentricity and inclination of low-mass planets}",
      journal = {\mnras},
         year = 2019,
        month = jun,
       volume = {485},
       number = {4},
        pages = {5035-5049},
          doi = {10.1093/mnras/stz718},
archivePrefix = {arXiv},
       eprint = {1903.04470},
 primaryClass = {astro-ph.EP},
       adsurl = {https://ui.adsabs.harvard.edu/abs/2019MNRAS.485.5035F}
}

@ARTICLE{1980goldreich,
       author = {{Goldreich}, P. and {Tremaine}, S.},
        title = "{Disk-satellite interactions.}",
      journal = {\apj},
         year = 1980,
        month = oct,
       volume = {241},
        pages = {425-441},
          doi = {10.1086/158356},
       adsurl = {https://ui.adsabs.harvard.edu/abs/1980ApJ...241..425G}
}

@ARTICLE{2024gangardt,
       author = {{Gangardt}, Daria and {Trani}, Alessandro Alberto and {Bonnerot}, Cl{\'e}ment and {Gerosa}, Davide},
        title = "{pAGN: the one-stop solution for AGN disc modelling}",
      journal = {\mnras},
         year = 2024,
        month = jun,
       volume = {530},
       number = {4},
        pages = {3689-3705},
          doi = {10.1093/mnras/stae1117},
archivePrefix = {arXiv},
       eprint = {2403.00060},
 primaryClass = {astro-ph.HE},
       adsurl = {https://ui.adsabs.harvard.edu/abs/2024MNRAS.530.3689G}
}

@ARTICLE{2025gerling-dunsmore,
       author = {{Gerling-Dunsmore}, Hannalore J. and {Begelman}, Mitchell C. and {Simon}, Jacob B. and {Armitage}, Philip J.},
        title = "{Magnetic Pressure Dominance Stabilizes AGN Disks Against Gravitational Instability}",
      journal = {arXiv e-prints},
         year = 2025,
        month = aug,
          eid = {arXiv:2508.16842},
        pages = {arXiv:2508.16842},
          doi = {10.48550/arXiv.2508.16842},
archivePrefix = {arXiv},
       eprint = {2508.16842},
 primaryClass = {astro-ph.HE},
       adsurl = {https://ui.adsabs.harvard.edu/abs/2025arXiv250816842G}
}

@ARTICLE{gilbaum2022,
       author = {{Gilbaum}, Shmuel and {Stone}, Nicholas C.},
        title = "{Feedback-dominated Accretion Flows}",
      journal = {\apj},
         year = 2022,
        month = apr,
       volume = {928},
       number = {2},
          eid = {191},
        pages = {191},
          doi = {10.3847/1538-4357/ac4ded},
archivePrefix = {arXiv},
       eprint = {2107.07519},
 primaryClass = {astro-ph.HE},
       adsurl = {https://ui.adsabs.harvard.edu/abs/2022ApJ...928..191G}
}

@ARTICLE{2022gondan,
       author = {{Gond{\'a}n}, L{\'a}szl{\'o} and {Kocsis}, Bence},
        title = "{Astrophysical gravitational-wave echoes from galactic nuclei}",
      journal = {\mnras},
         year = 2022,
        month = sep,
       volume = {515},
       number = {3},
        pages = {3299-3318},
          doi = {10.1093/mnras/stac1985},
archivePrefix = {arXiv},
       eprint = {2110.09540},
 primaryClass = {astro-ph.HE},
       adsurl = {https://ui.adsabs.harvard.edu/abs/2022MNRAS.515.3299G}
}

@ARTICLE{2021guilera,
       author = {{Guilera}, Octavio M. and {Miller Bertolami}, Marcelo M. and {Masset}, Frederic and {Cuadra}, Jorge and {Venturini}, Julia and {Ronco}, Mar{\'\i}a P.},
        title = "{The importance of thermal torques on the migration of planets growing by pebble accretion}",
      journal = {\mnras},
         year = 2021,
        month = nov,
       volume = {507},
       number = {3},
        pages = {3638-3652},
          doi = {10.1093/mnras/stab2371},
archivePrefix = {arXiv},
       eprint = {2108.04880},
 primaryClass = {astro-ph.EP},
       adsurl = {https://ui.adsabs.harvard.edu/abs/2021MNRAS.507.3638G}
}

@ARTICLE{2020hankla,
       author = {{Hankla}, Amelia M. and {Jiang}, Yan-Fei and {Armitage}, Philip J.},
        title = "{Local Simulations of Heating Torques on a Luminous Body in an Accretion Disk}",
      journal = {\apj},
         year = 2020,
        month = oct,
       volume = {902},
       number = {1},
          eid = {50},
        pages = {50},
          doi = {10.3847/1538-4357/abb4df},
archivePrefix = {arXiv},
       eprint = {2005.03785},
 primaryClass = {astro-ph.EP},
       adsurl = {https://ui.adsabs.harvard.edu/abs/2020ApJ...902...50H}
}

@ARTICLE{1982henrard,
       author = {{Henrard}, J.},
        title = "{Capture Into Resonance - an Extension of the Use of Adiabatic Invariants}",
      journal = {Celestial Mechanics},
         year = 1982,
        month = may,
       volume = {27},
       number = {1},
        pages = {3-22},
          doi = {10.1007/BF01228946},
       adsurl = {https://ui.adsabs.harvard.edu/abs/1982CeMec..27....3H}
}

@ARTICLE{1983henrard,
       author = {{Henrard}, J. and {Lemaitre}, A.},
        title = "{A Second Fundamental Model for Resonance}",
      journal = {Celestial Mechanics},
         year = 1983,
        month = jun,
       volume = {30},
       number = {2},
        pages = {197-218},
          doi = {10.1007/BF01234306},
       adsurl = {https://ui.adsabs.harvard.edu/abs/1983CeMec..30..197H}
}

@ARTICLE{2019hadden,
       author = {{Hadden}, Sam},
        title = "{An Integrable Model for the Dynamics of Planetary Mean-motion Resonances}",
      journal = {\aj},
         year = 2019,
        month = dec,
       volume = {158},
       number = {6},
          eid = {238},
        pages = {238},
          doi = {10.3847/1538-3881/ab5287},
archivePrefix = {arXiv},
       eprint = {1909.05264},
 primaryClass = {astro-ph.EP},
       adsurl = {https://ui.adsabs.harvard.edu/abs/2019AJ....158..238H}
}

@ARTICLE{2024hopkins,
       author = {{Hopkins}, Philip F. and {Squire}, Jonathan and {Su}, Kung-Yi and {Steinwandel}, Ulrich P. and {Kremer}, Kyle and {Shi}, Yanlong and {Grudic}, Michael Y. and {Wellons}, Sarah and {Faucher-Giguere}, Claude-Andre and {Angles-Alcazar}, Daniel and {Murray}, Norman and {Quataert}, Eliot},
        title = "{FORGE'd in FIRE II: The Formation of Magnetically-Dominated Quasar Accretion Disks from Cosmological Initial Conditions}",
      journal = {The Open Journal of Astrophysics},
         year = 2024,
        month = mar,
       volume = {7},
          eid = {19},
        pages = {19},
          doi = {10.21105/astro.2310.04506},
archivePrefix = {arXiv},
       eprint = {2310.04506},
 primaryClass = {astro-ph.HE},
       adsurl = {https://ui.adsabs.harvard.edu/abs/2024OJAp....7E..19H}
}

@ARTICLE{2017inayoshi,
       author = {{Inayoshi}, Kohei and {Tamanini}, Nicola and {Caprini}, Chiara and {Haiman}, Zolt{\'a}n},
        title = "{Probing stellar binary black hole formation in galactic nuclei via the imprint of their center of mass acceleration on their gravitational wave signal}",
      journal = {\prd},
         year = 2017,
        month = sep,
       volume = {96},
       number = {6},
          eid = {063014},
        pages = {063014},
          doi = {10.1103/PhysRevD.96.063014},
archivePrefix = {arXiv},
       eprint = {1702.06529},
 primaryClass = {astro-ph.HE},
       adsurl = {https://ui.adsabs.harvard.edu/abs/2017PhRvD..96f3014I}
}

@ARTICLE{2017jimenez,
       author = {{Jim{\'e}nez}, Mar{\'\i}a Alejandra and {Masset}, Fr{\'e}d{\'e}ric S.},
        title = "{Improved torque formula for low- and intermediate-mass planetary migration}",
      journal = {\mnras},
         year = 2017,
        month = nov,
       volume = {471},
       number = {4},
        pages = {4917-4929},
          doi = {10.1093/mnras/stx1946},
archivePrefix = {arXiv},
       eprint = {1707.08988},
 primaryClass = {astro-ph.EP},
       adsurl = {https://ui.adsabs.harvard.edu/abs/2017MNRAS.471.4917J}
}

@ARTICLE{2018kanagawa,
       author = {{Kanagawa}, Kazuhiro D. and {Tanaka}, Hidekazu and {Szuszkiewicz}, Ewa},
        title = "{Radial Migration of Gap-opening Planets in Protoplanetary Disks. I. The Case of a Single Planet}",
      journal = {\apj},
         year = 2018,
        month = jul,
       volume = {861},
       number = {2},
          eid = {140},
        pages = {140},
          doi = {10.3847/1538-4357/aac8d9},
archivePrefix = {arXiv},
       eprint = {1805.11101},
 primaryClass = {astro-ph.EP},
       adsurl = {https://ui.adsabs.harvard.edu/abs/2018ApJ...861..140K}
}

@ARTICLE{2011kocsis,
       author = {{Kocsis}, Bence and {Yunes}, Nicol{\'a}s and {Loeb}, Abraham},
        title = "{Observable signatures of extreme mass-ratio inspiral black hole binaries embedded in thin accretion disks}",
      journal = {\prd},
         year = 2011,
        month = jul,
       volume = {84},
       number = {2},
          eid = {024032},
        pages = {024032},
          doi = {10.1103/PhysRevD.84.024032},
archivePrefix = {arXiv},
       eprint = {1104.2322},
 primaryClass = {astro-ph.GA},
       adsurl = {https://ui.adsabs.harvard.edu/abs/2011PhRvD..84b4032K}
}

@ARTICLE{2004laughlin,
       author = {{Laughlin}, Gregory and {Steinacker}, Adriane and {Adams}, Fred C.},
        title = "{Type I Planetary Migration with MHD Turbulence}",
      journal = {\apj},
         year = 2004,
        month = jun,
       volume = {608},
       number = {1},
        pages = {489-496},
          doi = {10.1086/386316},
archivePrefix = {arXiv},
       eprint = {astro-ph/0308406},
 primaryClass = {astro-ph},
       adsurl = {https://ui.adsabs.harvard.edu/abs/2004ApJ...608..489L}
}

@ARTICLE{2014lega,
       author = {{Lega}, E. and {Crida}, A. and {Bitsch}, B. and {Morbidelli}, A.},
        title = "{Migration of Earth-sized planets in 3D radiative discs}",
      journal = {\mnras},
         year = 2014,
        month = may,
       volume = {440},
       number = {1},
        pages = {683-695},
          doi = {10.1093/mnras/stu304},
archivePrefix = {arXiv},
       eprint = {1402.2834},
 primaryClass = {astro-ph.EP},
       adsurl = {https://ui.adsabs.harvard.edu/abs/2014MNRAS.440..683L}
}

@ARTICLE{2003levin,
       author = {{Levin}, Yuri},
        title = "{Formation of massive stars and black holes in self-gravitating AGN discs, and gravitational waves in LISA band}",
      journal = {arXiv e-prints},
         year = 2003,
        month = jul,
          eid = {astro-ph/0307084},
        pages = {astro-ph/0307084},
archivePrefix = {arXiv},
       eprint = {astro-ph/0307084},
 primaryClass = {astro-ph},
       adsurl = {https://ui.adsabs.harvard.edu/abs/2003astro.ph..7084L}
}

@ARTICLE{1986lin,
       author = {{Lin}, D.~N.~C. and {Papaloizou}, John},
        title = "{On the Tidal Interaction between Protoplanets and the Protoplanetary Disk. III. Orbital Migration of Protoplanets}",
      journal = {\apj},
         year = 1986,
        month = oct,
       volume = {309},
        pages = {846},
          doi = {10.1086/164653},
       adsurl = {https://ui.adsabs.harvard.edu/abs/1986ApJ...309..846L}
}

@ARTICLE{2013lu,
       author = {{Lu}, J.~R. and {Do}, T. and {Ghez}, A.~M. and {Morris}, M.~R. and {Yelda}, S. and {Matthews}, K.},
        title = "{Stellar Populations in the Central 0.5 pc of the Galaxy. II. The Initial Mass Function}",
      journal = {\apj},
         year = 2013,
        month = feb,
       volume = {764},
       number = {2},
          eid = {155},
        pages = {155},
          doi = {10.1088/0004-637X/764/2/155},
archivePrefix = {arXiv},
       eprint = {1301.0540},
 primaryClass = {astro-ph.SR},
       adsurl = {https://ui.adsabs.harvard.edu/abs/2013ApJ...764..155L}
}

@ARTICLE{1995malhotra,
       author = {{Malhotra}, Renu},
        title = "{The Origin of Pluto's Orbit: Implications for the Solar System Beyond Neptune}",
      journal = {\aj},
         year = 1995,
        month = jul,
       volume = {110},
        pages = {420},
          doi = {10.1086/117532},
archivePrefix = {arXiv},
       eprint = {astro-ph/9504036},
 primaryClass = {astro-ph},
       adsurl = {https://ui.adsabs.harvard.edu/abs/1995AJ....110..420M}
}

@article{2001Martini,
   title={Quasar Clustering and the Lifetime of Quasars},
   volume={547},
   ISSN={1538-4357},
   url={http://dx.doi.org/10.1086/318331},
   DOI={10.1086/318331},
   number={1},
   journal={The Astrophysical Journal},
   publisher={American Astronomical Society},
   author={Martini, Paul and Weinberg, David H.},
   year={2001},
   month=jan, pages={12–26} }

@ARTICLE{2014mckernan,
       author = {{McKernan}, B. and {Ford}, K.~E.~S. and {Kocsis}, B. and {Lyra}, W. and {Winter}, L.~M.},
        title = "{Intermediate-mass black holes in AGN discs - II. Model predictions and observational constraints}",
      journal = {\mnras},
         year = 2014,
        month = jun,
       volume = {441},
       number = {1},
        pages = {900-909},
          doi = {10.1093/mnras/stu553},
archivePrefix = {arXiv},
       eprint = {1403.6433},
 primaryClass = {astro-ph.GA},
       adsurl = {https://ui.adsabs.harvard.edu/abs/2014MNRAS.441..900M}
}

@ARTICLE{2017meiron,
       author = {{Meiron}, Yohai and {Kocsis}, Bence and {Loeb}, Abraham},
        title = "{Detecting Triple Systems with Gravitational Wave Observations}",
      journal = {\apj},
         year = 2017,
        month = jan,
       volume = {834},
       number = {2},
          eid = {200},
        pages = {200},
          doi = {10.3847/1538-4357/834/2/200},
archivePrefix = {arXiv},
       eprint = {1604.02148},
 primaryClass = {astro-ph.HE},
       adsurl = {https://ui.adsabs.harvard.edu/abs/2017ApJ...834..200M}
}

@ARTICLE{2007morbidelli,
       author = {{Morbidelli}, Alessandro and {Crida}, Aur{\'e}lien},
        title = "{The dynamics of Jupiter and Saturn in the gaseous protoplanetary disk}",
      journal = {\icarus},
         year = 2007,
        month = nov,
       volume = {191},
       number = {1},
        pages = {158-171},
          doi = {10.1016/j.icarus.2007.04.001},
       adsurl = {https://ui.adsabs.harvard.edu/abs/2007Icar..191..158M}
}

@ARTICLE{2004nelson,
       author = {{Nelson}, Richard P. and {Papaloizou}, John C.~B.},
        title = "{The interaction of giant planets with a disc with MHD turbulence - IV. Migration rates of embedded protoplanets}",
      journal = {\mnras},
         year = 2004,
        month = may,
       volume = {350},
       number = {3},
        pages = {849-864},
          doi = {10.1111/j.1365-2966.2004.07406.x},
archivePrefix = {arXiv},
       eprint = {astro-ph/0308360},
 primaryClass = {astro-ph},
       adsurl = {https://ui.adsabs.harvard.edu/abs/2004MNRAS.350..849N}
}

@ARTICLE{2007oishi,
       author = {{Oishi}, Jeffrey S. and {Mac Low}, Mordecai-Mark and {Menou}, Kristen},
        title = "{Turbulent Torques on Protoplanets in a Dead Zone}",
      journal = {\apj},
         year = 2007,
        month = nov,
       volume = {670},
       number = {1},
        pages = {805-819},
          doi = {10.1086/521781},
archivePrefix = {arXiv},
       eprint = {astro-ph/0702549},
 primaryClass = {astro-ph},
       adsurl = {https://ui.adsabs.harvard.edu/abs/2007ApJ...670..805O}
}

@ARTICLE{2013okuzumi,
       author = {{Okuzumi}, Satoshi and {Ormel}, Chris W.},
        title = "{The Fate of Planetesimals in Turbulent Disks with Dead Zones. I. The Turbulent Stirring Recipe}",
      journal = {\apj},
         year = 2013,
        month = jul,
       volume = {771},
       number = {1},
          eid = {43},
        pages = {43},
          doi = {10.1088/0004-637X/771/1/43},
archivePrefix = {arXiv},
       eprint = {1305.1889},
 primaryClass = {astro-ph.EP},
       adsurl = {https://ui.adsabs.harvard.edu/abs/2013ApJ...771...43O}
}

@ARTICLE{2010paardekooper,
       author = {{Paardekooper}, S. -J. and {Baruteau}, C. and {Crida}, A. and {Kley}, W.},
        title = "{A torque formula for non-isothermal type I planetary migration - I. Unsaturated horseshoe drag}",
      journal = {\mnras},
         year = 2010,
        month = jan,
       volume = {401},
       number = {3},
        pages = {1950-1964},
          doi = {10.1111/j.1365-2966.2009.15782.x},
archivePrefix = {arXiv},
       eprint = {0909.4552},
 primaryClass = {astro-ph.EP},
       adsurl = {https://ui.adsabs.harvard.edu/abs/2010MNRAS.401.1950P}
}

@ARTICLE{2013paardekooper,
       author = {{Paardekooper}, Sijme-Jan and {Rein}, Hanno and {Kley}, Willy},
        title = "{The formation of systems with closely spaced low-mass planets and the application to Kepler-36}",
      journal = {\mnras},
         year = 2013,
        month = oct,
       volume = {434},
       number = {4},
        pages = {3018-3029},
          doi = {10.1093/mnras/stt1224},
archivePrefix = {arXiv},
       eprint = {1304.4762},
 primaryClass = {astro-ph.EP},
       adsurl = {https://ui.adsabs.harvard.edu/abs/2013MNRAS.434.3018P}
}

@ARTICLE{2006Quillen,
       author = {{Quillen}, Alice C.},
        title = "{Reducing the probability of capture into resonance}",
      journal = {\mnras},
         year = 2006,
        month = feb,
       volume = {365},
       number = {4},
        pages = {1367-1382},
          doi = {10.1111/j.1365-2966.2005.09826.x},
archivePrefix = {arXiv},
       eprint = {astro-ph/0507477},
 primaryClass = {astro-ph},
       adsurl = {https://ui.adsabs.harvard.edu/abs/2006MNRAS.365.1367Q}
}

@ARTICLE{2011quillen,
       author = {{Quillen}, Alice C.},
        title = "{Three-body resonance overlap in closely spaced multiple-planet systems}",
      journal = {\mnras},
         year = 2011,
        month = dec,
       volume = {418},
       number = {2},
        pages = {1043-1054},
          doi = {10.1111/j.1365-2966.2011.19555.x},
archivePrefix = {arXiv},
       eprint = {1106.0156},
 primaryClass = {astro-ph.EP},
       adsurl = {https://ui.adsabs.harvard.edu/abs/2011MNRAS.418.1043Q}
}

@ARTICLE{2009rein,
       author = {{Rein}, H. and {Papaloizou}, J.~C.~B.},
        title = "{On the evolution of mean motion resonances through stochastic forcing: fast and slow libration modes and the origin of HD 128311}",
      journal = {\aap},
         year = 2009,
        month = apr,
       volume = {497},
       number = {2},
        pages = {595-609},
          doi = {10.1051/0004-6361/200811330},
archivePrefix = {arXiv},
       eprint = {0811.1813},
 primaryClass = {astro-ph},
       adsurl = {https://ui.adsabs.harvard.edu/abs/2009A&A...497..595R}
}

@ARTICLE{2004sano,
       author = {{Sano}, Takayoshi and {Inutsuka}, Shu-ichiro and {Turner}, Neal J. and {Stone}, James M.},
        title = "{Angular Momentum Transport by Magnetohydrodynamic Turbulence in Accretion Disks: Gas Pressure Dependence of the Saturation Level of the Magnetorotational Instability}",
      journal = {\apj},
         year = 2004,
        month = apr,
       volume = {605},
       number = {1},
        pages = {321-339},
          doi = {10.1086/382184},
archivePrefix = {arXiv},
       eprint = {astro-ph/0312480},
 primaryClass = {astro-ph},
       adsurl = {https://ui.adsabs.harvard.edu/abs/2004ApJ...605..321S}
}

@ARTICLE{schlosman1989,
       author = {{Shlosman}, Isaac and {Begelman}, Mitchell C.},
        title = "{Evolution of Self-Gravitating Accretion Disks in Active Galactic Nuclei}",
      journal = {\apj},
         year = 1989,
        month = jun,
       volume = {341},
        pages = {685},
          doi = {10.1086/167526},
       adsurl = {https://ui.adsabs.harvard.edu/abs/1989ApJ...341..685S}
}

@ARTICLE{2019secunda,
       author = {{Secunda}, Amy and {Bellovary}, Jillian and {Mac Low}, Mordecai-Mark and {Ford}, K.~E. Saavik and {McKernan}, Barry and {Leigh}, Nathan W.~C. and {Lyra}, Wladimir and {S{\'a}ndor}, Zsolt},
        title = "{Orbital Migration of Interacting Stellar Mass Black Holes in Disks around Supermassive Black Holes}",
      journal = {\apj},
         year = 2019,
        month = jun,
       volume = {878},
       number = {2},
          eid = {85},
        pages = {85},
          doi = {10.3847/1538-4357/ab20ca},
archivePrefix = {arXiv},
       eprint = {1807.02859},
 primaryClass = {astro-ph.HE},
       adsurl = {https://ui.adsabs.harvard.edu/abs/2019ApJ...878...85S}
}

@ARTICLE{2020secunda,
       author = {{Secunda}, Amy and {Bellovary}, Jillian and {Mac Low}, Mordecai-Mark and {Ford}, K.~E. Saavik and {McKernan}, Barry and {Leigh}, Nathan W.~C. and {Lyra}, Wladimir and {S{\'a}ndor}, Zsolt and {Adorno}, Jose I.},
        title = "{Orbital Migration of Interacting Stellar Mass Black Holes in Disks around Supermassive Black Holes. II. Spins and Incoming Objects}",
      journal = {\apj},
         year = 2020,
        month = nov,
       volume = {903},
       number = {2},
          eid = {133},
        pages = {133},
          doi = {10.3847/1538-4357/abbc1d},
archivePrefix = {arXiv},
       eprint = {2004.11936},
 primaryClass = {astro-ph.HE},
       adsurl = {https://ui.adsabs.harvard.edu/abs/2020ApJ...903..133S}
}

@ARTICLE{1984sessin,
       author = {{Sessin}, W. and {Ferraz-Mello}, S.},
        title = "{Motion of two planets with periods commensurable in the ratio 2{\ensuremath{:}}1 solutions of the hori auxiliary system}",
      journal = {Celestial Mechanics},
         year = 1984,
        month = apr,
       volume = {32},
       number = {4},
        pages = {307-332},
          doi = {10.1007/BF01229087},
       adsurl = {https://ui.adsabs.harvard.edu/abs/1984CeMec..32..307S}
}

@ARTICLE{1973shakura,
       author = {{Shakura}, N.~I. and {Sunyaev}, R.~A.},
        title = "{Black holes in binary systems. Observational appearance.}",
      journal = {\aap},
         year = 1973,
        month = jan,
       volume = {24},
        pages = {337-355},
       adsurl = {https://ui.adsabs.harvard.edu/abs/1973A&A....24..337S}
}

@ARTICLE{sirko2003,
       author = {{Sirko}, Edwin and {Goodman}, Jeremy},
        title = "{Spectral energy distributions of marginally self-gravitating quasi-stellar object discs}",
      journal = {\mnras},
         year = 2003,
        month = may,
       volume = {341},
       number = {2},
        pages = {501-508},
          doi = {10.1046/j.1365-8711.2003.06431.x},
archivePrefix = {arXiv},
       eprint = {astro-ph/0209469},
 primaryClass = {astro-ph},
       adsurl = {https://ui.adsabs.harvard.edu/abs/2003MNRAS.341..501S}
}

@ARTICLE{2005thompson,
       author = {{Thompson}, Todd A. and {Quataert}, Eliot and {Murray}, Norman},
        title = "{Radiation Pressure-supported Starburst Disks and Active Galactic Nucleus Fueling}",
      journal = {\apj},
         year = 2005,
        month = sep,
       volume = {630},
       number = {1},
        pages = {167-185},
          doi = {10.1086/431923},
archivePrefix = {arXiv},
       eprint = {astro-ph/0503027},
 primaryClass = {astro-ph},
       adsurl = {https://ui.adsabs.harvard.edu/abs/2005ApJ...630..167T}
}

@ARTICLE{2004tanaka,
       author = {{Tanaka}, Hidekazu and {Ward}, William R.},
        title = "{Three-dimensional Interaction between a Planet and an Isothermal Gaseous Disk. II. Eccentricity Waves and Bending Waves}",
      journal = {\apj},
         year = 2004,
        month = feb,
       volume = {602},
       number = {1},
        pages = {388-395},
          doi = {10.1086/380992},
       adsurl = {https://ui.adsabs.harvard.edu/abs/2004ApJ...602..388T}
}

@ARTICLE{2025tomayo,
       author = {{Tamayo}, Daniel and {Hadden}, Samuel},
        title = "{A Unified, Physical Framework for Mean Motion Resonances}",
      journal = {\apj},
         year = 2025,
        month = jun,
       volume = {986},
       number = {1},
          eid = {11},
        pages = {11},
          doi = {10.3847/1538-4357/adc1c4},
archivePrefix = {arXiv},
       eprint = {2410.21748},
 primaryClass = {astro-ph.EP},
       adsurl = {https://ui.adsabs.harvard.edu/abs/2025ApJ...986...11T}
}

@ARTICLE{toomre1964,
       author = {{Toomre}, A.},
        title = "{On the gravitational stability of a disk of stars.}",
      journal = {\apj},
         year = 1964,
        month = may,
       volume = {139},
        pages = {1217-1238},
          doi = {10.1086/147861},
       adsurl = {https://ui.adsabs.harvard.edu/abs/1964ApJ...139.1217T}
}

@ARTICLE{2020velasco,
       author = {{Velasco Romero}, David A. and {Masset}, Fr{\'e}d{\'e}ric S.},
        title = "{Dynamical friction with radiative feedback - II. High-resolution study of the subsonic regime}",
      journal = {\mnras},
         year = 2020,
        month = jun,
       volume = {495},
       number = {2},
        pages = {2063-2074},
          doi = {10.1093/mnras/staa1215},
archivePrefix = {arXiv},
       eprint = {2004.13422},
 primaryClass = {astro-ph.EP},
       adsurl = {https://ui.adsabs.harvard.edu/abs/2020MNRAS.495.2063V}
}

@ARTICLE{1986wisdom,
       author = {{Wisdom}, J.},
        title = "{Canonical Solution of the Two Critical Argument Problem}",
      journal = {Celestial Mechanics},
         year = 1986,
        month = feb,
       volume = {38},
       number = {2},
        pages = {175-180},
          doi = {10.1007/BF01230429},
       adsurl = {https://ui.adsabs.harvard.edu/abs/1986CeMec..38..175W}
}

@ARTICLE{2024wu,
       author = {{Wu}, Yinhao and {Chen}, Yi-Xian and {Lin}, Douglas N.~C.},
        title = "{Chaotic Type I migration in turbulent discs}",
      journal = {\mnras},
         year = 2024,
        month = feb,
       volume = {528},
       number = {1},
        pages = {L127-L132},
          doi = {10.1093/mnrasl/slad183},
archivePrefix = {arXiv},
       eprint = {2311.15747},
 primaryClass = {astro-ph.EP},
       adsurl = {https://ui.adsabs.harvard.edu/abs/2024MNRAS.528L.127W}
}

@ARTICLE{2025xu,
       author = {{Xu}, Wen-Long and {Li}, Yu-Zhe and {Chen}, Yi-Gu and {Li}, Hui and {Lei}, Wei-Hua},
        title = "{Estimating the Lensing Probability for Binary Black Hole Mergers in AGN disk by Using Mismatch Threshold}",
      journal = {arXiv e-prints},
         year = 2025,
        month = may,
          eid = {arXiv:2505.07114},
        pages = {arXiv:2505.07114},
          doi = {10.48550/arXiv.2505.07114},
archivePrefix = {arXiv},
       eprint = {2505.07114},
 primaryClass = {astro-ph.HE},
       adsurl = {https://ui.adsabs.harvard.edu/abs/2025arXiv250507114X}
}

@ARTICLE{2017yang,
       author = {{Yang}, Huan and {Casals}, Marc},
        title = "{General relativistic dynamics of an extreme mass-ratio binary interacting with an external body}",
      journal = {\prd},
         year = 2017,
        month = oct,
       volume = {96},
       number = {8},
          eid = {083015},
        pages = {083015},
          doi = {10.1103/PhysRevD.96.083015},
archivePrefix = {arXiv},
       eprint = {1704.02022},
 primaryClass = {gr-qc},
       adsurl = {https://ui.adsabs.harvard.edu/abs/2017PhRvD..96h3015Y}
}

@ARTICLE{2011yunes,
       author = {{Yunes}, Nicol{\'a}s and {Kocsis}, Bence and {Loeb}, Abraham and {Haiman}, Zolt{\'a}n},
        title = "{Imprint of Accretion Disk-Induced Migration on Gravitational Waves from Extreme Mass Ratio Inspirals}",
      journal = {\prl},
         year = 2011,
        month = oct,
       volume = {107},
       number = {17},
          eid = {171103},
        pages = {171103},
          doi = {10.1103/PhysRevLett.107.171103},
archivePrefix = {arXiv},
       eprint = {1103.4609},
 primaryClass = {astro-ph.CO},
       adsurl = {https://ui.adsabs.harvard.edu/abs/2011PhRvL.107q1103Y}
}

@book{MurrayDermott1999,
  author    = {Murray, C.~D. and Dermott, S.~F.},
  title     = {Solar System Dynamics},
  publisher = {Cambridge University Press},
  year      = {1999},
  address   = {Cambridge, UK},
  isbn      = {9780521575973},
  doi       = {10.1017/CBO9781139174817},
  url       = {https://doi.org/10.1017/CBO9781139174817},
  note      = {Online publication date: June 2012}
}

@ARTICLE{Volpi2024,
       author = {{Volpi}, Mara and {Libert}, Anne-Sophie},
        title = "{The effects of general relativity on close-in radial-velocity-detected exosystems}",
      journal = {\aap},
         year = 2024,
        month = mar,
       volume = {683},
          eid = {A193},
        pages = {A193},
          doi = {10.1051/0004-6361/202346727},
       adsurl = {https://ui.adsabs.harvard.edu/abs/2024A&A...683A.193V}
}

@ARTICLE{Richardson1988,
       author = {{Richardson}, D.~L. and {Kelly}, T.~J.},
        title = "{Two-body motion in the post-Newtonian approximation.}",
      journal = {Celestial Mechanics},
         year = 1988,
        month = jan,
       volume = {43},
       number = {1-4},
        pages = {193-210},
          doi = {10.1007/BF01234566},
       adsurl = {https://ui.adsabs.harvard.edu/abs/1988CeMec..43..193R}
}

@ARTICLE{Migaszewski2009,
       author = {{Migaszewski}, Cezary and {Go{\'z}dziewski}, Krzysztof},
        title = "{Secular dynamics of a coplanar, non-resonant planetary system under the general relativity and quadrupole moment perturbations}",
      journal = {\mnras},
         year = 2009,
        month = jan,
       volume = {392},
       number = {1},
        pages = {2-18},
          doi = {10.1111/j.1365-2966.2008.14025.x},
archivePrefix = {arXiv},
       eprint = {0809.5248},
 primaryClass = {astro-ph},
       adsurl = {https://ui.adsabs.harvard.edu/abs/2009MNRAS.392....2M}
}

@ARTICLE{Batygin2013,
       author = {{Batygin}, K. and {Morbidelli}, A.},
        title = "{Analytical treatment of planetary resonances}",
      journal = {\aap},
         year = 2013,
        month = aug,
       volume = {556},
          eid = {A28},
        pages = {A28},
          doi = {10.1051/0004-6361/201220907},
archivePrefix = {arXiv},
       eprint = {1305.6513},
 primaryClass = {astro-ph.EP},
       adsurl = {https://ui.adsabs.harvard.edu/abs/2013A&A...556A..28B}
}

@ARTICLE{Masset2017,
       author = {{Masset}, Fr{\'e}d{\'e}ric S.},
        title = "{Coorbital thermal torques on low-mass protoplanets}",
      journal = {\mnras},
         year = 2017,
        month = dec,
       volume = {472},
       number = {4},
        pages = {4204-4219},
          doi = {10.1093/mnras/stx2271},
archivePrefix = {arXiv},
       eprint = {1708.09807},
 primaryClass = {astro-ph.EP},
       adsurl = {https://ui.adsabs.harvard.edu/abs/2017MNRAS.472.4204M}
}

@ARTICLE{Adams2006,
       author = {{Adams}, Fred C. and {Laughlin}, Gregory},
        title = "{Relativistic Effects in Extrasolar Planetary Systems}",
      journal = {International Journal of Modern Physics D},
         year = 2006,
        month = jan,
       volume = {15},
       number = {12},
        pages = {2133-2140},
          doi = {10.1142/S0218271806009479},
archivePrefix = {arXiv},
       eprint = {astro-ph/0611861},
 primaryClass = {astro-ph},
       adsurl = {https://ui.adsabs.harvard.edu/abs/2006IJMPD..15.2133A}
}

@ARTICLE{Pucacco2024,
       author = {{Pucacco}, Giuseppe},
        title = "{Dynamical stability of the Laplace resonance}",
      journal = {Celestial Mechanics and Dynamical Astronomy},
         year = 2024,
        month = dec,
       volume = {136},
       number = {6},
          eid = {51},
        pages = {51},
          doi = {10.1007/s10569-024-10221-3},
archivePrefix = {arXiv},
       eprint = {2410.12768},
 primaryClass = {astro-ph.EP},
       adsurl = {https://ui.adsabs.harvard.edu/abs/2024CeMDA.136...51P}
}

@ARTICLE{Bhaskar2022,
       author = {{Bhaskar}, Hareesh Gautham and {Li}, Gongjie and {Lin}, Douglas N.~C.},
        title = "{Black Hole Mergers through Evection Resonances}",
      journal = {\apj},
         year = 2022,
        month = aug,
       volume = {934},
       number = {2},
          eid = {141},
        pages = {141},
          doi = {10.3847/1538-4357/ac7b26},
archivePrefix = {arXiv},
       eprint = {2204.07282},
 primaryClass = {astro-ph.HE},
       adsurl = {https://ui.adsabs.harvard.edu/abs/2022ApJ...934..141B}
}

@ARTICLE{Li2023,
       author = {{Li}, Gongjie and {Bhaskar}, Hareesh Gautham and {Kocsis}, Bence and {Lin}, Douglas N.~C.},
        title = "{Secular Spin-Orbit Resonances of Black Hole Binaries in AGN Disks}",
      journal = {\apj},
         year = 2023,
        month = jun,
       volume = {950},
       number = {1},
          eid = {48},
        pages = {48},
          doi = {10.3847/1538-4357/acccf1},
archivePrefix = {arXiv},
       eprint = {2202.11739},
 primaryClass = {astro-ph.HE},
       adsurl = {https://ui.adsabs.harvard.edu/abs/2023ApJ...950...48L}
}

@ARTICLE{Bhaskar2023,
       author = {{Bhaskar}, Hareesh Gautham and {Li}, Gongjie and {Lin}, Douglas},
        title = "{Enhanced Black Hole Mergers in Active Galactic Nucleus Disks due to Precession-induced Resonances}",
      journal = {\apj},
         year = 2023,
        month = aug,
       volume = {952},
       number = {2},
          eid = {98},
        pages = {98},
          doi = {10.3847/1538-4357/acda8f},
archivePrefix = {arXiv},
       eprint = {2303.12539},
 primaryClass = {astro-ph.HE},
       adsurl = {https://ui.adsabs.harvard.edu/abs/2023ApJ...952...98B}
}

@ARTICLE{Weisserman2023,
       author = {{Weisserman}, D. and {Becker}, J.~C. and {Vanderburg}, A.},
        title = "{Kepler-80 Revisited: Assessing the Participation of a Newly Discovered Planet in the Resonant Chain}",
      journal = {\aj},
         year = 2023,
        month = mar,
       volume = {165},
       number = {3},
          eid = {89},
        pages = {89},
          doi = {10.3847/1538-3881/acac80},
archivePrefix = {arXiv},
       eprint = {2212.08695},
 primaryClass = {astro-ph.EP},
       adsurl = {https://ui.adsabs.harvard.edu/abs/2023AJ....165...89W}
}

@ARTICLE{Leleu2021,
       author = {{Leleu}, A. and {Alibert}, Y. and {Hara}, N.~C. and {Hooton}, M.~J. and {Wilson}, T.~G. and {Robutel}, P. and {Delisle}, J. -B. and {Laskar}, J. and {Hoyer}, S. and {Lovis}, C. and {Bryant}, E.~M. and {Ducrot}, E. and {Cabrera}, J. and {Delrez}, L. and {Acton}, J.~S. and {Adibekyan}, V. and {Allart}, R. and {Allende Prieto}, C. and {Alonso}, R. and {Alves}, D. and {Anderson}, D.~R. and {Angerhausen}, D. and {Anglada Escud{\'e}}, G. and {Asquier}, J. and {Barrado}, D. and {Barros}, S.~C.~C. and {Baumjohann}, W. and {Bayliss}, D. and {Beck}, M. and {Beck}, T. and {Bekkelien}, A. and {Benz}, W. and {Billot}, N. and {Bonfanti}, A. and {Bonfils}, X. and {Bouchy}, F. and {Bourrier}, V. and {Bou{\'e}}, G. and {Brandeker}, A. and {Broeg}, C. and {Buder}, M. and {Burdanov}, A. and {Burleigh}, M.~R. and {B{\'a}rczy}, T. and {Cameron}, A.~C. and {Chamberlain}, S. and {Charnoz}, S. and {Cooke}, B.~F. and {Corral Van Damme}, C. and {Correia}, A.~C.~M. and {Cristiani}, S. and {Damasso}, M. and {Davies}, M.~B. and {Deleuil}, M. and {Demangeon}, O.~D.~S. and {Demory}, B. -O. and {Di Marcantonio}, P. and {Di Persio}, G. and {Dumusque}, X. and {Ehrenreich}, D. and {Erikson}, A. and {Figueira}, P. and {Fortier}, A. and {Fossati}, L. and {Fridlund}, M. and {Futyan}, D. and {Gandolfi}, D. and {Garc{\'\i}a Mu{\~n}oz}, A. and {Garcia}, L.~J. and {Gill}, S. and {Gillen}, E. and {Gillon}, M. and {Goad}, M.~R. and {Gonz{\'a}lez Hern{\'a}ndez}, J.~I. and {Guedel}, M. and {G{\"u}nther}, M.~N. and {Haldemann}, J. and {Henderson}, B. and {Heng}, K. and {Hogan}, A.~E. and {Isaak}, K. and {Jehin}, E. and {Jenkins}, J.~S. and {Jord{\'a}n}, A. and {Kiss}, L. and {Kristiansen}, M.~H. and {Lam}, K. and {Lavie}, B. and {Lecavelier des Etangs}, A. and {Lendl}, M. and {Lillo-Box}, J. and {Lo Curto}, G. and {Magrin}, D. and {Martins}, C.~J.~A.~P. and {Maxted}, P.~F.~L. and {McCormac}, J. and {Mehner}, A. and {Micela}, G. and {Molaro}, P. and {Moyano}, M. and {Murray}, C.~A. and {Nascimbeni}, V. and {Nunes}, N.~J. and {Olofsson}, G. and {Osborn}, H.~P. and {Oshagh}, M. and {Ottensamer}, R. and {Pagano}, I. and {Pall{\'e}}, E. and {Pedersen}, P.~P. and {Pepe}, F.~A. and {Persson}, C.~M. and {Peter}, G. and {Piotto}, G. and {Polenta}, G. and {Pollacco}, D. and {Poretti}, E. and {Pozuelos}, F.~J. and {Queloz}, D. and {Ragazzoni}, R. and {Rando}, N. and {Ratti}, F. and {Rauer}, H. and {Raynard}, L. and {Rebolo}, R. and {Reimers}, C. and {Ribas}, I. and {Santos}, N.~C. and {Scandariato}, G. and {Schneider}, J. and {Sebastian}, D. and {Sestovic}, M. and {Simon}, A.~E. and {Smith}, A.~M.~S. and {Sousa}, S.~G. and {Sozzetti}, A. and {Steller}, M. and {Su{\'a}rez Mascare{\~n}o}, A. and {Szab{\'o}}, Gy. M. and {S{\'e}gransan}, D. and {Thomas}, N. and {Thompson}, S. and {Tilbrook}, R.~H. and {Triaud}, A. and {Turner}, O. and {Udry}, S. and {Van Grootel}, V. and {Venus}, H. and {Verrecchia}, F. and {Vines}, J.~I. and {Walton}, N.~A. and {West}, R.~G. and {Wheatley}, P.~J. and {Wolter}, D. and {Zapatero Osorio}, M.~R.},
        title = "{Six transiting planets and a chain of Laplace resonances in TOI-178}",
      journal = {\aap},
         year = 2021,
        month = may,
       volume = {649},
          eid = {A26},
        pages = {A26},
          doi = {10.1051/0004-6361/202039767},
archivePrefix = {arXiv},
       eprint = {2101.09260},
 primaryClass = {astro-ph.EP},
       adsurl = {https://ui.adsabs.harvard.edu/abs/2021A&A...649A..26L}
}

@ARTICLE{Johnson2006,
       author = {{Johnson}, Eric T. and {Goodman}, Jeremy and {Menou}, Kristen},
        title = "{Diffusive Migration of Low-Mass Protoplanets in Turbulent Disks}",
      journal = {\apj},
         year = 2006,
        month = aug,
       volume = {647},
       number = {2},
        pages = {1413-1425},
          doi = {10.1086/505462},
archivePrefix = {arXiv},
       eprint = {astro-ph/0603235},
 primaryClass = {astro-ph},
       adsurl = {https://ui.adsabs.harvard.edu/abs/2006ApJ...647.1413J}
}

@ARTICLE{Grishin2024,
       author = {{Grishin}, Evgeni and {Gilbaum}, Shmuel and {Stone}, Nicholas C.},
        title = "{The effect of thermal torques on AGN disc migration traps and gravitational wave populations}",
      journal = {\mnras},
         year = 2024,
        month = may,
       volume = {530},
       number = {2},
        pages = {2114-2132},
          doi = {10.1093/mnras/stae828},
archivePrefix = {arXiv},
       eprint = {2307.07546},
 primaryClass = {astro-ph.HE},
       adsurl = {https://ui.adsabs.harvard.edu/abs/2024MNRAS.530.2114G}
}

@ARTICLE{Gilbaum2025,
       author = {{Gilbaum}, Shmuel and {Grishin}, Evgeni and {Stone}, Nicholas C. and {Mandel}, Ilya},
        title = "{How to Escape from a Trap: Outcomes of Repeated Black Hole Mergers in Active Galactic Nuclei}",
      journal = {\apjl},
         year = 2025,
        month = mar,
       volume = {982},
       number = {1},
          eid = {L13},
        pages = {L13},
          doi = {10.3847/2041-8213/adb7dc},
archivePrefix = {arXiv},
       eprint = {2410.19904},
 primaryClass = {astro-ph.HE},
       adsurl = {https://ui.adsabs.harvard.edu/abs/2025ApJ...982L..13G}
}

@ARTICLE{2002lee,
       author = {{Lee}, Man Hoi and {Peale}, S.~J.},
        title = "{Dynamics and Origin of the 2:1 Orbital Resonances of the GJ 876 Planets}",
      journal = {\apj},
         year = 2002,
        month = mar,
       volume = {567},
       number = {1},
        pages = {596-609},
          doi = {10.1086/338504},
       adsurl = {https://ui.adsabs.harvard.edu/abs/2002ApJ...567..596L}
}

@ARTICLE{2007terquem,
       author = {{Terquem}, Caroline and {Papaloizou}, John C.~B.},
        title = "{Migration and the Formation of Systems of Hot Super-Earths and Neptunes}",
      journal = {\apj},
         year = 2007,
        month = jan,
       volume = {654},
       number = {2},
        pages = {1110-1120},
          doi = {10.1086/509497},
archivePrefix = {arXiv},
       eprint = {astro-ph/0609779},
 primaryClass = {astro-ph},
       adsurl = {https://ui.adsabs.harvard.edu/abs/2007ApJ...654.1110T}
}

@ARTICLE{2006cresswell,
       author = {{Cresswell}, P. and {Nelson}, R.~P.},
        title = "{On the evolution of multiple protoplanets embedded in a protostellar disc}",
      journal = {\aap},
         year = 2006,
        month = may,
       volume = {450},
       number = {2},
        pages = {833-853},
          doi = {10.1051/0004-6361:20054551},
       adsurl = {https://ui.adsabs.harvard.edu/abs/2006A&A...450..833C}
}

@ARTICLE{2011lissauer,
       author = {{Lissauer}, Jack J. and {Ragozzine}, Darin and {Fabrycky}, Daniel C. and {Steffen}, Jason H. and {Ford}, Eric B. and {Jenkins}, Jon M. and {Shporer}, Avi and {Holman}, Matthew J. and {Rowe}, Jason F. and {Quintana}, Elisa V. and {Batalha}, Natalie M. and {Borucki}, William J. and {Bryson}, Stephen T. and {Caldwell}, Douglas A. and {Carter}, Joshua A. and {Ciardi}, David and {Dunham}, Edward W. and {Fortney}, Jonathan J. and {Gautier}, III, Thomas N. and {Howell}, Steve B. and {Koch}, David G. and {Latham}, David W. and {Marcy}, Geoffrey W. and {Morehead}, Robert C. and {Sasselov}, Dimitar},
        title = "{Architecture and Dynamics of Kepler's Candidate Multiple Transiting Planet Systems}",
      journal = {\apjs},
         year = 2011,
        month = nov,
       volume = {197},
       number = {1},
          eid = {8},
        pages = {8},
          doi = {10.1088/0067-0049/197/1/8},
archivePrefix = {arXiv},
       eprint = {1102.0543},
 primaryClass = {astro-ph.EP},
       adsurl = {https://ui.adsabs.harvard.edu/abs/2011ApJS..197....8L}
}

@ARTICLE{2024dai,
       author = {{Dai}, Fei and {Goldberg}, Max and {Batygin}, Konstantin and {van Saders}, Jennifer and {Chiang}, Eugene and {Choksi}, Nick and {Li}, Rixin and {Petigura}, Erik A. and {Gilbert}, Gregory J. and {Millholland}, Sarah C. and {Dai}, Yuan-Zhe and {Bouma}, Luke and {Weiss}, Lauren M. and {Winn}, Joshua N.},
        title = "{The Prevalence of Resonance Among Young, Close-in Planets}",
      journal = {\aj},
         year = 2024,
        month = dec,
       volume = {168},
       number = {6},
          eid = {239},
        pages = {239},
          doi = {10.3847/1538-3881/ad83a6},
archivePrefix = {arXiv},
       eprint = {2406.06885},
 primaryClass = {astro-ph.EP},
       adsurl = {https://ui.adsabs.harvard.edu/abs/2024AJ....168..239D}
}

@ARTICLE{2017izidoro,
       author = {{Izidoro}, Andre and {Ogihara}, Masahiro and {Raymond}, Sean N. and {Morbidelli}, Alessandro and {Pierens}, Arnaud and {Bitsch}, Bertram and {Cossou}, Christophe and {Hersant}, Franck},
        title = "{Breaking the chains: hot super-Earth systems from migration and disruption of compact resonant chains}",
      journal = {\mnras},
         year = 2017,
        month = sep,
       volume = {470},
       number = {2},
        pages = {1750-1770},
          doi = {10.1093/mnras/stx1232},
archivePrefix = {arXiv},
       eprint = {1703.03634},
 primaryClass = {astro-ph.EP},
       adsurl = {https://ui.adsabs.harvard.edu/abs/2017MNRAS.470.1750I}
}

@ARTICLE{2021izidoro,
       author = {{Izidoro}, Andr{\'e} and {Bitsch}, Bertram and {Raymond}, Sean N. and {Johansen}, Anders and {Morbidelli}, Alessandro and {Lambrechts}, Michiel and {Jacobson}, Seth A.},
        title = "{Formation of planetary systems by pebble accretion and migration. Hot super-Earth systems from breaking compact resonant chains}",
      journal = {\aap},
         year = 2021,
        month = jun,
       volume = {650},
          eid = {A152},
        pages = {A152},
          doi = {10.1051/0004-6361/201935336},
archivePrefix = {arXiv},
       eprint = {1902.08772},
 primaryClass = {astro-ph.EP},
       adsurl = {https://ui.adsabs.harvard.edu/abs/2021A&A...650A.152I}
}

@ARTICLE{2022goldberg,
       author = {{Goldberg}, Max and {Batygin}, Konstantin},
        title = "{Architectures of Compact Super-Earth Systems Shaped by Instabilities}",
      journal = {\aj},
         year = 2022,
        month = may,
       volume = {163},
       number = {5},
          eid = {201},
        pages = {201},
          doi = {10.3847/1538-3881/ac5961},
archivePrefix = {arXiv},
       eprint = {2203.00801},
 primaryClass = {astro-ph.EP},
       adsurl = {https://ui.adsabs.harvard.edu/abs/2022AJ....163..201G}
}

@ARTICLE{2015delisle,
       author = {{Delisle}, J.-B. and {Correia}, A.~C.~M. and {Laskar}, J.},
        title = "{Stability of resonant configurations during the migration of planets and constraints on disk-planet interactions}",
      journal = {\aap},
         year = 2015,
        month = jul,
       volume = {579},
          eid = {A128},
        pages = {A128},
          doi = {10.1051/0004-6361/201526285},
archivePrefix = {arXiv},
       eprint = {1506.03750},
 primaryClass = {astro-ph.EP},
       adsurl = {https://ui.adsabs.harvard.edu/abs/2015A&A...579A.128D}
}

@ARTICLE{2023goldberg,
       author = {{Goldberg}, Max and {Batygin}, Konstantin},
        title = "{Dynamics and Origins of the Near-resonant Kepler Planets}",
      journal = {\apj},
         year = 2023,
        month = may,
       volume = {948},
       number = {1},
          eid = {12},
        pages = {12},
          doi = {10.3847/1538-4357/acc9ae},
archivePrefix = {arXiv},
       eprint = {2211.16725},
 primaryClass = {astro-ph.EP},
       adsurl = {https://ui.adsabs.harvard.edu/abs/2023ApJ...948...12G}
}

@ARTICLE{2023rosotti,
       author = {{Rosotti}, Giovanni P.},
        title = "{Empirical constraints on turbulence in proto-planetary discs}",
      journal = {\nar},
         year = 2023,
        month = jun,
       volume = {96},
          eid = {101674},
        pages = {101674},
          doi = {10.1016/j.newar.2023.101674},
archivePrefix = {arXiv},
       eprint = {2302.01433},
 primaryClass = {astro-ph.EP},
       adsurl = {https://ui.adsabs.harvard.edu/abs/2023NewAR..9601674R}
}

@BOOK{2007ferrazmello,
       author = {{Ferraz-Mello}, Sylvio},
        title = "{Canonical Perturbation Theories - Degenerate Systems and Resonance}",
         year = 2007,
       volume = {345},
          doi = {10.1007/978-0-387-38905-9},
       adsurl = {https://ui.adsabs.harvard.edu/abs/2007ASSL..345.....F}
}

@ARTICLE{1986henon,
       author = {{Henon}, M. and {Petit}, J.-M.},
        title = "{Series Expansions for Encounter-Type Solutions of Hill's Problem}",
      journal = {Celestial Mechanics},
         year = 1986,
        month = jan,
       volume = {38},
       number = {1},
        pages = {67-100},
          doi = {10.1007/BF01234287},
       adsurl = {https://ui.adsabs.harvard.edu/abs/1986CeMec..38...67H}
}

@ARTICLE{2012rein,
       author = {{Rein}, Hanno},
        title = "{Period ratios in multiplanetary systems discovered by Kepler are consistent with planet migration}",
      journal = {\mnras},
         year = 2012,
        month = nov,
       volume = {427},
       number = {1},
        pages = {L21-L24},
          doi = {10.1111/j.1745-3933.2012.01337.x},
archivePrefix = {arXiv},
       eprint = {1208.3583},
 primaryClass = {astro-ph.EP},
       adsurl = {https://ui.adsabs.harvard.edu/abs/2012MNRAS.427L..21R}
}

@ARTICLE{2011okuzumi,
       author = {{Okuzumi}, Satoshi and {Hirose}, Shigenobu},
        title = "{Modeling Magnetorotational Turbulence in Protoplanetary Disks with Dead Zones}",
      journal = {\apj},
         year = 2011,
        month = dec,
       volume = {742},
       number = {2},
          eid = {65},
        pages = {65},
          doi = {10.1088/0004-637X/742/2/65},
archivePrefix = {arXiv},
       eprint = {1108.4892},
 primaryClass = {astro-ph.EP},
       adsurl = {https://ui.adsabs.harvard.edu/abs/2011ApJ...742...65O}
}

@ARTICLE{2026moncrieff,
       author = {{Moncrieff}, Jordan W.~N. and {Grishin}, Evgeni and {Trani}, Alessandro A. and {Panther}, Fiona H. and {Pietrosanti}, Olga},
        title = "{Not all roads lead to merger: AGN disc properties influence the interactions of highly unequal mass black holes}",
      journal = {\mnras},
         year = 2026,
        month = jan,
       volume = {545},
       number = {3},
          eid = {staf2217},
        pages = {staf2217},
          doi = {10.1093/mnras/staf2217},
archivePrefix = {arXiv},
       eprint = {2511.09129},
 primaryClass = {astro-ph.HE},
       adsurl = {https://ui.adsabs.harvard.edu/abs/2026MNRAS.545f2217M}
}

@ARTICLE{2025trani,
       author = {{Trani}, Alessandro Alberto and {Di Cintio}, Pierfrancesco},
        title = "{Turbulent drag on stellar mass black holes embedded in disks of active galactic nuclei}",
      journal = {\aap},
         year = 2025,
        month = oct,
       volume = {703},
          eid = {A6},
        pages = {A6},
          doi = {10.1051/0004-6361/202555879},
archivePrefix = {arXiv},
       eprint = {2506.02173},
 primaryClass = {astro-ph.HE},
       adsurl = {https://ui.adsabs.harvard.edu/abs/2025A&A...703A...6T}
}
\bibliographystyle{aasjournalv7}

%% the collaboration and author truncation commands are used. 


\end{document}